\documentclass[12pt, a4paper]{article}

\usepackage[utf8]{inputenc}
\usepackage[english]{babel}

\usepackage[toc,page]{appendix}

\usepackage{amsmath} 
\usepackage{amssymb} 
\usepackage{amsthm} 
\usepackage{relsize,exscale} 
\usepackage{amsfonts}
\usepackage{mathrsfs}
\usepackage{dsfont}
\usepackage{enumerate}
\usepackage{interval}
\usepackage[shortlabels]{enumitem}
\usepackage{esint}
\usepackage{graphics}
\usepackage{caption}
\usepackage{subcaption}
\usepackage{mathtools}
\usepackage{multirow}
\usepackage{comment}
\usepackage{hhline}
\usepackage{array}

\DeclareFontFamily{U}{mathx}{\hyphenchar\font45}
\DeclareFontShape{U}{mathx}{m}{n}{
      <5> <6> <7> <8> <9> <10>
      <10.95> <12> <14.4> <17.28> <20.74> <24.88>
      mathx10
      }{}
\DeclareSymbolFont{mathx}{U}{mathx}{m}{n}
\DeclareFontSubstitution{U}{mathx}{m}{n}
\DeclareMathAccent{\widecheck}{0}{mathx}{"71}
\DeclareMathAccent{\wideparen}{0}{mathx}{"75}

\usepackage{hyperref}
\hypersetup{
    colorlinks,
    citecolor=black,
    filecolor=black,
    linkcolor=black,
    urlcolor=blue
}

\usepackage[font=small,labelfont=bf]{caption}

\newtheoremstyle{droit}
{}
{}
{\normalfont}
{}
{\boldface}
{}
{\newline}
{}
\newtheoremstyle{italique}
{}
{}
{\itshape}
{}
{\boldface}
{}
{\newline}
{}

\newtheorem*{thm*}{Theorem}
\newtheorem{thm}{Theorem}[section]
\newtheorem*{prop*}{Proposition}
\newtheorem{prop}[thm]{Proposition}
\newtheorem*{lem*}{Lemma}
\newtheorem{lem}[thm]{Lemma}
\newtheorem*{cor*}{Corollary}
\newtheorem{cor}[thm]{Corollary}

\theoremstyle{definition}
\newtheorem{ex}[thm]{Example}

\newtheorem{rema}[thm]{Remark}
\newtheorem{defi}[thm]{Definition}
\newtheorem*{defi*}{Definition}

{\begin{enumerate}[\textup{(}i\textup{)}] }
{\end{enumerate}}

\newcommand{\N}{\mathbb{N}}
\newcommand{\Z}{\mathbb{Z}}

\newcommand{\abv}{\mathbf{A}}
\newcommand{\bel}{\mathbf{B}}

\newcommand{\pws}[1]{\mathcal{P}\left( #1 \right)}

\newcommand{\face}[1]{\triangle\left(#1\right)}
\newcommand{\cface}[1]{\bigtriangledown\left(#1\right)}
\newcommand{\kface}[2]{\triangle_{#2}\left(#1\right)}
\newcommand{\kcface}[2]{\bigtriangledown_{#2}\left(#1\right)}
\newcommand{\rk}{\operatorname{rk}}
\newcommand{\Rk}{\operatorname{Rk}}

\newcommand{\E}{\mathcal{E}}
\newcommand{\V}{\mathcal{V}}

\newcommand{\dual}[1]{\overline{#1}}
\newcommand{\bual}[1]{\widetilde{#1}}
\newcommand{\bdual}[2]{\bual{#1}^{#2}}
\newcommand{\cdual}[2]{\overline{#1}^{#2}}

\newcommand{\all}{\dual{\emptyset}}
\newcommand{\sphi}{\rho}
\newcommand{\cphi}{\pi}
\newcommand{\red}{\mathrel{\raisebox{1pt}{$\prec$}}}
\newcommand{\der}{\mathrel{\raisebox{1pt}{$\succ$}}}
\newcommand{\col}{\vdash}
\newcommand{\loc}{\dashv}
\newcommand{\cpa}{\leftthreetimes}
\newcommand{\apc}{\rightthreetimes}
\newcommand{\refl}{\mathbin{\top}} 
\newcommand{\ccup}[1]{\stackrel{#1}{\cup}}

\newcommand{\bdiv}[1]{\widecheck{#1}}

\newcommand{\kz}{K^{[0]}}
\newcommand{\ko}{K^{[1]}}
\newcommand{\kt}{K^{[2]}}
\newcommand{\kg}{K^{(1)}}
\newcommand{\dg}[1]{\mathcal{G}_{\dual{#1}}}

\newcommand{\kr}{K^{[r]}}
\newcommand{\kR}{K^{[R]}}
\newcommand{\kb}{\dual{K}}

\newcommand{\vb}{\dual{v}}

\newcommand{\xb}{\dual{x}}

\newcommand{\kbb}{\dual{\dual{K}}}

\newcommand{\ce}[1]{\dual{#1}}

\newcommand{\nsc}{\mathsf{C}}
\newcommand{\cob}{\mathsf{Cob}}
\newcommand{\mlc}{\mathsf{B}}

\newcommand{\Hom}{\mathsf{Hom}}

\newcommand{\Cob}{\mathscr{C}}
\newcommand{\sts}{\mathcal{S}}
\newcommand{\ists}{\sts^{\top}}
\newcommand{\osts}{\sts^{\perp}}
\newcommand{\tim}{\mathtt{T}}
\newcommand{\parity}{\mathtt{P}}
\newcommand{\charge}{\mathtt{C}}
\newcommand{\rev}[1]{{#1}^{-1}}
\newcommand{\seqs}{\mathcal{Q}}
\newcommand{\Dom}{\mathfrak{D}}
\newcommand{\bra}[1]{ \langle #1 | }
\newcommand{\ket}[1]{ | #1 \rangle  }
\newcommand{\braket}[2]{\langle #1 | #2 \rangle}
\newcommand{\ktbr}[2]{\ket{#1}\bra{#2}}

\newcommand{\ins}{\leq_{\text{in}}}
\newcommand{\outs}{\leq_{\text{out}}}
\newcommand{\homc}{\Hom_{\Cob_d}}

\newcommand{\G}{\mathcal{G}}

\newcommand{\A}{\mathcal{A}}
\newcommand{\B}{\mathcal{B}}

\newcommand{\id}{\operatorname{id}}

\newcommand{\im}{\mathfrak{Im}}
\newcommand{\abs }[1]{ \left| #1 \right| }
\newcommand{\dans}{\longrightarrow}

\author{Maxime Savoy}

\title{Combinatorial Cobordism Theory}

\begin{document}

\maketitle

\begin{abstract}
We introduce a formalism based on a combinatorial notion of cell complex subject to an inclusion-reversing duality operation. Our main goal is to open the way for a functorial definition of field theories in a context where no manifold or topological structure is assumed. This is achieved via a discrete notion of cobordism for which a composition operation is defined. Our main theorem enables the composition of cobordisms by showing that certain sequences of maps between cell complexes are in bijective correspondence with a cell complex of dimension one higher. As a result we obtain a category whose morphisms are cobordisms having a causal structure generalizing that of Causal Dynamical Triangulations as well as dualities inherited from the duality map defined on cell complexes.
\end{abstract}



\pagebreak

\tableofcontents

\newpage
\section*{Introduction}

\addcontentsline{toc}{section}{Introduction}

Motivated by the physics of space-time at low scales, this paper proposes a novel framework of discrete geometry based on a combinatorial notion of \textit{cell complex.} By combinatorial we mean that a \textit{cell} of a cell complex simply corresponds to a subset of a given finite set of \textit{vertices}, and each cell is only assigned a rank (or dimension). The structure of cell complexes is then purely characterized using conditions on how cells are included in one another according to their rank and does not require to specify the topology associated to each cell.

Our main goal is to identify a minimal set of assumptions on cell complexes leading to a notion of Quantum Field Theory (QFT) taking values on the resulting discrete structure and defined via a functor between categories. Resting on ideas of Witten, similar functorial field theories were originally defined in the work of Atiyah \cite{ati88} for the case of Topological Quantum Field Theory, and Segal \cite{seg04} for the case of Conformal Field Theory and were also used in various other contexts. Such field theories usually correspond to a functor from a category of cobordisms (or bordisms, from which we make no distinction here) to the category of vector spaces (and this is as far as categorical notions will be used in the present work).

\begin{figure}[!h]
\centering
\includegraphics[scale=0.3]{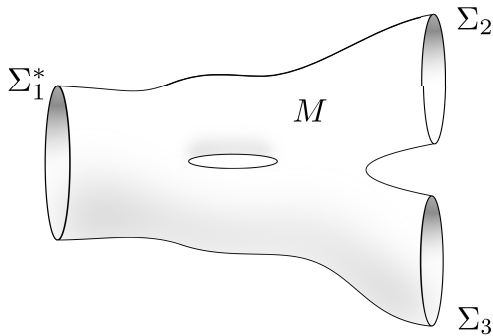}
\caption{\label{figtqft} A two-dimensional cobordism $M$ between an ingoing component $\Sigma_1^*$ and outgoing components $\Sigma_2$ and $\Sigma_3.$}
\end{figure}

Cobordisms were initially studied by René Thom (\cite{tho54}) as defining an equivalence relation on closed oriented manifolds of dimension $n,$ where $\Sigma \cong \Sigma'$  whenever there exists a smooth oriented compact manifold $M$ of dimension $n+1,$ a cobordism, such that $\partial M = \Sigma \sqcup \Sigma',$ as for example illustrated in Figure \ref{figtqft}. In our context a cobordism more generally stands for a geometry of a given dimension having a definite notion of boundary split in two types of components: ingoing and outgoing. A cobordism can then be interpreted as interpolating between an initial "state of geometry" associated with the ingoing boundary components to a finial state of geometry associated with the outgoing ones, and can be seen as a possible transition between the two states. The definition of field theory we aim for hence essentially requires a notion of cobordism that, seen as a mapping between states of geometry, satisfies the composition law of morphisms in a category.
This in turn requires an appropriate notion of state of geometry such that if a given state is both the final state of a cobordism $C_1$ and the initial state of a cobordism $C_2$ then the composition $C_2 \circ  C_1$ (i.e. the union $C_2 \cup C_1$) is a cobordism from the initial state of $C_1$ to the final state of $C_2.$

In the final section of this paper we introduce a category whose morphisms correspond to cobordisms of fixed finite dimension and whose structure is based on the notion of cell complexes introduced first. 
The cobordisms of this category are "causal" in a sense used also in the context of Causal Dynamical Triangulations (CDT), see e.g. \cite{lol20} for a  recent physical review or \cite{dj15} for a mathematical treatment. In simple terms, a causal structure here means that it is decomposed into "slices", where each slice can be compared to space at fixed time according to some choice of (discrete) time coordinate. The structure of causal cobordisms thus generalizes that of CDT by incorporating discrete geometries having more general cells than simplices, but it is also endowed with an additional duality operation we discuss in the next paragraph.

An important feature of causal cobordisms is that they are characterized using sequences of maps between cell complexes that we call geometric sequences. This characterization is based on the main result in this work, the Correspondence Theorem \ref{thmcorresp}, which establishes a bijection between certain geometric sequences and cell complexes we call slices, generalizing the notion of slice from CDT. The latter result (in its dual version) is key in defining a composition operation on causal cobordisms and finding an appropriate notion of state of geometry.

\paragraph*{A duality is built-in at the base of the formalism.}

\begin{figure}[!h]
\centering
\includegraphics[scale=0.35]{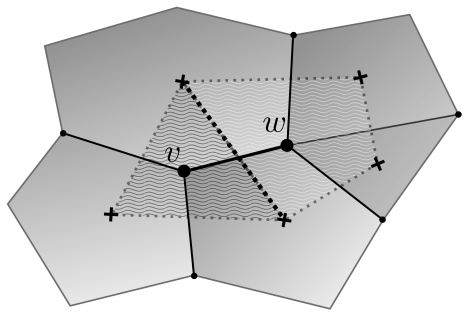}
\caption{\label{dcell} Dual cells in a cell complex of rank 2 having $v$ and $w$ as vertices.  The duals of maximal cells are singletons represented by crosses. The dual cell associated with the vertex $v$ is a set containing three crosses which intersects the dual cell associated with the vertex $w$ on two crosses constituting the cell dual to the edge $\{v,w\}.$ }
\end{figure}

For the sake of clarity, let us explicit a few elementary facts about the notion of cell complexes used here. As mentioned earlier, a cell is simply a subset of a finite set of vertices to which one assigns a non-negative integer called \textit{rank}. Cell complexes satisfy certain axioms implying in particular that vertices have rank 0 and the rank of a cell $x$ in fact corresponds to the number of cells (including $x$) contained in $x$ and containing any fixed vertex in $x.$ We will focus on cell complexes satisfying also a number of basic assumptions well motivated by applications in physics: edges, i.e. cells of rank 1, contain two vertices, and cells are connected, meaning that any two vertices in a cell can be connected by a path of edges. Cell complexes satisfying the latter conditions will be called \textit{local} if they moreover satisfy that their underlying graph is connected. We will also assume that all maximal cells in a cell complex, i.e. cells having no other cell containing it, have same rank defining the \textit{rank} or dimension of the cell complex.
The image of a cell $x$ under the \textit{duality map} is then simply defined as the set $\dual{x}$ of maximal cells containing $x.$ Figures \ref{dcell} shows examples of dual cells in a cell complex of dimension 2.

Using these minimal notions of cell and dual cell, we argue in the first section that these two collections of cells can be seen as equivalent encodings of a given discrete geometry. This point of view motivated our framework in which the duality map is a bijection on a wide class of cell complexes having empty boundary that we call \textit{closed} cell complexes and which play an important role in this work. Consequently, this defines an involution, i.e. the dual of the dual of closed cell complex is isomorphic to the primal cell complex. If a closed cell complex is obtained as the face lattice of some CW-decomposition of a manifold, then its image under the duality map corresponds to the face lattice of the dual CW-complex leading to Poincaré Duality, but our framework doesn't require such geometrical realization.
 
Not only does this duality emerge at the base of our formalism, but it also drove its development: as explained in Section \ref{SecCobs}, we were lead to consider a notion of cobordism by looking for a structure allowing to extend the duality map on closed cell complexes to local cell complexes with a non-empty boundary.
It also potentially represents an interesting leverage in order to make touch with more standard notions of geometry used in physics, as it allows to single out cell complexes that are self-dual - or locally so - a property typically satisfied by the square or hypercubic lattices widely used in statistical mechanics or Lattice QCD. These areas of research include many applications of the dual lattice or of more general discrete spaces to study certain transformations on fields defined on it, the simplest example being arguably the Kramers-Wannier duality of the two-dimensional Ising model. The latter duality has for example been treated as a generalized Fourier transform (\cite{mck64}) or linked with the electromagnetic duality (\cite{ft19}) and such connections might shed light on how the duality on cell complexes would couple with fields defined on causal cobordisms, however this is a topic for further research.


\paragraph*{On connections with lattices, quantum logic and other combinatorial notions of cell complexes.}

Our definition of cell complexes is a somewhat simpler and more explicit formulation of the definition  of combinatorial cell complexes introduced by Basak in \cite{bas10} (we chose to drop the term "combinatorial" for brevity). The main goal of Basak's work is to show a version of Poincaré duality for combinatorial cell complexes based on the duality operation discussed above, hence some of our basic definitions are the same. However he focuses on closed complexes satisfying certain orientability conditions and no such assumption will be needed in our context. An other definition of combinatorial cell complexes was introduced in \cite{asc96} using a rather general formalism primarily designed to discuss relations with manifolds and CW-complexes using assumptions on the topology associated to the cells of the complex.

In order for the duality map to define a bijection on closed cell complexes, it is imposed as an axiom that the intersection of two cells is a cell (or is empty). 
We close part \ref{secdualncc} by explaining how the latter axiom also leads to a close relation between cell complexes and lattices, and more specifically the notion orthocomplemented lattice. Such lattices were considered by Von Newman and Birkhoff \cite{bvn36} who initiated a line of research, called quantum logic, aiming to formalize the logical structure underlying possible measurements on a quantum system. Interestingly, following this connection with quantum logic (seeing cell complexes as analogous to orthocomplemented lattices) suggest the interpretation that simplicial complexes are analogous to Boolean lattices (or "locally Boolean"), and thus describe the subclass of quantum systems that behave classically. We see this as a motivation to study cell complexes to look for a natural notion of discrete "quantum geometry" encompassing more general structures than triangulations. The field of quantum logic was further developed in the sixties (see e.g. Mackey \cite{mac63} or Jauch \cite{jau68}) but has also faced a number of criticisms (mentioned for example in \cite{cg01}). We also remarked that the distinction between cell complexes and orthocomplemented lattices could potentially solve certain issues raised about the physical interpretation of the latter. These connections were however only discovered in a later stage of this research and would deserve further explorations.


\paragraph*{Why renouncing to a manifold structure while looking for applications to the physics of Quantum Gravity?}



In a nutshell, the search of a theory of Quantum Gravity in the path integral approach aims to give a precise meaning to the formal notion of integration over possible space-time geometries with given boundary data. The ordinarily large size of the space of possible geometries is a major source of difficulties, it is hence natural to regularize this problem by using discrete geometries.
One could also take the view that a fundamental description of space-time would require a discrete notion of geometry and that manifolds only model space-time at larger scales.

Regardless of the choice of point of view, in order to define an integration over discrete geometries two possible strategies stand out as relatively tractable: either fixing a discretization and integrate over possible metrics on it (in which case it is relatively hard to recover the symmetries of General Relativity) or sum over a class of discrete geometries each of which is associated a fixed metric. Following the second strategy, one faces the task of choosing an appropriate class of possible discrete geometries to sum over. It is often convenient or argued as sufficient to restrict oneself to triangulations, as in the case of Dynamical Triangulation or Tensor models (\cite{adj97}, \cite{gr12}).
The point is that the study of these models lead to consider sums over triangulations without constraint on their topology, but simply constructed from a collection of simplices of a given dimension $d$ by identifying faces of dimension $d-1.$ Such construction is encoded in the (colored) "dual graph", for which each vertex corresponds to a $d$-dimensional simplex and two vertices are linked by an edge whenever the corresponding simplices share a $(d-1)$-dimensional face. Seeing triangulations as duals to the graphs obtained as the Feynman diagrams associated to a certain field theory is the idea behing the approach of Tensor models (also called Group Field Theories, \cite{fre05}), which aim is to generalize the rigorous results obtained in dimension two with matrix models (\cite{fra04}). 
The geometries considered in these models hence emerge from purely combinatorial structures encoded in the terms of a Feynman expansion, therefore restricting to those having the structure of a manifold does not come as such a natural condition to impose (especially in dimensions higher than 2).

The cell complexes discussed in the present paper include triangulations (or more precisely the associated face lattice) and structures dual to that, which underlying graph corresponds to the generalized Feynman diagrams mentioned above. Hence our framework allows to consider a rather general notion of field prior to fixing or recovering a manifold structure, but it also puts triangulations and their duals on an equal footing, as well as many other types of discrete geometries such as lattices used in statistical mechanics or other polyhedral complexes. 






\paragraph*{This paper is self-contained, and organised as follows.} The first section is a step-by-step introduction to the basic notions underlying this framework and states the simplest version of the duality for closed cell complexes. We also present a novel combinatorial definition of a subdivision of a cell complex widely generalizing the common notion of barycentric subdivision. A subdivision is defined via a map called reduction and we also introduce maps called collapses that are seen as dual to reductions. 

Section \ref{secbdry} develops the machinery needed for the last section. The notion of uniformity is introduced, characterizing a large class of cell complexes whose boundary can be described using a sequence of a reduction and a collapse. Such sequence is at the base of the definition of geometric sequence that comes next and the main result of this paper, the Correspondence Theorem \ref{thmcorresp}, show that there is a bijection between certain geometric sequences and a class of uniform cell complexes called slices.

Section \ref{SecCobs} defines cobordisms and their duality, generalizing the duality obtained for closed cell complexes. It also uses geometric sequences and the Correspondence Theorem to define a composition operation on cobordisms. This leads to the definition of a category whose morphisms are cobordisms constructed using slices called causal cobordisms.

Certain proofs have been alleviated of some details which can be found in the authors thesis \cite{Sav22}.

\newpage
\section{Laying down the formalism} \label{SecNotions}

This work is based on a combinatorial notion of cell complexes, abbreviated cc, which is introduced in part \ref{secdefcc} of this section together with some simple examples. An other central protagonist of our formalism is the duality map introduced in part \ref{secdualncc} for the case of cc with empty boundary, and which would be extended to cc with boundary in Section \ref{SecCobs}. The part \ref{susecmidsection} then introduces an other important cc called midsection used throughout this article. We finish this first section by defining a notion of subdivision for cc, together with the operation dual to it, which is the cement for the construction of causal cobordisms, the end-goal of this work. Let us first start by clarifying some notation and terminology used in what follows.

\subsection{Notation}

In these notes $\N = \{0,1,2, \dots\}$ are the natural numbers and $\N^* = \{1,2,\dots \}.$ We usually use the letters $i,j,k,n,r$ to denote indices in $\N.$ We use $:=$ to denote an equality where the left hand side is defined by the right hand side and we use italic words whenever we \textit{define a notion for the first time.}

We use the notation $\abs{S}$ for the number of elements in a set $S.$  The set of subsets of $S$ is denoted by $\pws{S}$ and the disjoint union of the sets $S$ and $S'$ is denoted by $S \sqcup S'$  (used purely to emphasize that $S$ and $S'$ are disjoint). The set $S \setminus S'$ denotes the difference between $S$ and $S',$ i.e. the elements in $S$ that are not in $S'.$
A \textit{map} $f: A \dans B$ is a function with domain equal to $A$ and the composition of two maps $f:A \dans B,$ $g: B \dans C$ is denoted by $g \circ f.$

Let $(P, \leq)$ be a \textit{partially ordered set} or \textit{poset}, which means that $\leq$ is a reflexive antisymmetric transitive binary relation on the set $P.$ 
The set define the sets of cells respectively \textit{above} and \textit{below} a cell $x$ in $P:$
$$ \abv(x) := \{ y \in P ~|~ x \subset y \}, \quad \bel(x) := \{ y \in P ~|~ y \subset x \}. $$
One sometimes write the latter sets respectively $\abv_P(x)$ and $\bel_P(x)$ for in case of ambiguity on the poset $P.$
An element in $P$ is \textit{maximal} if it is not below any other element and \textit{minimal} if it is not above any other element. 
If $T \subset P$ an \textit{upper bound of $T$} is an element $a \in P$ such that $b \leq a$ for all $b \in T$ and a \textit{lower bound of $T$} is an element $a \in P$ such that $a \leq b$ for all $b \in T.$ If there is a unique upper bound of $T$ which is below all upper bounds of $T$ then this element is called the \textit{least upper bound of $T$} and is denoted by $\bigvee T$. Similarly if there is a unique lower bound of $T$ which is above all lower bounds of $T$ then this element is called the \textit{greatest lower bound of $T$} and is denoted by $\bigwedge T.$  If $T= \{a,b\},$ then we write $\bigvee T = a \vee b $ and $\bigwedge T = a \wedge b.$

\subsection{Cell Complexes (cc)} \label{secdefcc}

The motivation for considering the following definition of cc can be thought of as looking for a minimalistic notion of geometry that conserves enough structure to retain a meaningful notion of field, formalized axiomatically using cobordisms (or bordisms) as explained in the introduction. That is, one should in particular be able to identify what the boundary of a given geometry is, and this is done via a notion of dimension as a non-negative integer.

We thus start with the simplest possible structure, a finite set $S,$ which elements we call \textit{vertices}. Vertices can be though of as the fundamental units of the description of the geometry we are formalizing, whether considered complete or approximate. In order to build a complex out of these vertices, one then considers a family of \textit{cells}, i.e. non-empty subsets of vertices, with the assumption that every vertex, conventionally identified with the corresponding singleton, is a cell. The natural way to assign a rank (or dimension) to these cells is to endow them with the structure of a ranked (or graded) poset defined as follows.

\begin{defi}[Ranked poset, face, co-face]
Let $(P, \subset)$ be a poset together with a \textit{rank function} \mbox{$\rk = \rk_P: P \rightarrow \N$}.
We define the set of \textit{faces} of $x \in P$ by
$$ \face{x} := \{ y \in K ~|~ y \subset x, ~\rk(y) = \rk(x) -1\}$$
and the set of \textit{co-faces} of $x$ by 
$$ \cface{x} :=\{ y \in K ~|~ x \subset y, ~ \rk(y) = \rk(x) +1\}.$$ 
Then we say that $(P,\subset, \rk)$ is a \textit{ranked poset} if $\rk(x) = 0$ whenever $x$ is a minimal element and the following axioms are satisfied:
\begin{enumerate}[label=\textbf{\roman*}),topsep=1pt, parsep=2pt, itemsep=1pt]
\item \label{cccrank} $\rk$ is \textit{strictly compatible} with the partial order $\subset:$ for all $x,y \in P,$ if $x \subsetneq y$ then $\rk(x) < \rk(y),$
\item \label{cccenough} there is \textit{no gap} in the value of the rank function: for all $x,y \in P,$ if $x \subsetneq y$ then $x$ is a face of a cell contained in $y,$
\end{enumerate}
We denote by $A^{[r]}$ the set of $r$-cells in a subset $A$ of a ranked poset $(P, \rk).$
\end{defi}
 
In addition to the structure of ranked poset, a fundamental assumption we make on our collection of cells is that the intersection of two cells is also a cell or is empty. As explained later in part \ref{secdualncc}, this assumption is key in order to obtain a duality on posets whose elements are cells, i.e. defined as subsets of a finite set of vertices.

Finally the "diamond property", named after shape of the Hasse diagram of the four cells it involves, constitutes the last axiom of the definition of cc stated thereafter.  This property is also used in the context of abstract polytopes (\cite{ms02}) and allows to have a well behaved notion of boundary $\partial K$ of a cc $K$ (defined in point \ref{exbdry} of Example \ref{exampcc}) in the sense that $\partial( \partial K)= \emptyset$ holds for the class of cc satisfying the assumptions needed for our purposes.

\begin{defi}[Cell complex (cc)] \label{defccc}
Let $S$ be a finite set and let $K \subset \pws{S} \setminus \{\emptyset\} $ together with a rank function \mbox{$\rk = \rk_K: K \rightarrow \N$}. We call an element of $K$  a \textit{cell} and an element of $\kr$ an \textit{r-cell.} 
The data $(K, \rk)$  defines a \textit{cell complex,} or cc for short, if it is a ranked poset such that the two additional axioms are satisfied:
\begin{enumerate}[label=\textbf{\roman*}),topsep=1pt, parsep=2pt, itemsep=1pt]
\setcounter{enumi}{2}
\item \label{cccinter} $K \cup \{ \emptyset \}$ is \textit{closed under intersection}: for all $x,y \in K,$ $x \cap y \in K \cup \{\emptyset\}.$
\item \label{cccdiamond} $K$ satisfies the \textit{diamond property}: for all $x,y \in K,$ if $x \subset y$ and $\rk(x) = \rk(y) - 2$ then there are exactly two cells $z_1,z_2$ such that
$$ \cface{x} \cap \face{y} =\{z_1,z_2\}.$$ 
\end{enumerate}
\end{defi}

Let us now introduce some terminology related to cc (and more generally to finite ranked posets). The \textit{rank} $\Rk(K)$ of a cc $K$ is the maximum value among the ranks of its cells, and we sometimes call a $R$-cc a cc of rank $R.$ The \textit{$k$-skeleton} of a cc $K$ is defined as $$K^{(k)} := \bigcup_{r=0}^{k} \kr.$$ A cc for which all maximal cells have same rank is called \textit{pure.}
A cc $J$ is a \textit{sub-cc} of $K,$ noted $J \leq K,$ if $J \subset K$ and $\rk_J = \rk_K |_J.$

We called \textit{edges} the set the cells of rank $1$ in a cc. An edge in a cc necessarily contains more than two vertices as the definition of cc asserts that all singletons are vertices (i.e. have rank $0$), and we call \textit{graph-based} a cc for which all edges contain no more than two vertices.

\textbf{In what follows, every cc is pure and graph-based unless otherwise stated.}

\begin{ex} \label{exampcc}
We shall discuss a number of basic examples, among which simplicial complexes, graphs and the boundary of a cc.
\begin{enumerate}[label=\arabic*),leftmargin= 5pt, itemindent=20pt, topsep=0pt, parsep=0pt, itemsep=1pt]
\item \textbf{Simplicial complexes:} Let $S$ be a finite set. A collection $\Delta \subset \pws{S}$ is called a \textit{simplicial complex} if it is closed under taking the subset of a cell, i.e. if $ x \subset y \in \Delta$ then $x \in \Delta.$ 
An element $x \in \Delta$ is called a \textit{simplex} and we define the rank function of $\Delta$ by
$$ \rk_\Delta(x) = \abs{x} - 1.$$

It is a good warm up exercise to check that $(\Delta, \rk_\Delta)$ satisfies the four axioms of a cell complex. We could have also defined a simplicial complex to be a cc $K$ with rank function $\rk_K$ satisfying closeness under taking the subset of a cell. In this case, as a consequence of Axiom \ref{cccenough}, the rank function satisfies $\rk_K(x) = \abs{x} - 1.$

The \textit{face lattice} of a triangulation of a manifold, i.e. the poset defined as the set of simplices in the triangulation ordered by inclusion, is an example of an "abstract" simplicial complex as defined here. But the geometric realization of such simplicial complexes (defined e.g. in \cite{rs72}) does not in general have the topology of a manifold.
\item \textbf{Graphs:} A graph is defined as a simplicial complex of rank 1. We denote by $\V$ the set of vertices of a graph and $\E$ its set of edges.
In particular a graph $\G$ is a 1-cc where we use the notation $\V = \G^{[0]}$ and $\E = \G^{[1]}$ and there cannot be multiple edges between two vertices nor self-edge (linking a vertex with itself) in a graph. As expected, if a cc $K$ is graph-based then $\kg$ is a graph and we sometimes use the notation $\E$ for $\ko,$ the set of edges of $K,$ when there is no ambiguity on the cc $K.$

A graph is \textit{connected} if any two vertices can be linked by a \textit{path,} i.e. a sequence of edges such that two consecutive edges intersect on a vertex, called a crossed vertex. A path is \textit{simple} if it doesn't cross a vertex twice and a \textit{cycle} is a path starting from and ending at the same vertex. A simple cycle is called a \textit{loop.}
\item \label{exrestrict} Let $K$ be a cc and let $A \subset \kz$ (or sometimes $A \subset L^{[0]}$ for some cc $L$ containing $K.$) Then the \textit{restriction of $K$ to $A$} 
$$\bel(A) := \{ x \in K ~|~ x \subset A\}$$
is clearly a cc. In particular, for every cell $x$ in a cc $K,$ $\bel(x)$ is a cc of rank $\rk(x)$ having $x$ as its only maximal cell.
\item \label{exbdry} The \textit{boundary} of a cc $K$ is defined as the set of cells contained in a sub-maximal cell having only one maximal cell above itself. In other word the boundary of a $R$-cc $K$ is defined as $$\partial K := \bigcup_{\substack{y \in K^{[R-1]} \\ \abs{\cface{y}} = 1}} \bel(y).$$
One can easily check that $(\partial K, \rk_K|{\partial K})$ is a cc and it is explained in Remark \ref{rembdry} that under certain hypothesis we have $\partial ( \partial K) = \emptyset$ as a consequence of the diamond property \ref{cccdiamond}.
\end{enumerate}
\end{ex}

As we are primarily interested in considering cc in the context of physics, it is natural to restrict our focus on cc satisfying a certain notion of locality which is simply defined as follows.

\begin{defi}[Local cell complex (lcc)]
A cc $K$ is \textit{local} if it is \textit{connected}, that is to say $\kg$ is connected as a graph, and \textit{cell-connected}, i.e. $\bel(x)$ is connected for all $x \in K.$ Local cc is sometimes be abbreviated lcc.
\end{defi}

As we assumed cc to be graph-based, edges are constrained to be sets of two vertices, and as a consequence of the diamond property \ref{cccdiamond}, one also has that a 2-cell in a local cc necessarily satisfies that their faces form a loop. Indeed if $x$ is a 2-cell in a lcc $K,$ the latter axiom implies that for every vertex $v\in x$ we have
$$ \cface{v} \cap \face{x} =\{e,e'\}.$$
Iterating this argument on the vertices in $e$ and $e'$ let us grow a path made of faces of $x$ that has no other possibility than being simple and closing itself into a cycle.

Let us also introduce homomorphisms of cc. In part \ref{ssecredncol} we consider particular types of homomorphisms of cc that can be interpreted either as defining a subdivision or a collapse of a cc and which plays a central role in the later sections. 

\begin{defi}[cc-homomorphism, cc-isomorphism]
A map $\phi: K \dans L$ is called a homomorphism of cc, or \textit{cc-homomorphism}, if it is a poset homomorphism, i.e. $\phi(x) \subset \phi(y)$ whenever $x \subset y,$ such that for each cell in the image of $\phi$ there is a cell in its pre-image having same rank, i.e. $\phi^{-1}(y)^{[\rk(y)]} \neq \emptyset$ for all $y \in \phi(J).$ A \textit{cc-isomorphism} is a bijective cc-homomorphism $\phi$ such that $\phi^{-1}$ is a cc-homomorphism. We use the notation $K \cong L$ to indicate that $K$ and $J$ are cc-isomorphic.
\end{defi}

\begin{figure}[!h]
\centering
\includegraphics[scale=0.47]{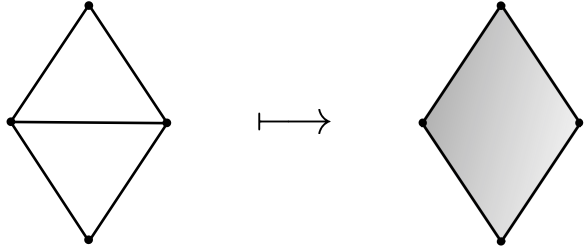}
\caption{\label{figexccchom} Example of a bijective poset homomorphism from a cc of rank 1 to a cc of rank 2 which is not a cc-homomorphism and such that the inverse is not a poset homomorphism. The outer edges from the 1-cc are "mapped to themselves" and the middle edge is mapped to the 2-cell.}
\end{figure}

Figure \ref{figexccchom} shows an non-example of a cc-homomorphism and illustrates the fact that a bijective poset homomorphism is not necessarily a cc-isomorphism. Lemma \ref{lemisom} is a simple consequence of Axioms \ref{cccrank} and \ref{cccenough} and gives two other ways to characterize a cc-isomorphism that will be of use later.

\begin{lem}\label{lemisom}
Let $K,J$ be cc and $\phi: J \dans K$ be a poset homomorphism. The following three statements are equivalent.
\begin{enumerate} [label=\textbf{\arabic*}),topsep=2pt, parsep=2pt, itemsep=1pt]
\item \label{lemiso1}$\phi$ is a cc-isomorphism.
\item \label{lemiso2}$\phi$ is bijective and $\phi(x) \in K^{[\rk_J(x)]}$ for all $x \in J.$
\item \label{lemiso3}$\phi$ is bijective and $\phi^{-1}$ is a poset homomorphism.
\end{enumerate}
\end{lem}

\subsection{Duality map for closed cc} \label{secdualncc}

The dual of a poset $P$ is a standard notion defined by $P^* = \{p^* | p \in P\}$ and $p^* \leq_{P^*} q^*$ if and only if $q \leq_P p.$ It provides an \textit{anti-isomorphism} of posets, i.e. a bijection reversing the partial order, which is therefore \textit{involutive:} $(P^*)^*$ is isomorphic to $P.$  Our goal is to restrict this duality to the set of cc, which in particular includes the additional assumption that cells are defined as subsets of a set of vertices, i.e. are characterized by the minimal elements it contains. Therefore it should also be the case for dual cells, meaning that dual cells can be defined as subsets of maximal cells of the "primal" cc \footnote{Hence it is more than assuming the poset to be an inclusion order, i.e. defined as the subset-inclusion relation on a set of objects, as every poset is isomorphic to an inclusion order. The condition used here to restrict the duality from posets to cc can be formulated using the notion of 2-dimension of a poset i.e. the size of the smallest set of objects such that the poset is an inclusion order on these objects. One then considers duals of posets $P$ such that both the 2-dimension of $P$ is equal to the number of minimal elements in $P$ and the 2-dimension of the dual $P^*$ is equal to the number of maximal elements in $P.$}. As we assume cc to be pure, all maximal cells have the same rank equal to the rank of the cc, hence we can use the following explicit definition of the dual of a cell, or more generally of any set of vertices.

\begin{defi}[Dual set] \label{defdualset}
The \textit{dual set} of a subset $A$ of vertices in a cc $K$ of rank $R$ is defined by
$$\dual{A} := \{ z \in K^{[R]} ~|~ A \subset z \}.$$
We sometimes use the notation $\cdual{A}{K}$ for $\ce{A}$ to specify the cc relative to which the dual set is defined. 
\end{defi}

We explain in details in part \ref{subsecbdiv} how this definition of dual cell differs from a more standard definition used in the context of simplicial complexes. From the latter definition, one is naturally lead to introduce the following poset.

\begin{defi}[Dual of a cc] \label{defdualccc}
We define the \textit{dual} of a cc $K$ to be the poset $\left( \kb, \subset \right)$ by 
\begin{equation*}
\kb := \{ \ce{x} ~|~ x \in K \}.
\end{equation*}
In this case $K$ is sometimes called \textit{primal} with respect to $\kb.$
\end{defi}

In other words, $\kb$ is the image of $K$ through the \textit{duality map}
$$ x \longmapsto \dual{x}.$$
We will more generally use the notation $\dual{S}:= \{\dual{x} ~|~ x \in S\}$ for any subset $S$ of cells in $K.$ 

This duality map can also be seen as a natural operation to define on cc from the following simple observation.  
Specifying a cell of a cc consists in differentiating between vertices contained and those not contained in it, therefore both sets of vertices in way encode the same data. Looking at the complement of a cell $x$ in the set of vertices clearly represents a shift from a local to a non-local description, however restricting our view to vertices outside $x$ but only "up to one maximal cell away" from $x,$ i.e. such that there exists a maximal cell containing both $x$ and a vertex not in $x,$ can be considered as a local analogue of the notion of complement. In fact this amounts to specifying the maximal cells containing $x,$ which indeed is our the definition of the dual cell $\dual{x}.$

As will be shown at the end of this part, the notion of closed cc introduced next in fact constitutes a first set of assumptions defining a structure on cc invariant under the duality map.

\begin{defi}[Non-singular and closed cc]
A cc is \textit{non-singular} if every sub-maximal cell is contained in at most two maximal cells and a cc is \textit{closed} if it is non-singular and has empty boundary, i.e. every sub-maximal cell is contained in exactly two maximal cells.
\end{defi}

It's should be intuitive that, in order for the dual of a cc to be a cc, the primal cc has to be non-singular, as this assumption is dual to that of graph-based cc. One can also easily see that if a sub-maximal cell $y$ is included in only one maximal cell $z$ then $\dual{y} = \dual{z}$ which also spoils the injectivity of the duality map. In fact introducing the notion of cobordism in Section \ref{SecCobs} will constitute a way around this fact, and will allow to define a duality map over a class of cc with non-empty boundary.

The fact that cells of a cc are characterized by the vertices it contains in addition implies that every cell $x$ having two or more faces, i.e. of rank 1 or higher, is the smallest upper bound of its faces, that is to say
\begin{equation}\label{equbigwedge}
x = \bigvee \face{x} = \bigcup_{w \in \face{w}} y.
\end{equation}
The identity (\ref{equbigwedge}) follows from observation that for all $v \in x$ there is a face $w$ of $x$ containing $v,$ hence $x \subset y$ for all upper bound $y$ of $\face{x},$ i.e. $x$ is indeed the smallest upper bound of $\face{x}.$ It isn't necessarily verified in more general posets, which explains why the definition of combinatorial cell complexes from \cite{bas10} includes (\ref{equbigwedge}) as an axiom in the definition of combinatorial cell complexes.

A closed cc, having empty boundary, satisfies that all non-maximal cells have at least two cofaces, i.e. the condition dual to that satisfied in a cc that all cells of rank 1 or higher have two or more faces. This yields the identity dual to (\ref{equbigwedge}) (shown in Lemma \ref{lemcface}), for $x$ a cell in a closed cc:
\begin{equation*}
x = \bigwedge \cface{x} = \bigcap_{y \in \cface{x}} y.
\end{equation*}

It is then seen that the bijectivity of the duality map is achieved under the assumption of closeness.

\begin{lem}\label{claim2propdual}
If $K$ is a closed cc then the duality map $K \dans \kb$ is an anti-isomorphism of posets, i.e. the duality map is bijective and for all $x,y \in K$ we have
$$x \subsetneq y \quad \text{ if and only if } \quad \ce{y} \subsetneq \ce{x}.$$

\begin{proof}
It is a simple consequence of the following statement: if $x,y \in K$ are such that there exists $v \in y \setminus x$ then $\ce{x}$ contains a maximal cell which does not contain $v,$ meaning $\ce{x} \not\subset \ce{y}.$ This can be shown by induction on $n := R - \rk(x)$ and using the fact that $ x = \bigwedge \cface{x}$ by Lemma \ref{lemcface}.
\end{proof}
\end{lem}

We can therefore introduce the \textit{dual rank function} $\rk_{\kb} : \kb \dans \N$ defined by  
$$\rk_{\kb}(\xb) := \Rk(K) - \rk_K(x)$$
well defined by the previous lemma and such that $\rk_{\kb}(\dual{z}) = 0$ for all maximal cells $z \in K.$

Moreover $(\kb, \rk_{\kb})$ is a ranked poset as the strict compatibility \ref{cccrank} and no gap property \ref{cccenough} are easily derived from the same axioms for $(K, \rk_K):$ the first is clear from Lemma \ref{claim2propdual} and the second follows from the observation that $\dual{x}$ is a face of $\dual{y}$ if and only if $y$ is a face of $x,$ and the (recursive) application of the statement of \ref{cccenough} for the cells in $(K, \rk_K).$

\begin{figure}[!h]
\centering
\includegraphics[scale=0.47]{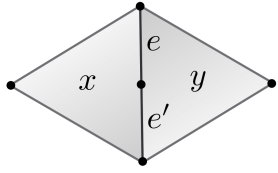}
\caption{\label{figcccinter} This is an example of two cells $x,y$ of rank 2 with four faces whose intersection is the union of two edges $e$ and $e'.$ The dual sets associated to $e$ and $e'$ as defined for a cc of rank 2  both equal to the two-set $\{x,y\}.$ It therefore spoils the injectivity of the dual map and explains the inclusion of axiom \ref{cccinter} in the definition of cc. }
\end{figure}

One can then easily point out why Axiom \ref{cccinter}, ensuring that the intersection of two cells is a cell (or is empty), is required in order for the duality map to be bijective. This can be seen using an simle example illustrated in Figure \ref{figcccinter} where $x,y$ are maximal cells of rank 2 intersecting on two of their faces. 
Note that this condition is not necessarily imposed in other notions of cell complexes (for example the notion of "ball complex" defined in \cite{rs72}). 

The next result is the first exhibiting a class of cc on which the duality map defines an involution. We refine this result in the next part to include locality of cc and it is generalized in Section \ref{SecCobs} to include cc with non-empty boundary.

\begin{prop} \label{propdual}
If $K$ is a closed $R$-cc then so is $\kb$ and $$\kbb \cong K.$$
\begin{proof}
As discussed before $(\kb, \rk_{\kb})$ is a ranked poset, and Axiom \ref{cccinter} for $\kb$ is a consequence of the first identity in (\ref{equduality}).  Moreover the diamond property \ref{cccdiamond} follows from the observation that by Lemma \ref{claim2propdual} we have $\dual{\cface{x}} = \face{\dual{x}}$ for all cell $x$ of in a closed cc and this shows that $\kb$ is a cc.

$\kb$ is also pure and satisfies $\Rk(\kb) = \Rk(K)$ for the following reasons: the maximal cells of $\kb$ of rank $R$ are the images of vertices of $K$ under the duality map and since every cell of $K$ contains at least one vertex, we have that every cell of $\kb$ is contained in $\vb$ for some $v \in K.$ Since $K$ is graph-based, Lemma \ref{claim2propdual} implies that every sub-maximal cell in $\kb$ is contained in exactly two maximal cells, hence $\kb$ is non-singular and has empty boundary.

By Lemma \ref{claim2propdual}, the map $x \mapsto \dual{\dual{x}}$ is a poset isomorphism from $K$ to $\kbb,$ which defines a cc-isomorphism by Lemma \ref{lemisom} since $\rk_{\kbb}(\ce{\ce{x}}) = \rk_K(x).$ 
\end{proof}
\end{prop}

Let us present a few examples of dual cc.

\begin{ex} \label{exampclosedcc}
We first look at the duals of simplicial complexes and also discuss the case of the square lattice.
\begin{enumerate}[label=\arabic*),leftmargin= 5pt, itemindent=20pt, topsep=0pt, parsep=0pt, itemsep=1pt]
\item \textbf{Simple cc:} A closed cc $K$ is said to be \textit{simple} if $\kb$ is simplicial. In dimension two, a closed cc is simple if and only if every vertex is included in 3 edges, which by Axiom \ref{cccinter} implies that every pair of edges containing a vertex is contained in a 2-cell. One can show that the boundary of a tetrahedra (i.e. a simplicial cc of rank 3 with one maximal cell) is the only example of a cc which is both simple and simplicial, and it is also self-dual.
\item \label{exampsqlattice} \textbf{Square lattice:} Consider a portion of the (face lattice of) the square lattice $\Z^2$ of size $4 \times 5,$ seen as a cc $L$ of rank 2 with 2-cells corresponding to the 12 unit squares it contains. We can turn $L$ into a closed cc $K$ by identifying opposite parts of the boundary and obtain the cc pictured in Figure \ref{figlattice}. This is also an example of self-dual cc, that is we have $\kb \cong K.$

\begin{figure}[!h]
\centering
\includegraphics[scale=0.47]{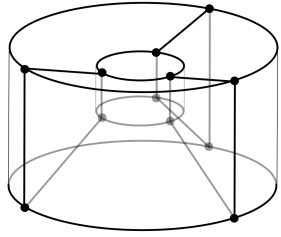}
\caption{\label{figlattice} The cc $K$ illustrated here can be seen as the boundary of a 3-cc having one 3-cell with the shape of a torus.}
\end{figure}
\end{enumerate}
\end{ex}

We close this part by briefly mentioning connections to lattices and quantum logic.
An important consequence of Axiom \ref{cccinter} is that the \textit{smallest upper bound} or \textit{join} $x \vee y$ of two cells $x,y$ in a cc $K$ such that there exists a cell containing both $x$ and $y$ exists: if both $w$ and $w'$ are minimal elements in $\abv(x \cup y)$ then $w \cap w' \in \abv(x \cup y)$ and cannot be strictly below $w$ and $w'$ hence $w =w'.$ We then have that a maximal cell of $K$ contains $x \vee y$ if and only if it contains both $x$ and $y.$ Applying a similar reasoning for the case $x \cap y \neq \emptyset$ we get the identities
\begin{equation}\label{equduality}
\dual{x} \cap \dual{y} = \dual{x \vee y}, \quad \dual{x} \vee \dual{y} = \dual{x \cap y}.
\end{equation}
If no cell contains both $x$ and $y,$ then, equivalently, no maximal cell contains the two cells and therefore $\dual{x} \cap \dual{y} = \emptyset.$ It is thus convenient to introduce a symbol $\all$ such that $x \subset \all$ for all $x \in K,$ and
$$x \vee y = \all \quad \Leftrightarrow \quad \dual{x} \cap \dual{y} = \emptyset.$$
As a result, the poset $L := (K \cup \{\emptyset, \all \}, \subset)$ has the structure of a (ranked) \textit{lattice}, where $\cap = \wedge$ defines the meet operation and $\vee$ defines the join operation. The dual lattice then corresponds to $\dual{L} := (\kb \cup \{ \emptyset, \all\})$ in which $\emptyset$ and $\all$ respectively contains and is contained in every element and in which the relations (\ref{equduality}) hold.

The relation between the lattice structure and formal logic is made by seeing the inclusion order as the partial order induced by implications between propositions and where the $\wedge$ corresponds to the conjunction "AND" and $\vee$ to the disjunction "OR".
The relations (\ref{equduality}) are then comparable to DeMorgan's laws, where the duality operation plays the role of a negation operation 
but with an important twist: the meet/AND and join/OR operations are \textit{not} defined between a cell in $L$ and a cell in $\dual{L}.$

A similar lattice structure interpreted as a formal logic is at the base of quantum logic (\cite{bvn36}). Although interestingly, the lattice of quantum logic does not emerge as a structure associated to the geometry of space-time, but merely from the relations among propositions associated to a physical system that does not depend upon the state of the system, in other words, its kinematic structure (see \cite{jau68}). In summary, propositions are identified with subspaces of a Hilbert space and taking the orthogonal complement of a sub-space is analogous to the negation operation. This yields a lattice structure with a idempotent automorphism called an orthocomplemented lattice.

The lack of distributivity of $\vee$ over $\wedge$ and vice-versa (and also its non-modularity) in such general lattices leads to difficulties in truly interpreting it as describing a logical structure, unlike the case of Boolean lattices. As mentioned in \cite{cg01}, an other criticism comes from the fact that closeness under intersection (or conjunction) of the lattice of quantum logic is too strong in that it implies non-physical statements when seen as propositions associated to a quantum system. The structure of primal and dual lattices we see in our case could precisely represent a weakening of the structure of orthocomplemented lattice preventing such incoherences. Seeing the negation or complement operation as mapping to a different space for example corroborates a more recent discussion on the relations between physics, topology, logic and computations \cite{bs10} where, when formulated in the language of category, the negation is seen as a functor to the opposite category. 

On a more conceptual level, the latter discussion would suggest to see the relation between primal and dual lattices from our formalism as akin to the relation between conjugated variables in quantum mechanics, where the Born duality is defined via a Fourier transform which switches between complementary representations of a quantum system. This famously yields the Heisenberg Uncertainty Principle, implying for example that both the position and momentum representations cannot simultaneously and exactly describe a quantum system. 

\subsection{Midsections and pinches} \label{susecmidsection}

In the previous parts we saw several examples of cc constructed from the cells of an other cc: the restriction of Example \ref{exampcc} \ref{exrestrict}, the boundary of Example \ref{exampcc} \ref{exbdry} and the duality map. We will now see yet an other cc, called midsection, constructed from the cells intersecting but not included in a given cell or a given sub-complex of a cc.
Midsections are used widely in the next sections where it will become apparent that it is generalization of an existing notion of midsection used in the context of CDT (see e.g. \cite{dj15}).

Let us start by introducing collars, named after an notion used for example in the context of polyhedra (c.f. \cite{rs72}), but defined here purely in terms of finite sets.

\begin{defi}[Collar] \label{defcollar}
Let $A$ be a finite set and $\A$ be a collection of subsets of $A.$ If $B \subset A$ we define the \textit{collar of $B$ in $\A,$} by
$$ \A_B := \{ x \in \A ~|~ \emptyset \neq x \cap B \neq x \}.$$
Moreover, if $C \in \A,$ we use the following definition and notation for the collar of $B$ in $\A$ restricted to $C:$
$$ \A_B^C := \{ x \in \A_B ~|~ x \subset C \}.$$
By abuse of notation, in case $\B$ is also a collection of subsets of $A,$ the notation $\A_\B$ is used to denote $\A_{B},$ where $B := \cup_{x \in \B} x$ is the set of elements of $A$ contained in a set in $\B.$
\end{defi}

For instance if $K \in \nsc$ and $J \leq K$ we have $$K_J =  \{ x \in K ~|~ \emptyset \neq x \cap J^{[0]} \neq x \}.$$ 
Since we use the notation $\E$ for the set of edges in $K$ when there is no ambiguity on what the cell complex $K$ is, we often use the following notation for $x \in K:$ $$\E_J^x = \{ e \in \ko ~|~ \emptyset \neq e \cap J^{[0]} \neq e, \quad e \subset x \}.$$

The key fact that allows us to construct cells from the collar of a given subset of cells is the following (Lemma \ref{lemmidsec}): if $K$ be a local cell complex and $J \leq K$ then the map
\begin{equation*}
x \longmapsto \E_J^x, \quad x \in K_J
\end{equation*} 
is an injective poset homomorphism such that the inverse is a poset homomorphism.
The latter indeed ensures that the midsection introduced in the following definition is a ranked poset. 

\begin{defi}[Midsection]
Let $K$ be a local cell complex and $J \leq K.$ We define the \textit{midsection of $J$ in $K$} to be $(M_J^K, \rk_{M_J^K})$ where
$$M_J^K := \{\E_J^x  ~|~ x \in K_J \} \quad \text{and} \quad \rk_{M_J^K}(\E_J^x) = \rk_K(x) - 1.$$
\end{defi}

\begin{figure}[!h]
\centering
\includegraphics[scale=0.42]{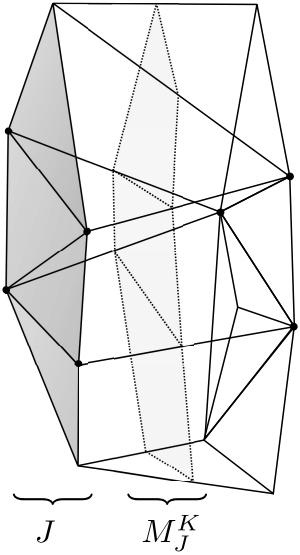}
\caption{\label{figmidsec} Midsection $M_J^K$ associated to the boundary part $J$ of a 3-cc $K$ composed of three maximal cells (the center one having dots on its vertices).}
\end{figure}

An example of midsection is illustrated in Figure \ref{figmidsec}, and the next proposition shows that the midsection associated with a sub-complex is a cc under rather general assumptions.

\begin{prop} \label{propmidsec}
Let $K$ be a local cell complex and $J \lneq K$ such that $x \not \subset J^{[0]}$ for all $x \in K \setminus J$ and such that $K_J \neq \emptyset.$ Then $(M_J^K, \rk_{M_J^K})$ is a cc and if $K$ is pure then so is $M_J^K$ and $\Rk(M_J^K) = \Rk(K)-1.$ Moreover if $J \cap x$ is connected for all $x \in \kt_J$ then $M_J^K$ is graph-based.
\begin{proof}
Since $K$ is cell-connected, $J \neq K,$ and by the condition $x \not \subset J^{[0]}$ for all $x \in K \setminus J,$ there is at least one edge of $K$ having one vertex in $J^{[0]}$ and one vertex in $\kz \setminus J^{[0]},$ therefore $K_J \neq \emptyset.$

By definition of $\E_J^x$ and $\rk_{M_J^K},$ the elements in $M_J^K$ of rank $0,$ i.e. vertices, are indeed 1-element sets and every set $\E_J^x$ is indeed a set of vertices of $M_J^K.$ By the fact that $x,y \in K,$ if $x \in K_J$ and $x \subset y$ then $y \in K_J,$ the axioms of cc for $M_J^K$ follow as a direct consequence of Lemma \ref{lemmidsec} and the axioms for $K.$

Let $R = \Rk(K).$ By purity of $K,$ every cell in $x \in K_J$ is contained in a $R$-cell of $K,$ which must then belong to $K_J$ and constitute an $R-1$ cell in $M_J^K$ containing $\E_J^x,$ showing that $M_J^K$ is also pure.

As mentioned when introducing local cc, we have that $\bel(x)$ is a cycle for every 2-cell $x$ of $K.$ Therefore, for every edge $\E^x_J$ of $M_J^K,$ i.e. for every $x \in \kt_J,$ if $\bel(x)$ is connected then there are exactly two edges in $\ko_J$ included in $\bel(x),$ and therefore there are exactly two vertices of $M_J^K$ in $\E_J^x.$ 
\end{proof}
\end{prop}

From the previous result, one has that the midsection $M_{\bel(x)}^K$ associated with the restriction of a local cc $K$ to a non-maximal cell $x$  is a graph based cc, that we call the \textit{local figure} of $x$ in $K$ (in analogy with the notion of vertex figure defined in the context of polytopes). Indeed the condition that every cell in $K \setminus \bel(x)$ cannot have all its vertices in $x$ is trivially verified and the condition that $\bel(x) \cap y$ is connected for all $y \in \kt$ is a direct consequence of Axiom \ref{cccinter}. 

Let us observe that in a non-singular cc $K$ the cells in $K \setminus \partial K$ satisfy that every non-maximal cell has 2 or more co-faces. We can therefore use the same arguments as in the proof of Lemma \ref{claim2propdual} to show that the duality map is bijective on $K \setminus \partial K$ and see that $(\dual{K \setminus \partial K}, \rk_{\kb}|_{\dual{K \setminus \partial K}})$ defines a ranked poset. The non-singularity of $K$ also guarantees that the 1-skeleton of $\dual{K \setminus \partial K}$ is a graph. We can thus introduce the following notion, dual to that of connectedness of a cc.

\begin{defi}
A cc $K$ is \textit{strongly connected} if $\dual{K \setminus \partial K}$ is connected.
\end{defi}

And the following notion of pinch allows to define the notion dual to that of cell-connectedness.

\begin{defi}[Pinch and non-pinching cc]
A \textit{pinch}, or \textit{$K$-pinch,} in a local cc $K$ is a cell $x \in K$ whose local figure is \textit{not} strongly-connected. $K$ is said to be \textit{non-pinching} if it contains no $K$-pinch and no $\partial K$-pinch.
\end{defi}

\begin{figure}[h!]
\centering
\includegraphics[scale=0.55]{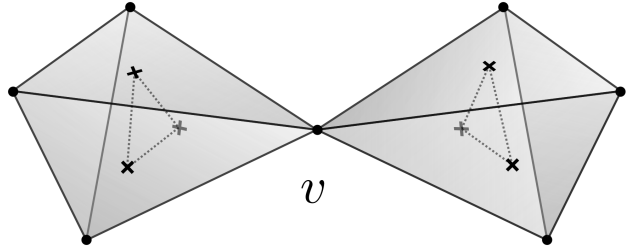}
\caption{\label{pinchpic} There are three different ways to look at this illustration. First, one can see it as a portion of a closed 2-cell complex $K$ and the central vertex $v$ gives an example of a $K$-pinch. The dual local figure $\dual{M_{\bel(v)}^K}$ (which is a closed cc in this example) is then constituted of two disconnected simplices of rank 2 (i.e. triangles) whose vertices correspond to the 2-cells of $K$ containing $v.$ Second, one could look at this illustration as a portion of a non-singular 3-cell complex $L,$ in which $v$ is included in two simplices of rank 3 (i.e. tetrahedra) considered as two different 3-cells of $L.$ In this case $v$ is both a $L$-pinch and a $\partial L$-pinch. The third way to look at this illustration is to consider that the two tetrahedra of $L$ containing $v$ form a single cell of a non-singular 3-cell complex $L'.$ In this case $v$ is a $\partial L'$-pinch, but not a $L'$-pinch.}
\end{figure}

Figure \ref{pinchpic} illustrates a simple example of a pinch. As explained in the caption, this illustration can also help understand the reason for adding in our definition of a non-pinching cc $K$ the condition that $K$ also do not contain any $\partial K$-pinch.

The next remark explains why the notion of non-pinching cc constitutes a class of cc well suited for our purposes.

\begin{rema} \label{rembdry}
Let us note the two following facts about non-pinching cc.
\begin{enumerate}[label=\arabic*),leftmargin= 5pt, itemindent=20pt, topsep=0pt, parsep=0pt, itemsep=1pt]
\item It follows from the observation that a maximal cell belongs to the local figure of all its vertices that if a closed cc is connected and non-pinching then it is strongly connected. 
\item One can show (Lemma \ref{rembdrycc}) that if $K$ is a non-pinching cc then $\partial K$ is closed. This means in particular that in this case $\partial( \partial K ) = \emptyset.$
\end{enumerate}
\end{rema}


Let us finally introduce sets of cc we often use in the remaining of this work, one reason being, as mentioned afterwards, that the duality we showed in the previous part naturally restricts to this class of cc.

\begin{defi}[$\nsc$ and $\mlc$]
We define $\nsc$ to be the set of non-singular non-pinching local (hence strongly connected by Remark \ref{rembdry}) cc and $\nsc^R$ for the set of elements in $\nsc$ of rank $R.$ 

We define $\mlc$ to be the set of closed elements in $\nsc$ and similarly $\mlc^R$ is the set of elements in $\mlc$ of rank $R.$

We use the notation $\mlc_\sqcup$ (respectively $\mlc^R_\sqcup$) to denote the set of (possibly empty) disjoint unions of elements in $\mlc$ (respectively $\mlc^R$).
\end{defi}

As a consequence, the boundary of an element in $\nsc^R$ is in $\mlc_\sqcup^{R-1},$ and conversely one can show (Lemma \ref{lemconcomp}) that a closed $(R-1)$-sub-cc of the boundary of an element in $\nsc^R$ is a dijoint union of connected component of its boundary. We finish with the following corollary:

\begin{cor}\label{cordualityclosedcc}
If $K \in \mlc^R$ then so is $\kb.$
\begin{proof}
If $x \in K$ then the 1-skeleton of the cc $ \bel_{\kb}(\dual{x})$ is connected if and only if the local figure of $x$ is strongly connected, hence a closed cc is non-pinching if and only if $\kb$ is cell connected. By remark \ref{rembdry} $K$ is strongly connected, hence $\kb$ is connected.
\end{proof}
\end{cor}

\subsection{Subdividing and collapsing closed cc} \label{ssecredncol}

The goal of this part is to give meaning to the operation of subdividing a cc, i.e. a purely combinatorial structure not defined from a discretization of a continuous space like a manifold where the topology of the latter space put constraints on what a valid subdivision can be. We start by giving an abstract definition of barycentric subdivision to give a foretaste of how such an operation on purely combinatorial structure can be defined. The more general definition of a subdivision of a closed cc is given in part \ref{sssecredncol} via a map called reduction. The map dual to that of a reduction will be introduced as a collapse and these two maps are in fact cc-homomorphisms that have just the right properties to describe the structure near the boundary of a large class of cc, called uniform cc, as shown in Section \ref{secbdry}. 

\subsubsection{Barycentric subdivision and dual cells} \label{subsecbdiv}

Let us start by 
recalling that a \textit{totally ordered subset} of a poset $P$ is a sub-poset $T$ of $P$ such that for all $x,y \in T,$ either $x \leq y$ or $y \leq x.$ 
The barycentric subdivision of a poset (and by extension of a any cc), also called order poset in some contexts, is then simply defined as follows.

\begin{defi}[Barycentric subdivision] \label{defbdiv}
Let $(P, \subset)$ be a poset. The \textit{barycentric subdivision} of $P$ is the poset, noted $\bdiv{P},$ of totally ordered subsets of $P.$ 
\end{defi}

\begin{figure}[!h]
\centering
\includegraphics[scale=0.45]{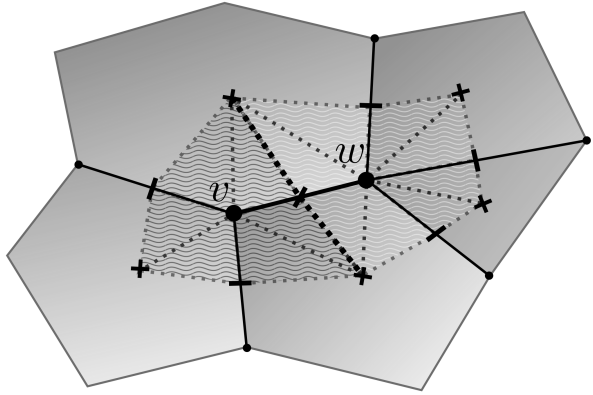}
\caption{\label{dualcellpic}This illustration represents certain cells of the barycentric subdivision $\bdiv{K}$ of a 2-cc $K$ and gives three examples of cells in $\kb^B.$ Being a simplicial complex, each 2-cell of the barycentric subdivision is a triangle and the vertices of the barycentric subdivision are represented using three different symbols depending of the rank of the cells it represent: a vertex of $\bdiv{K}$ of the form $\{C\}$ where $C \in \kt$ is represented by a cross, a vertex of the form $\{e'\}$ where $e' \in \ko$ is represented by a dash and a vertex of the form $\{v\}$ simply corresponds to a vertex of $K.$ There are two 2-cells of $\kb^B,$ the first $\dual{v}^B$ is filled the black wavy pattern, the other $\dual{w}^B$ is filled with the white wavy pattern. We also have the 1-cell $\dual{e}^B,$ the dual of $e = \{v,w\},$ represented by the two thick dashed lines. As a comparison, the dual cell $\dual{v}$ and $\dual{w}$ used in our framework are illustrated in Figure \ref{dcell}.}
\end{figure}

We give simple examples of cells in the barycentric subdivision of a 2-cc in Figure \ref{dualcellpic}. 
For the case of a simplicial complex obtained from a triangulations of a manifold, this definition corresponds to the face lattice of the (first derived) barycentric subdivision of the triangulation as defined for example in \cite{bry02}.
The barycentric subdivision of a cc is a simplicial complex (and therefore also a cc) a subset of a totally ordered set is also totally ordered.

As a direct consequence of Lemma \ref{claim2propdual}, the barycentric subdivision is isomorphic to the dual barycentric subdivision via the cc-isomorphism
\begin{equation} \label{mappropdual}
\bdiv{K} \ni \{x_1 \subsetneq \dots \subsetneq x_k\} \longmapsto \{\ce{x_k} \subsetneq \dots \subsetneq \ce{x_1}\} \in \bdiv{\kb}.
\end{equation}

The barycentric subdivision is sometimes used to define a notion of dual cell, as for example in \cite{bry02}. This definition can also be defined in a purely combinatorial manner, but it corresponds to different cells than those from Definition \ref{defdualset} we chose to use here, as it leads to the involution property from Proposition \ref{propdual}. 
Let us therefore explain how this other notion of dual cell is defined, and why it is less suited for our framework.
Let $K$ be a closed cc. Consider the sets $\abv_L(w),\bel_L(w)$ in the cases where $L= \bdiv{K}$ and $w = \{x\} \in \bdiv{K}^{[0]}$ and the case $L = \bdiv{\kb}$ and $w = \{\dual{x}\} \in \bdiv{\kb}^{[0]}$ respectively. These sets can be more explicitly expressed as   
$$\abv_{\bdiv{K}}(\{x\}) = \{ \{y_1 \subsetneq \dots \subsetneq y_k\} \in \bdiv{K} ~|~ x \subset y_1 \},$$
and
$$ \bel_{\bdiv{\kb}}(\{\ce{x}\}) = \{ \{\ce{y_1} \subsetneq \dots \subsetneq \ce{y_k} \} \in \bdiv{\kb} ~|~  \ce{y_k} \subset \ce{x} \}.$$ 
Using the map (\ref{mappropdual}) we get that $\abv_{\bdiv{K}}(\{x\})$ and $\bel_{\bdiv{\kb}}(\{\ce{x}\})$ are isomorphic ranked posets, hence 
$$ \abv_{\bdiv{K}}(\{x\}) \cong  \bel_{\bdiv{\kb}}(\{\ce{x}\}) = \bdiv{\bel_{\kb}(\dual{x})}.$$

For the case of a simplicial complex $K,$ the map $K \dans \pws{\bdiv{K}}$ defined by
$$ x \longmapsto \dual{x}^B := \bdiv{\bel_{\kb}(\dual{x})} \cong \{\{y_1 \subsetneq \dots \subsetneq y_k \} \in \bdiv{K} ~|~ x \subset y_1 \}$$
is equivalent to the definition of dual cell used in \cite{bry02} (hence our choice for the letter $B$ in the notation $\dual{x}^B$). The dual cell $\dual{x}^B$ is therefore a sub-simplicial complex of $\bdiv{K}.$ 

We can then define the poset $\kb^B := \{ \dual{x}^B ~|~ x \in K \}$ with rank function 
$$\rk_{\kb^B}(\dual{x}^B) := \Rk(\bdiv{\bel_{\kb}(\dual{x})}) = \rk_{\kb}(\ce{x}).$$ 
It is possible to show that the map $ \phi(\dual{x}^B) := \ce{x}$ defines an isomorphism of posets from $\kb^B$ to $\kb$ such that $\rk_{\kb}(\phi(\dual{x}^B)) = \rk_{\kb^B}(\dual{x}^B).$ Nevertheless $\kb^B$ is not a cc since the cells in $\kb^B$ are not subsets of $(\kb^B)^{[0]}$ as these cells also contain elements in $\bdiv{K}^{[0]} \setminus (\kb^B)^{[0]}.$

\subsubsection{Reductions and collapses} \label{sssecredncol}

\begin{figure}[!h]
\centering
\includegraphics[scale=0.72]{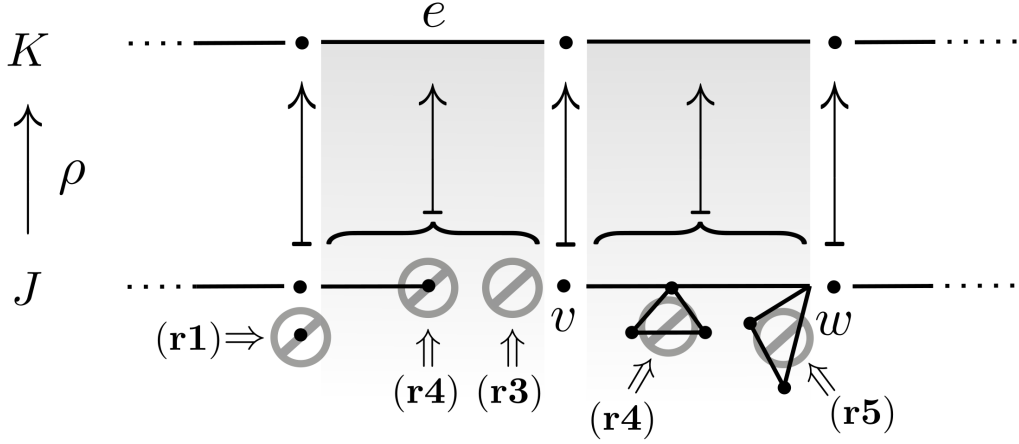}
\caption{\label{figcondred} In this picture, we illustrate what undesirable features the conditions \ref{cond1red}, \ref{cond3red}, \ref{cond4red} and \ref{cond5red} of a reduction are preventing, in the case of a reduction $\sphi: J \dans K$ between two 1-dimensional cell complexes. The vertical arrows indicate which cells in $J$ are mapped to each cell in $K,$ for example all the cells contained in the grey areas are mapped to an edge in $K.$ More explicitly, condition \ref{cond1red} prevents more than two vertices to be mapped to the same vertex. Condition \ref{cond3red} ensures that the vertex $v \in J$ is contained in an edge in the pre-image of the edge $e \in K$ since $\sphi(v) \in e.$ Condition \ref{cond4red} does not allow branchings or boundaries in the pre-image of a cell and condition \ref{cond5red} with $x = w \in J^{[0]}$ ensures in particular that the type of branching on $w$ shown on the figure does not occur or that a cell is not subdivided multiple times.}
\end{figure}

In order to define a generalization of the notion of barycentric subdivision including cc which are not simplicial complexes, we use a map $\sphi:J \dans K$ satisfying certain conditions allowing to interpret $J$ as a subdivision of $K.$ For simplification, we only define $\sphi$ in the case where $J$ and $K$ are closed cc but it can be generalized to arbitrary cc, see \cite{Sav22} for a more general treatment. The meaning of each condition in the following definition of reduction is illustrated in Figure \ref{figcondred}. 

\begin{defi}[Reduction and subdivision]\label{defreduction}
Let $J,K,$ be  closed cc. We say that a map $\sphi: J \dans K$ is a \textit{reduction} if $\sphi$ is a surjective poset homomorphism satisfying the following conditions. 
\begin{enumerate} [label=(\textbf{r\arabic*}),topsep=2pt, parsep=2pt, itemsep=1pt]
\item \label{cond1red} $\abs{\sphi^{-1}(v)} = 1,$ for all $v \in \kz$ (i.e. $\sphi|^{\kz}$ is injective).
\item \label{cond3red} $\abv( \sphi(x) )^{[r]} \subset \sphi(\abv(x)^{[r]}),$ for all $x \in J, ~r \geq 0;$
\item \label{cond4red} if $x \in J$ satisfies $\rk_J(x)= \rk_K(\sphi(x)) - 1$ then $\abs{ \cface{x} \cap \sphi^{-1}(\sphi(x))} = 2;$
\item \label{cond5red} if $x \in J$ satisfies $\rk_J(x)= \rk_K(\sphi(x))$ then 
$$\abs{ \cface{x} \cap \sphi^{-1}(y)} = 1, \quad \forall y \in \cface{\sphi(x)}.$$
\end{enumerate}
We shall use the notation $J \red_\sphi K$ if $\sphi: J \dans K$ is a reduction and call \textit{$J$ a subdivision of $K$} or equivalently say that \textit{$K$ is a reduction of $J,$} written $J \red K,$ if there exists a reduction $\sphi : J \dans K.$ A cell $y \in K$ is called \textit{subdivided by} $\sphi$ if $\abs{\sphi^{-1}(y)} > 1.$ For example, a subdivision does not subdivide vertices.
\end{defi}

If $\phi:J \dans K$ is a map between closed cc, then one can define the \textit{dual of $\phi$} to be the map
\begin{align}\label{equdualmap}
\dual{\phi}: \dual{J} &\dans \dual{K} \notag \\
\dual{x} &\longmapsto \dual{\phi}(\dual{x}) := \dual{\phi(x)},
\end{align}
well defined thanks to Lemma \ref{claim2propdual}.

By simply checking that each condition in the following definition of collapse is indeed the statement dual to the corresponding condition for reductions, one can see that collapses are indeed duals of reductions.

\begin{defi}[Collapse]
Let $J,K$ be closed cc. We say that a map $\cphi: J \dans K$ is a \textit{collapse} it is a surjective poset homomorphism and 
\begin{enumerate} [label=(\textbf{c\arabic*}),topsep=2pt, parsep=2pt, itemsep=1pt]
\item \label{cond1col}$\abs{\cphi^{-1}(z)} = 1,$ for all maximal cell $z \in K;$
\item \label{cond3col}$\bel(\cphi(x))^{[r]} \subset \cphi(\bel(x)^{[r]})$ for all $x \in J, ~r \geq 0;$
\item \label{cond4col}if $x \in J$ satisfies $\rk_J(x) = \rk_K(\cphi(x)) + 1$ then $\abs{ \face{x} \cap \cphi^{-1}(\cphi(x))} = 2;$
\item \label{cond5col}if $x \in J$ satisfies $\rk_J(x)= \rk_K(\cphi(x))$ then 
$$\abs{ \face{x} \cap \cphi^{-1}(y)} = 1, \quad \forall y \in \face{\cphi(x)}.$$
\end{enumerate}
We use the notation $J \col_\cphi K$ if $\cphi: J \dans K$ is a collapse and say that \textit{ $J$ is an expansion of $K$} or equivalently that \textit{$K$ is a collapse of $J,$} written $J \col K,$ if there exists a collapse $ \cphi : J \dans K.$
\end{defi}

\begin{figure}
\centering
\includegraphics[scale=0.3]{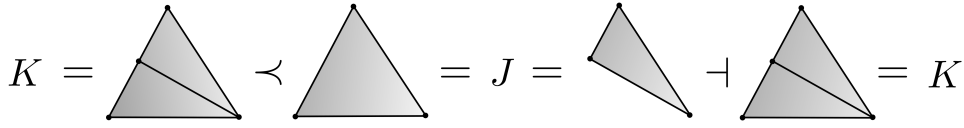}
\caption{\label{figexredncol} Example of cell complexes $J$ and $K$ such that $J \red K$ and $J \col K.$}
\end{figure}

Figure \ref{figexredncol} includes a simple example of reduction and collapse, and show that a cc can be both a subdivision and an expansion of a given cc. Also, $\bdiv{K} \red K.,$ i.e. the barycentric subdivision is indeed an example of subdivision (Lemma \ref{lembdiv}).

An important property of reductions of collapses which supports their relation with the notion of subdivision is the following result.

\begin{prop} \label{propprecpartialorder} 
$\red$ and $\col$ are  partial orders on the set of closed cc, up to cc-isomorphism.
\begin{proof}\renewcommand{\qedsymbol}{}
In Appendix \ref{appPropPrec}.
\end{proof}
\end{prop}

We close this part with two remarks that will be referred to later on, pointing particular at the fact that reductions and collapses are cc-homomorphisms between cc of same rank.

\begin{rema}\label{remcond3redncol} 
A consequence of condition \ref{cond3red} is that $\rk_K(\sphi(x)) \geq \rk_J(x),$ for a reduction \mbox{$\sphi: J \dans K$} and \ref{cond3col} implies that $\rk_K(\cphi(x)) \leq \rk_J(x)$ for a collapse $\cphi : J \dans K.$ This can be seen for example for the case of a reduction $\sphi$ by considering the instance $r = \rk_K(\sphi(x))$ in condition \ref{cond3red} according to which
$$ \{\sphi(x)\} = \abv( \sphi(x) )^{[r]} \subset \sphi(\abv(x)^{[r]}).$$
This implies $\abv(x)^{[r]} \neq \emptyset$ and therefore $\rk_J(x) \leq r = \rk_K(\sphi(x))$ for all $x \in J.$ The case of a collapse is similar. This argument also shows that conditions \ref{cond3red} and \ref{cond3col} in particular imply that reductions and collapses are cc-homomorphisms.
\end{rema}

\begin{rema} \label{remrksredncol} 
If $J \red K$ or $J \col K$ then $\Rk(J) = \Rk(K).$ Indeed, for the case $J \red K,$ $\sphi = \sphi_J^K$ is in particular a cc-homomorphism hence there exists a $\Rk(K)$-cell in $\sphi^{-1}(z),$ for all $z \in K^{[\Rk(K)]}$ and therefore $\Rk(J) \geq \Rk(K).$ Remark \ref{remcond3redncol} also implies $\rk_J(x) \leq \rk_K(\sphi(x))$ for all $x \in J$ and we get $\Rk(J) = \Rk(K).$ 
\end{rema}

\newpage
\section{Structure near the boundary and Correspondence Theorem} \label{secbdry}

This section focuses on the study of the relation between the cells in the boundary of a cc and the cells in their collar, i.e. linking the boundary to the associated midsection. This relation will be essential for Section \ref{SecCobs} where we discuss the duality of cells on the boundary, leading to the definition of cobordism, and give sufficient conditions for the union of two cobordisms sharing a boundary component to form a cell complex.

It is therefore sometimes convenient to use relative cc, i.e. couples $(K,J)$ where $J \leq K,$ to formulate the notions needed for our purposes in this and the next sections, which are discussed to a large extend in part \ref{ssecrelcc}. 

Part \ref{ssectrans} defines a poset called the transition made of certains cells arising "between" the boundary and the midsection. This poset is then used to formulate the assumption of uniformity, under which the transition is a cc.
We finish this part with an important result of this section involving uniformity (Proposition \ref{proptransit}) and making explicit the relation between a boundary component and the associated midsection and transition via a reduction and a collapse. 

It is shown in part \ref{sseccomporthrefl} that the reduction and collapse from Proposition \ref{proptransit} satisfy additional conditions that lead us to the definition of geometric sequence. We finish in part \ref{ssecslicecthm} by introducing slices and showing the main theorem of this work (Theorem \ref{thmcorresp}) establishing a bijection between slices and some geometric sequences. This result is what allows to define the composition of cobordisms introduced in the last section and slices are a basic building block in the definition of causal cobordisms.

\subsection{Relative cc} \label{ssecrelcc}

 A relative cc is simply one for which we single out a sub-cc and it will be mostly used in the case where the sub-cc is a connected component of the boundary. We will also need a relative notion of homomorphism defined as follows.

\begin{defi}[Relative cc and homomorphism] \label{defirelccc}
A couple $(K,J)$ of cc is called a \textit{relative cc} if $J \leq K.$
A \textit{relative cc-homomorphism} $\phi: (K,J) \dans (K',L)$ (or a \textit{homomorphism of $J$ relative to $K$}) is a cc-homomorphism $\phi : K \dans K'$  such that $\phi(J) = L$ and $\phi|_{K \setminus J}: K \setminus J \dans K' \setminus L$ is a poset isomorphism. 
\end{defi}

Typically a relative cc-homomorphism is used to define a map on a boundary component of a cc $K$ such that its image can still be seen as a sub-cc of $K,$ and we will primarily use the two following types of relative cc-homomorphisms.

\begin{defi}[Relative reduction and collapse] \label{defirelredncol}
Let $(K,J), (H,L)$ be relative cc. A \textit{ relative reduction (respectively a relative collapse)} or a \textit{reduction (respectively collapse) of $J$ relative to $K$} is a cc-homomorphism $\phi : (K,J) \dans (H,L)$  of $J$ relative to $K$ such that $\phi : K \dans H$ is a reduction (respectively a collapse). We will also write $(K,J) \red (H,L)$ and say that $(K,J)$ is \textit{a subdivision of $L$ relative to $H$} if there exists a reduction $\sphi : (K,J) \dans (H,L)$ of $J$ relative to $K.$ Similarly we will write $(K,J) \col (H,L)$ and say that $(K,J)$ is \textit{an expansion of $L$ relative to $H$} if there exists a collapse $\cphi: (K,J) \dans (H,L)$ of $J$ relative to $K.$
\end{defi}

Clearly Definition \ref{defirelredncol} implies in particular that a relative cc-homomorphism $\phi : (K,J) \dans (H,L)$ is a relative reduction (respectively a relative collapse) if and only if $\phi|_J : J \dans L$ is a reduction (respectively a collapse). By abuse of notation, we will sometimes also use the notation $\phi$ to designate $\phi|_J$ when $\phi : (K,J) \dans (H,L)$ is a relative homomorphism.

The property that a relative cc-homomorphism $\phi: (K,J) \dans (K',L)$ is a poset isomorphism when restricted to $K\setminus J$ reveals the fact that most of the emphasis on a relative cc $(K,J)$ is put on the sub-cc $J.$ Hence $K$ can be merely considered as a "background" in which $J$ is included, and this enables the following definition that will be convenient towards the end of this section.

\begin{defi}[$K_\phi$ and $K^\phi$] \label{defikphi}
Let $\phi: J \dans L$ be a surjective cc-homomorphism and let $H$ be a cc such that $ L \leq H.$ If there exists a cc $K \geq J$ and a relative cc-homomorphism $\phi^H : (K,J) \dans (H,L)$ such that $\phi^H|_J = \phi$ then $K$ is uniquely determined by $H$ and $\phi : J \dans L$ up to cc-isomorphism. We will therefore use the notation $K =: H^\phi.$

Similarly, let $\phi: J \dans L$ be a cc-homomorphism and let $K$ be a cc such that $ J \leq K.$ If there exists a cc $H \geq L$ and a relative cc-homomorphism $\phi_K : (K,J) \dans (H,L)$ such that $\phi_K|_J = \phi$ then $H$ is uniquely determined by $K$ and $\phi : J \dans L$ up to cc-isomorphism. We will therefore use the notation $H =: K_\phi.$
\end{defi}


Let us make the following remark about the previous definitions considering the example of a reduction  $\sphi : J \dans L.$ In general it is wrong that given a relative cc $(H ,L)$ there exists a cc $K$ such that $(K, J) \red_{\sphi} (H,L)$ is a relative reduction or that given a relative cc $(K,J)$ there exists a cc $H$ such that $(K, J) \red_{\sphi} (H,L)$ is a relative reduction.

One reason that this is wrong for the latter case is that it is possible for the reduction $\sphi$ to reduce two $r$-cells $x_1,x_2$ in $J$ each sharing a face $y_1,y_2 \in J^{[r-1]}$ with a $r$-cell in $y \in K_J$ into one cell $x =\sphi(x_1) = \sphi(x_2),$ hence $\{x_1,x_2\} \subset x \cap y,$ violating the intersection Axiom \ref{cccinter} for $H.$

Similarly, a reason for the first case is that a relative subdivision of $(H,L)$ can subdivide a cell $x$ of $L$ contained in a cell of $y \in H_L$ only if every cell $w$ of $L$ strictly containing $x$ is also subdivided in such a way that any intersection of an element in $\sphi^{-1}(y)$ with an element in $\sphi^{-1}(w)$ contains at most one cell, which is not the case in general.

As mentioned before, we are particularly interested in relative cc $(K,J)$ where $J$ is a connected component of the boundary of $K$ but, even in this case, there can be cells in the collar $K_J$ having all their vertices in $J.$ For example all vertices in a cc having a unique maximal cell are on the boundary. Hence it will be valuable to have the following notion of non-degenerate relative cc at hand when studying the midsection associated with a boundary component.

\begin{defi}[Non-degenerate relative cc]
We say that a relative cc $(K,J)$ is \textit{degenerate} if there exists $x \in K \setminus J$ such that $x \subset J^{[0]}.$ We say that $(K,J)$ is \textit{non-degenerate} if it is not degenerate. 
\end{defi}

When discussing the duality of cobordisms in the next section, the midsection associated with a boundary component will play an important role. As we are interested in preserving a framework where cc are local, one would like to ensure that the midsection is a local cc. However, even in a local and non-pinching cc, the midsection associated with a boundary component can be a non-local cc (such an example can be constructed using a cell as in Figure \ref{pinchpic} in the collar of a boundary component). We thus also need a refined relative notion of locality as defined next.

\begin{defi}[Local relative cc]
We say that a relative cc $(K,J)$ is \textit{local} if it is non-degenerate, $K$ is local, every connected component $J_0$ of $J$ satisfies that $\bel_{J_0}(x)$ is connected for all $x \in K_{J_0}$ and $M_{J_0}^K$ is local. In particular $(K,\emptyset)$ is local for any lcc $K.$
\end{defi} 

Finally, a particular type of relative cc we will use in the next section is the following notion of exactly collared relative cc. It will allow us to obtain a class of cobordisms on which the duality map acts as a involution.

\begin{defi}[Exactly collared relative cc]
Let $(K,J)$ be a non-degenerate cc. We say that $(K,J)$ is \textit{exactly collared} if the map from  $M_J^K$ to $J$ defined by $\E_J^x \mapsto J \cap x$ is a cc-isomorphism. In particular an exactly collared relative cc $(K,J)$ such that $K$ is local is a local relative cc and $(K,\emptyset)$ is also exactly collared for all lcc $K.$
\end{defi}

\subsection{Transition and uniformity} \label{ssectrans}

\begin{figure}[!h]
\centering
\includegraphics[scale=0.42]{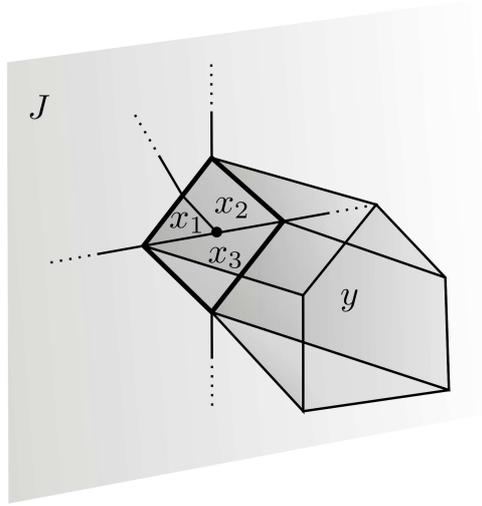}
\caption{\label{figtrans} Example of a 3-cell $y$ in the collar of the boundary component $J$ of a 3-cc.
}
\end{figure}

Consider for example a cc $K \in \nsc$ of rank 3 and a boundary component $J \leq \partial K.$ If we consider a maximal cell $y$ in the collar $K_J$ as illustrated in Figure \ref{figtrans}, we notice that the edges in $y$ and in the collar $K_J$ define a subset $w$ of the vertices in $y \cap J^{[0]}$ ($w$ contains the four vertices linked by thicker lines in the example considered). These types of subsets of vertices in $J$ is what defines the transition.

\begin{defi}[Transition] \label{defreducedbdry}
Let $(K,J)$ be a relative cc. For $x \in K_J,$ we define the \textit{reduced cell of $x$} to be
$$ x_J := \{ e \cap J^{[0]} ~|~ e \in \E_J^x \}.$$
We then define the \textit{transition of $(K,J)$} to be the poset
$$J(K) := \{ x_J ~|~ x \in K_J \}.$$
\end{defi}

The whole idea of this part will be to identify sufficient assumptions to prove Proposition \ref{proptransit} showing that there is a reduction from $J$ to $J(K)$ and a collapse from the $M_J^K$ to $J(K).$

In order to associate a rank to a cell $w$ of the transition there are two possibilities: either we consider a maximal cell in $\bel_J(y)$ (e.g. $x_i,~ i=1,2,3$) or we take $\rk_K(y) - 1,$ considering that $y$ is minimal with the property of containing $w.$ In order for the two choices to agree, we need an additional relative notion of purity introduced next.

Once the transition admits the structure of a ranked poset, one formulates the uniformity assumption as enforcing in particular that it is a cc. We will show that uniformity is sufficient to prove Proposition \ref{proptransit}.


In this part the two following subsets of the collar $K_J$ will be used repeatedly. First we will use the notation 
$$ K_J(A) := \{y \in K_J ~|~ A \subset y\}$$
where $A$ is simply a subset of $J^{[0]}.$ We also define $K^-_J(A)$ to be the set of minimal elements in $K_J(A).$
Second, it will also be convenient to refer to the set
$$K_J[x] = \{ y \in K_J ~|~ y_J = x\}$$
associated to a cell $x \in J(K)$ and denote by $K_J^-[x]$ the set of minimal elements in $K_J[x].$

\begin{defi}[Pure relative cc and relative rank]\label{defpurerelcc}
A relative cc  $(K,J)$ is said to be \textit{pure} if $J \cap x$ is pure for all $x \in K_J$ and, dually, for all $x \in J,$ all cells in $K^-_J(x)$ have equal rank $r(x).$ It is clear that in this case $r(x) \geq 1$ for all $x \in J$ and we define $\rk_J^K(x) := r(x) - 1$ to be the \textit{relative rank of $x \in J$ in $K.$} 
\end{defi}

\begin{figure}
\centering
\includegraphics[scale=0.4]{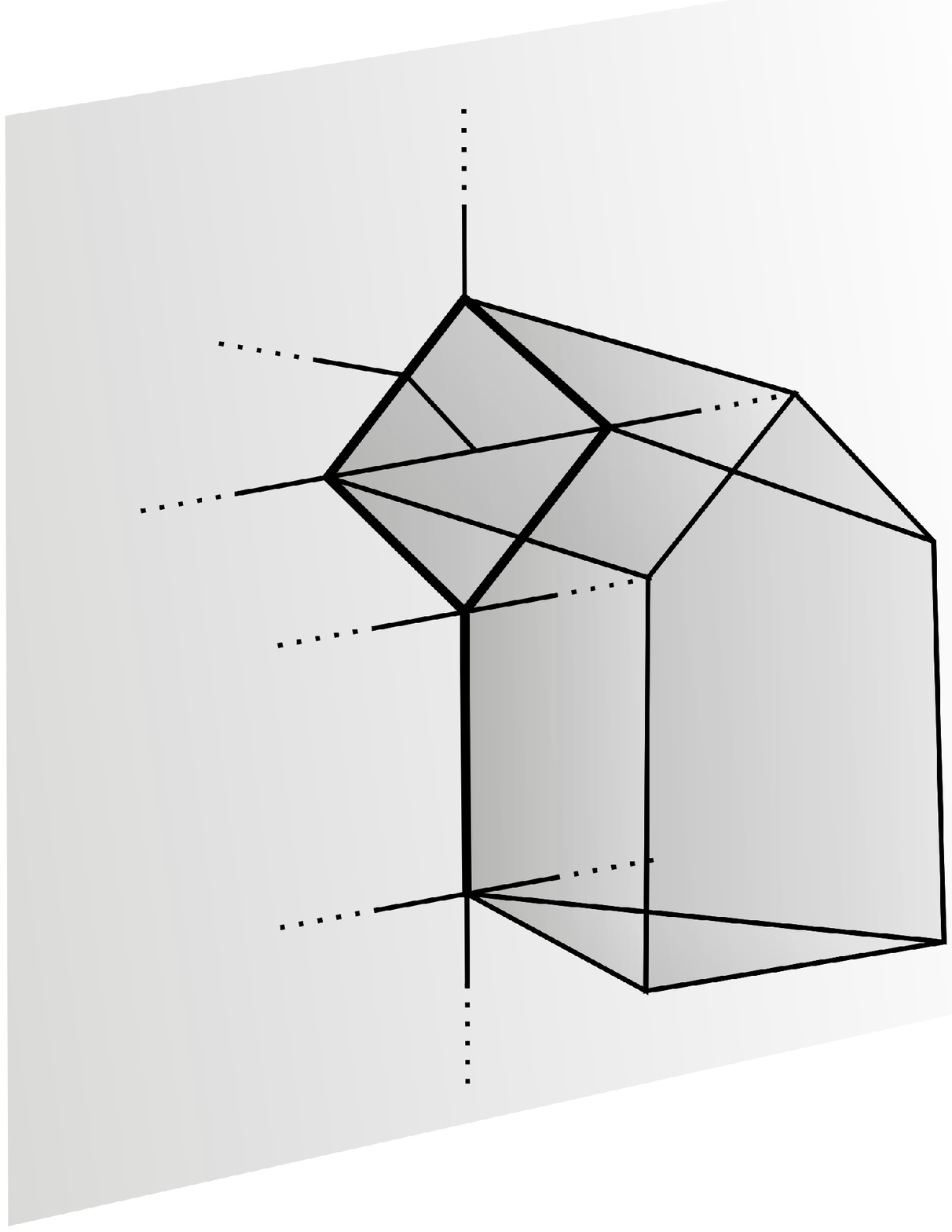}
\caption{\label{fignopure}  An example of a 3-cell $x \in K_J$ of a relative cc that does not satisfy the first condition for pure relative cc since $\bel_J(x),$ whose edges are represented with thicker lines, is not pure.}

\includegraphics[scale=0.4]{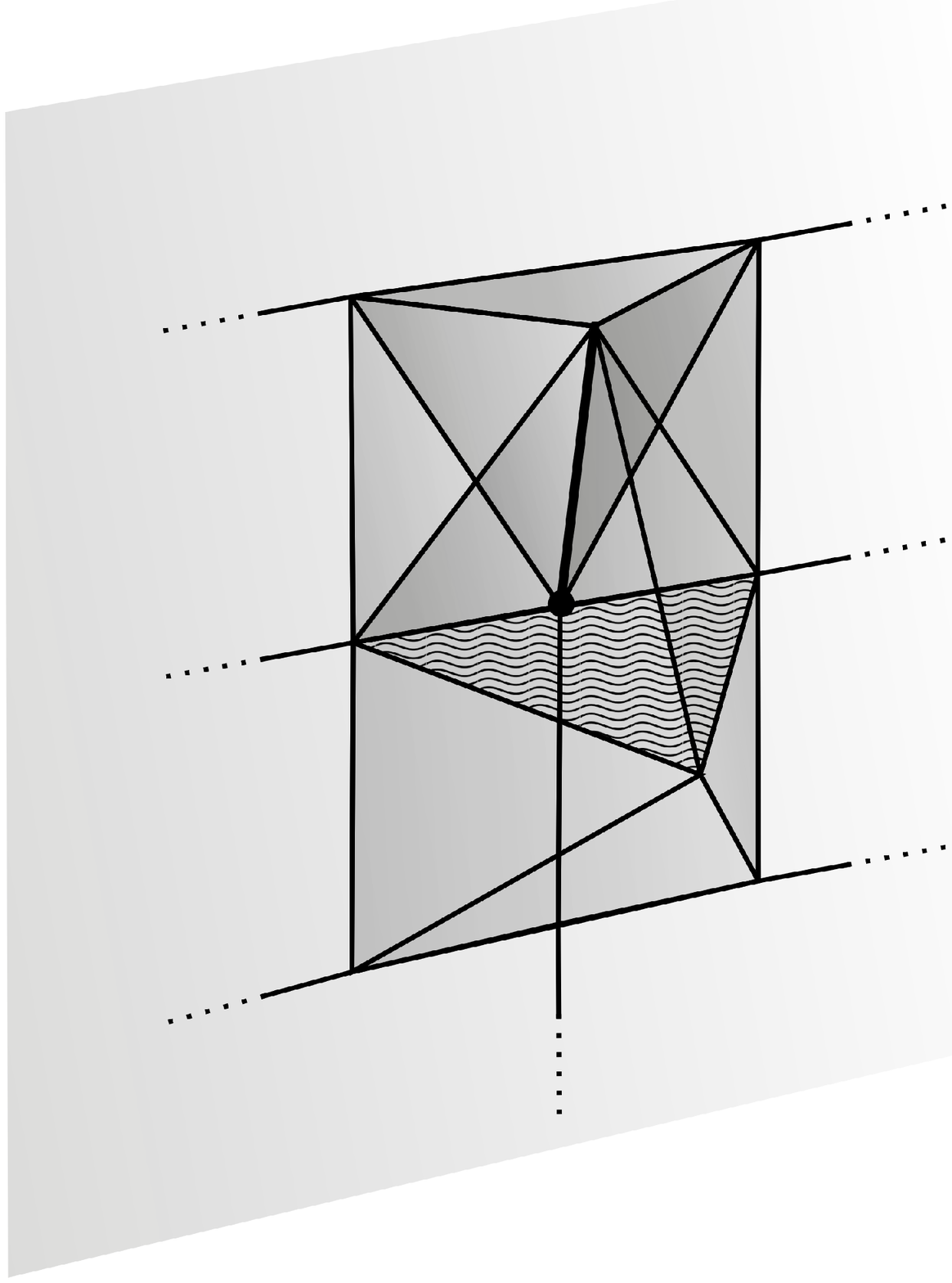}
\caption{\label{fignocopure}  An example of a cell $x \in J$ that does not satisfy the second condition of purity for the relative cc $(K,J),$ i.e. $x$ is the central vertex (black dot) and is both included in an edge (thicker line) and a 2-cell (wavy triangle) in $K_J.$ There are five 3-cells in this picture: four tetrahedra and one 3-cell below the wavy triangle having seven vertices.}
\end{figure}

Counterexamples of pure relative cc $(K,J)$ are illustrated in Figures \ref{fignopure} and \ref{fignocopure} (using as setup where the cc $K$ and $J$ are similar to those used in Figure \ref{figtrans}).
 
Reduced cells  are yet an other example of cells derived from the cell structure of a cc and in order for the transition to satisfy the axioms of cc, we need to ensure that reduced cells contain "enough vertices" in some sense. Indeed if $x \in K_J$ then $x_J \subset x \cap J^{[0]},$ but in general there can be vertices in $x \cap J^{[0]}$ not contained in an edge in $\E_J^x.$ 

We  therefore add an additional condition to our framework,  namely that the cells of a relative cc $(K,J)$ satisfy what we will refer to as the \textit{uniformity condition}:
\begin{equation} \label{conduniform}
J^{[0]} \cap x \subset J^{[0]} \cap y \quad \text{if and only if} \quad x_J \subset y_J, \quad \forall x,y \in K_J. \tag{\textbf{U}}
\end{equation}
This rather weak additional condition implies in particular (Lemma \ref{lemunifrelcc}) that the two ways to associated a rank to a cell of the transition discussed above are equivalent. More explicitly if $(K,J)$ be a pure relative cc satisfying (\ref{conduniform}) and  $x \in J(K)$ then for all $x' \in K_J^-[x]$ and all maximal cells $w \in \bel_J(x')$ we have
\begin{equation*}
\rk_J^K(x) := \rk_K(x') - 1 = \rk_K(w)
\end{equation*}
As a consequence the map
$$ J(K) \ni x_J \longmapsto \Rk(\bel_J(x)) $$
is well defined and we have $\Rk(\bel_J(x)) = \rk_K(x) - 1$ whenever $x \in K_J^-[x_J].$ We can therefore associate the rank function 
$$ \rk_{J(K)}( x_J) := \Rk(\bel_J(x))$$
to $J(K).$

\begin{defi}[Uniform relative cc]
A relative cc $(K,J)$ is called \textit{uniform} if the following four conditions are fulfilled:
\begin{itemize}[topsep=2pt, parsep=2pt, itemsep=3pt]
\item $(K,J)$ non-degenerate,
\item $(K,J)$ is pure,
\item the uniformity condition (\ref{conduniform}) holds for $(K,J),$
\item $(J(K), \rk_{J(K)}) \in \mlc.$
\end{itemize}
We will say that a cc $K \in \nsc$ is uniform if so is $(K, J)$ for all connected component $J$ of $\partial K.$
\end{defi}

One can easily check that these conditions are satisfied if $K \in \nsc$ has rank two or lower. It is however not clear to us whether the uniformity condition (\ref{conduniform}) and the assumption that $J(K) \in \mlc$ are independent from the assumptions of purity and non-degeneracy in dimension 3 or higher.

A few more technical manipulations (Lemma \ref{lemredcell}) show that if $(K,J)$ is uniform and $x \in J$ then there exists a unique cell $x_J \in J(K)$ such that
$$K_J^-[x_J] = K_J^-(x) \quad \text{and} \quad \rk_{J(K)}(x_J) = \rk_J^K(x).$$
This provides an generalization of the notion of \textit{reduced cell $x_J$} for cells  $x \in J.$ 

We close this part with the following important result, confirming that uniformity indeed is the assumption that unlocks the ability to characterize a boundary component of a cc and its neighbourhood using a reduction and a collapse.

\begin{prop}\label{proptransit}
Let $K \in \nsc^{R+1}$ and $J \leq \partial K,$ $J \in \mlc^R$ such that $(K,J)$ is uniform and local.
Then one has the two following results.
\begin{enumerate}[label=\arabic*) , topsep=2pt, parsep=2pt, itemsep=1pt]
\item \label{proptransitred} We have $J \red J(K)$ as the poset homomorphism $\sphi_J^K: J \dans J(K)$ defined by
$$\sphi_J^K(x) = x_J$$
is a reduction.
\item \label{proptransitcol} We have $M_J^K \col J(K)$ as the poset homomorphism $ \cphi_J^K: M_J^K \dans J(K)$ defined by $$\cphi_J^K(\E^x_J) = x_J$$ is a collapse.
\end{enumerate}

\begin{proof}
In Appendix \ref{appproptransit}.
\end{proof}
\end{prop}

\subsection{Geometric sequences} \label{sseccomporthrefl}

In the previous part we saw that assuming uniformity allows to relate the cells in a boundary component $J$ of a cc $K$ with the cells in the corresponding midsection $M_J^K$ using the reduction $\sphi_J^K : J \dans J(K)$ and the collapse $\cphi_J^K: M_J^K \dans J(K).$ In this part we show that these to maps satisfy additional conditions inherited from the axioms of the cc $K$ and we present technical results in preparation to the next part. An example of such condition can be obtained by considering two disjoint cells in the midsection $M_J^K$ having identical images through $\cphi_J^K$ in the transition. These correspond to two cells in the collar intersecting only on the boundary, therefore the image of these two cells cannot be subdivided by $\sphi_J^K$ as it would contradict the intersection axiom \ref{cccinter}. Similar scenarios of intersections between cells in the collar and in the boundary leads to the following notion of compatible cc-homomorphisms, a property that is shown to hold for the case of $\sphi_J^K$ and $\cphi_J^K$ in the corollary that comes next.

\begin{defi}[Compatible homomorphisms, $\cpa$]  \label{defcoorthom}
Let $j : J \dans I$  and $l : L \dans I$ be cc-homomorphisms. Then we say that $j$ is \textit{compatible with} $l,$ noted $j \cpa l,$ if the three following conditions are satisfied:
\begin{enumerate}[label=\arabic*) , topsep=2pt, parsep=2pt, itemsep=1pt]
\item \label{cpa} For all $w \in I$ either $\abs{j^{-1}(w)} = 1$ or $\abs{l^{-1}(w)}=1.$
\item \label{cpau} If $x,x' \in J$ are such that $x \vee x' \neq \all,$ then $$j(x \vee x') = j(x) \vee j(x').$$
\item \label{cpai} If $y,y' \in L$ are such that $y \cap y' \neq \emptyset$ then $$l(y \cap y') = l(y) \cap l(y').$$
\end{enumerate}
Note that if $j$ and $l$ are poset isomorphisms then $j$ is compatible with $l.$
\end{defi}

\begin{cor} \label{cortransitmidsec}
Let $K \in \nsc^R$ and $J \leq \partial K,$ $J \in \mlc^{R-1}$ such that $(K,J)$ is uniform and local then
$$J \red_{\sphi_J^K} J(K) \loc_{\cphi_J^K} M_J^K \quad \text{and} \quad \sphi_J^K \cpa \cphi_J^K.$$
\begin{proof}
The first statement is a direct consequence of Proposition \ref{proptransit}.
As for the first condition of compatibility, suppose by contradiction that there exists $x \in J(K)$ such that there are two distinct elements $x_1,x_2 \in J$ satisfying $(x_1)_J = (x_2)_J = x,$ as well as two distinct elements $\E_J^{y_1},\E_J^{y_2} \in M_J^K$ such that $(y_1)_J = (y_2)_J = x.$ By the conditions \ref{cond3red}, \ref{cond3col}, \ref{cond4red} and \ref{cond4col} we can assume that 
$$ r:= \rk_{J(K)}(x) = \rk_J(x_i)  = \rk_{M_J^K}(\E_J^{y_i})  = \rk_K(y_i)-1,$$
for $i=1,2$ and where the two last equalities come from the definition of the midsection $M_J^K.$ This implies that $y_1,y_2 \in K_J^-[x]$ and therefore $y_1 \cap y_2 \in J^{[r]}.$ But we have $\{x_1, x_2\} \subset \bel_J( (y_1 \cap y_2) )$ so this contradicts the intersection axiom \ref{cccinter} of $K.$

In order to prove the condition \ref{cpau} of compatibility, take $x,x' \in J$ such that $x \vee x' \neq \all.$ We need to show that 
$$\sphi_J^K(x \vee x') = (x \vee x')_J = x_J \vee x'_J = \sphi_J^K(x) \vee \sphi_J^K(x'),$$ where the map $J \ni x \mapsto x_J \in J(K)$ is defined in Lemma \ref{lemredcell}, which also states that $K_J^-[x_J] = K_J^-(x)$ and $\rk(y)-1=\rk(x)$ for all $y \in K_J^-(x).$ 

The inclusion $x_J \vee x'_J \subset (x \vee x')_J$ is clear (and in particular $x_J \vee x'_J \neq \all$) since if $y \in K_J^-(x \vee x')$ then in particular $x\subset y$ and $x' \subset y$ hence the definition of reduced cells easily leads $x_J \subset y_J$ and $x'_J \subset y_J$ and we have $$x_J \vee x'_J \subset y_J = (x \vee x')_J.$$ 

Now consider $w \in K_J^-[x_J \vee x'_J],$ hence we have that $x_J \subset w_J$ and $x'_J \subset w_J.$ The uniformity condition (\ref{conduniform}) implies that  $J^{[0]} \cap y' \subset J^{[0]} \cap w$ for all $y' \in K_J^-(x) \cup K_J^-(x')$ which in particular implies that $x \subset w$ and $x' \subset w.$  Hence we have $x \vee x' \subset w$ and this implies that $y_J = (x \vee x')_J \subset w_J$ which leads the desired equality.

The condition \ref{cpai} of compatibility is a direct consequence of Lemma \ref{lemmidsec} since if $\E_J^y,\E_J^{y'} \in M_J^K$ are such that $\E_J^y \cap \E_J^{y'} \neq \emptyset,$ then
\begin{align*}
\cphi_J^K(\E_J^y \cap \E_J^{y'}) &= \cphi_J^K(\E_J^{y \cap y'}) = (y \cap y')_J\\
&= \{ e \cap J^{[0]} ~|~ e \in \E_J^{y \cap y'} \} = \{e \cap J^{[0]} ~|~ e \in \E_J^y \cap \E_J^{y'}\}\\
&= \{e \cap J^{[0]} ~|~ e \in \E_J^y \} \cap \{e \cap J^{[0]} ~|~  e \in \E_J^{y'}\} \\
&= x_J \cap y_J = \cphi_J^K(\E_J^y) \cap \cphi_J^K(\E_J^{y'}).
\end{align*} 
\end{proof}
\end{cor}

Conversely, the compatibility condition allows to determine when a subdivision of a boundary component of a cc $K$ allows to define a relative subdivision of $K,$ as shown in the next proposition. This result is more involved than the previous corollary and constitutes an important simplification step in the proof of the Correspondence Theorem \ref{thmcorresp} shown in the next part.

\begin{prop} \label{propredbdry}
Let $K \in \nsc$ and let $J'$ be a connected component of $\partial K$ such that $(K , J')$ is uniform. If $\sphi : J \dans J'$ is a reduction such that $\sphi \circ \sphi_{J'}^K : J \dans J'(K)$ is compatible with $\cphi_{J'}^K : M_{J'}^K \dans J'(K)$ and $J$ is disjoint from $K$ then the cc $K^\sphi$ (introduced in Definition \ref{defikphi}) is an element in $\nsc$ and $J$ is a connected component of $\partial K^\sphi,$ in other words 
$$( K^\sphi, J) \red_\sphi (K, J')$$
defines a relative reduction. Moreover, we have that $\sphi_J^{K^\sphi} = \sphi \circ \sphi_{J'}^K$ and $(K^\sphi, J)$ is uniform.
\begin{proof}
In Appendix \ref{apppropredbdry}.
\end{proof}
\end{prop}

As we are also interested in seeing a cc $J \in \mlc$ as the boundary component of two different cc $K$ and $K'$ in order to study in which case their union is a cc, this leads to having two reductions of the form
\begin{equation} \label{equredseq}
 J(K) \der_{\sphi_J^K} J \red_{\sphi_J^{K'}} J(K').
\end{equation}
The main idea from here is to show that the following notion of reflectivity is indeed a sufficient assumption on $\sphi_J^K$ and $\sphi_J^{K'}$ in order for the union $K \cup K'$ to become a cc (as done in part \ref{sseccompcob}). For this we will pass by the dual notion to that called orthogonality also defined below.

\begin{defi}[Reflective ($\refl$) and Orthogonal ($\perp$) homomorphisms] \label{defrefnorth}
Let $j : I \dans J$ and $l : I \dans L$ be two cc-homomorphisms.
\begin{enumerate}[label=\arabic*) , topsep=2pt, parsep=2pt, itemsep=1pt]
\item \label{condoref} We say that $j$ and $l$ are \textit{reflective}, noted $j \refl l,$ if whenever two elements $w,w' \in I$ satisfy that both $j(w) \vee j(w') \neq \all$ and $l(w) \vee l(w') \neq \all$ then we have that $w \vee w' \neq \all,$ as well as the two equalities  
$$ j(w \vee w') = j(w) \vee j(w') \quad \text{ and } \quad l(w \vee w') = l(w) \vee l(w').$$
\item \label{condorth} Dually, we say that $j$ and $l$ are \textit{orthogonal}, noted $j \perp l,$ if whenever two elements $w,w' \in I$ satisfy that both $j(w) \cap j(w') \neq \emptyset$ and $l(w) \cap l(w') \neq \emptyset$ then we have that $w \cap w' \neq \emptyset,$  as well as the two equalities $$ j(w \cap w') = j(w) \cap j(w') \quad \text{ and } \quad l(w \cap w') = l(w) \cap l(w').$$
\end{enumerate}
We also have that if $j$ and $l$ are poset isomorphisms then $j$ and $l$ are reflective and orthogonal.
\end{defi}

The key point about the latter definitions, that will allow to go back and forth between the primal and dual perspectives, is that they satisfy the following properties.

\begin{lem} \label{lemdualcompnorth}
Let $I,J,L \in \mlc^R.$ Then the two following statements hold.
\begin{enumerate}[label=\Alph*) , topsep=2pt, parsep=2pt, itemsep=1pt]
\item \label{lemdualcomp} If $j: J \dans I$ and $\l : L \dans I$ are cc-homomorphisms then
$$ j \cpa l \quad \text{if and only if} \quad \dual{j} \apc \dual{l}.$$
\item \label{lemdualorth} If $j: I \dans J$ and $l : I \dans L$ are cc-homomorphisms then
$$ j \perp l \quad \text{if and only if} \quad \dual{j} \refl \dual{l}.$$
\end{enumerate}
\begin{proof}
For the claim \ref{lemdualcomp}, the first condition of the compatibility of $\dual{l}$ with $\dual{j}$ is obtained from the fact that if $A$ is subset of cells in a closed cc then $\abs{\dual{A}} = \abs{A}$ by Lemma \ref{claim2propdual}. As for the other conditions for compatibility and the claim \ref{lemdualorth}, they follow from the "DeMorgan" identities (\ref{equduality}).
\end{proof}
\end{lem}

From the notions of compatibility, reflectivity and orthogonality, it is natural to combine them together into the following notion of geometric sequence. These will turn out to be essentially the main constituents in the definition of the category of causal cobordisms in part \ref{sseccatcausalcob}.

\begin{defi}[Geometric sequence]
Let $K_1, \dots, K_N$ be distinct cc in $\mlc.$ We call a sequence of reductions and collapses a \textit{geometric sequence} if is is of the form
\begin{equation}\label{equgeomseq}
 K_1 \dots \longleftarrow K_i \longrightarrow K_{i+1} \longleftarrow K_{i+2} \longrightarrow  \dots K_N
\end{equation}
without constraint on whether the first and last map have respectively $K_1$ and $K_N$ as domain or codomain, and if it satisfies the conditions
\begin{enumerate}[label=\textbf{\alph*)} , topsep=2pt, parsep=2pt, itemsep=1pt]
\item if two maps have same codomain, then it is a reduction and a collapse and they are compatible,
\item if two cc-homomorphisms have same domain, then either they are two orthogonal collapses or two reflective reductions.
\end{enumerate}
\end{defi}

As noted in remark \ref{remrksredncol}, the fact that the $K_i$'s belong to the same geometric sequence implies that they all have the same rank. Moreover, we have the following corollary.

\begin{cor}
The dual of a geometric sequence (\ref{equgeomseq}), i.e. where each cc $K_i$ is replaced by $\dual{K_i}$ and each map is replaced by the dual map defined by (\ref{equdualmap}), is also a geometric sequence.
\begin{proof}
This is a simple consequence of Lemma \ref{lemdualcompnorth} and the fact that reductions and collapses are dual to one another.
\end{proof}
\end{cor}

As mentioned before, we are going to switch from two reflective reductions as in (\ref{equredseq}) to the dual sequence then composed of two orthogonal collapses. The next lemma is an important technical result that leads to the definition of augmented (ranked) poset, named after the cells of rank one higher than their pre-image it allows to define in the proof of the Correspondence Theorem \ref{thmcorresp}.

\begin{lem} \label{lemsqcup}
Let $M,J,L$ be mutually disjoint cc and let $\phi_J : M \dans J$ and $\phi_L:M \dans L$ be two orthogonal cc-homomorphisms. Then the map from $M$ to $\pws{J^{[0]} \sqcup L^{[0]}}$ defined by 
\begin{equation*}
x \mapsto \phi_J(x) \sqcup \phi_L(x) 
\end{equation*}
is an injective poset homorphism such that the inverse is a poset homomorphism. The poset
$$ (\phi_J \sqcup \phi_L)(M) := \{ \phi_J(m) \sqcup \phi_L(m) ~|~ m \in M\}$$
will be called the \textit{augmented poset associated to $\phi_J$ and $\phi_L$} and will be assigned the rank function
$$\rk_{(\phi_J \sqcup \phi_L)(M)}(x \sqcup y) = \rk_M(m(x,y)) + 1,$$
where $m(x,y)$ is the unique element in $M$ such that 
$$\phi_J(m(x,y)) \sqcup \phi_L(m(x,y)) = x \sqcup y.$$ 
\begin{proof}
Follows directly from the orthogonality condition \ref{condorth}.
\end{proof}
\end{lem}

\subsection{Slices and Correspondence Theorem} \label{ssecslicecthm}

We are now going to introduce slices as a certain cc with all its vertices on the boundary and see in the following proposition that slices have exactly two connected components. We then introduce the notion of slice sequence and show our main result of this section, the Correspondence Theorem, proving that slice sequences are in bijective correspondence with slices up to cc-isomorphism.

\begin{defi}[Slice]
A cc $S \in \nsc$ if $(S, J)$ is uniform and local for all connected component $J \leq \partial S$ and such that
$$S^{[0]} = (\partial S)^{[0]}.$$
We denote \textit{the set of slices} by $\nsc_s.$
\end{defi}

Figure \ref{figmidsec} is in fact an illustration of a portion of a slice of rank 3. The next result is also used to introduce the \textit{midsection associated to a slice} and is a good demonstration of how the assumptions of (relative) locality and non-degeneracy can be combined. Note also that the proof does not make use of the uniformity of the slice except for the non-degeneracy condition.

\begin{prop} \label{propslice}
Let $S$ be a slice. Then $\partial S = J \sqcup L$ where $J$ and $L$ are connected components, and $M_L^S = M_J^S.$ We thus define the \textit{midsection associated to a slice} $S$ by 
$$M^S := M_J^S = M_L^S.$$ 
\begin{proof}
Let $J$ be a connected component of $S.$ Since $(S, J)$ is in particular non-degenerate, we have that $L := \partial S \setminus J$ is non-empty.

We next show that $L$ is connected.  Suppose by contradiction that $L = L_1 \sqcup L_2$ where $L_i \in \mlc$ for $i= 1,2$ and let $R =\Rk(L) = \Rk(J).$ The case $R = 0$ is clear since the connectedness of $S$ implies that there is a path from $L_1$ to $L_2$ going through $J,$ which means that there is a vertex in $J$ included in two edges and this is a contradiction with the definition of a boundary cell, i.e. that a sub-maximal cell of the boundary - here a 0-cell - is included in only one maximal cell. The strategy of the proof for the case $R \geq 3$ is to show that our hypothesis implies that there exists an edge of $S$ having one vertex in $L_1$ and the other vertex in $L_2,$ which contradicts the non-degeneracy of $(S,L).$

By connectedness of $S,$ there exists a path $p$ in $S$ from $v_1 \in L_1^{[0]}$ to $v_2 \in L_2^{[0]}$ and we may suppose that $p$ is simple and intersects $L$ only on two vertices. Let $\{v_1,w\}$ be the first edge of $p.$ If $w \in L_2$ we are done, so we suppose that $w \in J^{[0]}.$ By our assumptions, $p$ enters $J$ from the vertex $v$ and leaves $J$ through an edge of the form $\{w',v_2\},$ where $w \in J^{[0]}.$ By locality of $(S,J)$ the midsection $M := M_{J_0}^S$ is connected, hence there exists a path $p_M $ in $M$ from $\{w,v_1\}$ to $\{w',v_2\}$ (seen as vertices of $M$). Each edge of the path $p_M$ corresponds to a 2-cell in $S_J$ which by locality intersects $J$ in a simple path $p_0,$ or a single vertex $v,$ in which case we set $p_0$ to be the empty path (based at the vertex $v$). So using each such 2-cell $C$ in $p_M$ we can successively deform $p$ using a path move of the form $m_C^{p_1}$ where $p_1 \subset p$ is the simple path with first edge $\{v_1,w\}$ followed by the edges in $p_0.$ At each iteration the resulting path might cross $L$ in more than two vertices. If it does so by also crossing through $L_2$ this would imply that there is an edge of $S$ crossing from $L_1$ to $L_2$ which would produce the desired contradiction. We can therefore assume that deforming $p$ can only increase the number of vertices it crosses in $L_1,$ in which case we can simply re-define $v_1$ to be the last vertex the path crosses in $L_1.$ In this way we can successively deform and shorten the path $p$ until the last step, corresponding to the last edge in $p_M$ which again produces an edge of $S$ crossing from $L_1$ to $L_2,$ which concludes our argument.

Proving that $M_L^S = M_J^S$ amounts to prove that $S_L = S_J.$ The latter is clear from the property $S^{[0]} = (\partial S)^{[0]}$  which implies that $\emptyset \neq x \cap J^{[0]} \neq x $ if and ony if $ \emptyset \neq x \cap L^{[0]} \neq x,$ for all $x \in S$ and this concludes the proof.
\end{proof}
\end{prop}

In \cite{dj15}, a notion of slice is defined and corresponds to a restriction of the present definition to simplicial complexes satisfying the condition of being homeomorphic to a cylinder (where the topology is defined via a notion of geometric realisation). A notion of midsection is also defined and studied essentially in dimension 2, for which it is a cell complexes made only of triangles and quadrangles (2-cell containing 4 vertices). Interestingly, in our framework both the boundary components of a slice and the associated midsection belong to the same class of cc. Moreover, in the context of causal cobordisms discussed in part \ref{sseccatcausalcob}, boundary components and midsections will be put on the same footing, being both a cc in a geometric sequence. 

In part \ref{sseccatcausalcob} we also see that general geometric sequences can be seen as constituted of two types of sub-sequences: the slice sequence introduced next, and the dual notion of connecting sequence introduced in part \ref{sseccompcob} when discussing the composition of cobordisms.

\begin{defi}[Slice sequence] \label{defsliceseq}
A geometric sequence of the form
$$J \red_{\sphi_J} J' \loc_{\cphi_J} M \col_{\cphi_L} L' \der_{\sphi_L} L$$
will be called a \textit{slice sequence.} More explicitly this implies that $J,J',M,L',L$ are elements in $\mlc^R$ for some $R \leq 0$ and that $\sphi_J, \cphi_J, \sphi_L, \cphi_L$ satisfy 
$$ \sphi_J \cpa \cphi_J, \quad \sphi_L \cpa \cphi_L, \quad  \cphi_J \perp \cphi_L.$$ 
\end{defi}

We can finally turn to our main result.

\begin{thm}[Correspondence] \label{thmcorresp}
There is a one-to-one correspondence between slice sequences and slices, up to cc-isomorphism. More precisely, a sequence
$$J \red_{\sphi_J} J' \loc_{\cphi_J} M \col_{\cphi_L} L' \der_{\sphi_L} L$$
is a slice sequence if and only if the poset $(S, \rk_S)$ defined by
$$S :=  (\cphi_J \sqcup \cphi_L)(M) \sqcup J \sqcup L,$$
$$\rk_S :=  \rk_{(\cphi_J \sqcup \cphi_L)(M)} + \rk_J + \rk_L, $$ 
is an element in $\nsc_s$ such that
$$\partial S = J \sqcup L, \quad M = M^S, \quad J(S) = J', \quad L(S) = L',$$
$$ \phi_I = \phi_I^S ~~\text{ for all }~~ \phi \in \{\sphi,\cphi\}, ~~ I \in \{J,L\}.$$

\begin{proof}
We first show that a slice sequence is uniquely defined from a slice. This can be obtained directly as a consequence of Corollary \ref{cortransitmidsec} if we also show that if $S$ is a slice with boundary components $J$ and $L$ then 
$$ \cphi_J^S \perp  \cphi_L^S.$$
 
In order to prove that $\cphi_J^S$ and $\cphi_L^S$ are orthogonal, i.e. satisfy the point \ref{condorth} of Definition \ref{defrefnorth}, let us consider $m_1,m_2 \in M^S$ such that $\cphi_J^S(m_1) \cap \cphi_J^S(m_2) \neq \emptyset$ and $\cphi_L^S(m_1) \cap \cphi_L^S(m_2) \neq \emptyset.$  By Proposition \ref{propslice} we have that $m_i = \E_J^{y_i} = \E_L^{y_i}$ for some $y_i \in S_J = S_L,$ for $i=1,2,$ and the definition of $\cphi_J^S$ in Proposition \ref{proptransit} implies that $\cphi_J^S(m_i) = (y_i)_J$ for $i=1,2.$ By assumption we therefore have that $(y_1)_J \cap (y_2)_J \neq \emptyset$ and $(y_1)_L \cap (y_2)_L \neq \emptyset,$ hence $y_1 \cap y_2 \neq \emptyset.$ Lemma \ref{lemmidsec} implies $\E_J^{y_1 \cap y_2} = \E_J^{y_1} \cap \E_J^{y_2}$ hence we get the following inclusion for $I=J,L:$
\begin{align*}
(y_1 \cap y_2)_I &= \{ e \cap J^{[0]} ~|~ e \in \E_J^{y_1 \cap y_2}\} = \{ e \cap J^{[0]} ~|~ e \in \E_J^{y_1} \cap \E_J^{y_2}\}\\
&\subset \{ e \cap J^{[0]} ~|~ e \in \E_J^{y_1}\} \cap \{ e \cap J^{[0]} ~|~ e \in  \E_J^{y_2}\} = (y_1)_I \cap (y_2)_I.
\end{align*}
To prove the reverse inclusion, and conclude that $\cphi_J^S$ and $\cphi_L^S$ are orthogonal, suppose by contradiction that there exists a vertex $v$ in $(y_1)_I \cap (y_2)_I \setminus	(y_1 \cap y_2)_I$ for $I = J$ or $I=L.$ If $\rk(y_1 \cap y_2) = 1$ we directly obtain a contradiction as in this case $v \in y_1 \cap  y_2 = (y_1 \cap y_2)_I.$ If $\rk(y_1 \cap y_2) \geq 2,$ we have in particular that $v$ is by assumption contained in an edge $e \subset \E_I^{y_1}$ and since $v \notin (y_1 \cap y_2)_I,$ there is no edge in $\E_I^{y_1 \cap y_2}$ containing $v.$ As a consequence, if $x$ is a minimal cell in $\abv(v) \cap S_I^{y_1 \cap y_2}$ then $\rk(x) \geq 2.$ But $x$ is then also a minimal cell in $S(v)$ and this contradicts the assumption that $(S,\partial S)$ is pure, since both $e$ and $x$ are minimal cells and $1 = \rk(e) \neq \rk(x).$

We then show how to obtain a slice from a slice sequence. By Proposition \ref{propredbdry}, it is sufficient to consider the case $J = J'$ and $L = L'$ since if, in this case, we obtain a slice $S$ (satisfying the desired conditions) the latter result using $\sphi = \sphi_I$ for  $I = J,L$ indeed leads $\sphi_I^S = \sphi_I$ and $I(S) = I'.$

Our goal is then to prove that $(S, \rk_S) \in \nsc_s$ where we recall that $(S, \rk_S)$ is defined by
$$S :=  (\cphi_J \sqcup \cphi_L)(M) \sqcup J \sqcup L,$$
$$\rk_S :=  \rk_{(\cphi_J \sqcup \cphi_L)(M)} + \rk_J + \rk_L, $$
and $(\cphi_J \sqcup \cphi_L)(M)$ is the augmented poset associated to $\cphi_J$ and $\cphi_L.$
As usual we start by proving that $S$ is a cc. By Lemma \ref{lemsqcup} the Axioms \ref{cccrank} and \ref{cccenough} are directly derived from the same axioms for the cc $J,L$ and $M.$ 

As for Axiom \ref{cccinter}, it is sufficient to show that if $m,m' \in M,$ then
$$x := (\cphi_J \sqcup \cphi_L)(m) \cap (\cphi_J \sqcup \cphi_L)(m') \in S \cup \{ \emptyset \}.$$
If either $(\cphi_J(m) \cap \cphi_J(m') = \emptyset$ or $(\cphi_L(m) \cap \cphi_L(m') = \emptyset$ then Axiom \ref{cccinter} for respectively $L$ and $J$ implies that $x \in L$ or $x \in J.$ If both $(\cphi_J(m) \cap \cphi_J(m')$ and $(\cphi_L(m) \cap \cphi_L(m')$ are non-empty, then as a consequence of the orthogonality of $\cphi_J$ and $\cphi_L$ we have that $m \cap m' \neq \emptyset$ and 
$$x = (\cphi_J \sqcup \cphi_L)( m \cap m') \in S.$$

In order to show Axiom \ref{cccdiamond} for $S,$ the only cases that are not directly obtained from Axiom \ref{cccdiamond} for $J,L$ or $M$ are inclusions of the form $w \subsetneq (\cphi_J \sqcup \cphi_L)(m),$ for some $m\in M$ and $w \in J \sqcup L$ such that 
$$\rk_S(w) = \rk_S((\cphi_J \sqcup \cphi_L)(m)) - 2 = \rk_M(m) - 1.$$ 
To deal with this cases we set $(\cphi_J \sqcup \cphi_L)(m):= x \sqcup y$ and assume without loss of generality that $w \in J,$ hence $w \subset x = \cphi_J(m).$ As noted in Remark \ref{remcond3redncol}, condition \ref{cond3col} implies that $\rk_J(\cphi_J(m)) \leq \rk_M(m).$ Therefore we have
$$ \rk_S(w) = \rk_J(w) \leq \rk_J(x)  \leq \rk_M(m)$$
and this leaves us with two possibilities: either $\rk_J(x) = \rk_M(m) - 1$ or $\rk_J(x) = \rk_M(m).$ In the first case we have $w = x,$ so that \ref{cond4col} implies
$$ \abs{\face{m} \cap (\cphi_J)^{-1}(x)} = 2.$$
By Lemma \ref{lemsqcup} an element $m' \in M$ belongs to $\face{m} \cap (\cphi_J)^{-1}(x)$ if and only if it satisfies that $\rk_S(\cphi_J(m') \sqcup \cphi_L(m')) = \rk_S(x) + 1$  and $\cphi_J(m') = x.$ Moreover any such $m'$ satisfies $\cphi_L(m') \subset y$ hence $m' \in \face{m} \cap (\cphi_J)^{-1}(x)$ if and only if
$$ \cphi_J(m') \sqcup \cphi_L(m') \in \cface{x} \cap \face{ x \sqcup y}.$$
Therefore we indeed have
$$ \abs{\face{m} \cap (\cphi_J)^{-1}(x)}  = \abs{ \cface{x} \cap \face{ x \sqcup y}}$$
which concludes the case when $x = w.$

The second case is when $w \subsetneq x = \cphi_J(m)$ and therefore $\rk_J(\cphi_J(m)) = \rk_M(m).$ In this case \ref{cond5col} implies
$$ \abs{\face{m} \cap (\cphi_J)^{-1}(x')} = 1 \quad \forall x' \in \face{x}.$$
Since $w \in \face{x}$ we have $\face{m} \cap (\cphi_J)^{-1}(w) = \{w'\}$ and therefore
$$ \cface{w} \cap \face{ x \sqcup y} = \{ x, \cphi_J(w') \sqcup \cphi_L(w')\}.$$
This concludes the proof that $S$ is a cc.

$S$ is graph-based since every edge in $(\cphi_J \sqcup \cphi_L)(M)$ is of the form $\cphi_J(v) \sqcup \cphi_L(v)$ for some vertex $v \in M^{[0]}$ and $\rk_J(\cphi_J(v)) = \rk_L(\cphi_L(v)) = 0$ by Remark \ref{remcond3redncol}.

$S$ is also clearly connected and cell-connected, as every cell $\cphi_J(m) \sqcup \cphi_L(m) \in (\cphi_J \sqcup \cphi_L)(M)$ satisfies that $\bel_J(\cphi_J(m))$ and $\bel_L(\cphi_L(m))$ are connected and can be linked by an edge of the form $\cphi_J(v) \sqcup \cphi_L(v)$ where $v \in m.$

By Remark \ref{remrksredncol} we have $\Rk(J) = \Rk(M) = \Rk(L) = R.$ It follows that $S$ is pure of rank $R+1$ since  the surjectivity of $\cphi_J$ and $\cphi_L$ implies that every cell in $J \sqcup L$ is contained in a cell of the form $\cphi_J(m) \sqcup \cphi_L(m)$ for some $m \in M^R$ and hence
$$ \rk_S(\cphi_J(m) \sqcup \cphi_L(m)) = \rk_M(m) + 1 = R + 1.$$

$S$ is non-branching as a consequence of Lemma \ref{lemsqcup} for the cells in $(\cphi_J \sqcup \cphi_L)(M)$ and since $J$ and $L$ are also non-branching. 

By Lemma \ref{lemsqcup} and the closeness of $M$ every $R$-cell in $S \setminus (J \sqcup L)$ is contained in two $(R+1)$-cells. Therefore every $R$ cell contained in one $(R+1)$-cell must be contained in either $J$ or $L$ and by the surjectivity of $\cphi_J$ and $\cphi_L$ it is the case for all maximal cells of $J$ and $L,$ hence $\partial S = J \sqcup L.$

$S$ is non-pinching since an $S$-pinch would imply an $M$-pinch by Lemma \ref{lemsqcup} and a $\partial S$-pinch would imply a $J$-pinch or an $L$-pinch by the previous derivation, all of which are excluded since $M,J,L$ are non-pinching.

As a consequence we indeed have that $S \in \nsc.$
It remains to show that $S$ is a slice. First, we have that $\kz = J^{[0]} \sqcup L^{[0]}$ since every cell in $(\cphi_J \sqcup \cphi_L)(M)$ has  rank higher than or equal to one. Clearly  $S_J = S_L = (\cphi_J \sqcup \cphi_L)(M)$ and by Lemma \ref{lemsqcup} this implies $M = M^S.$ $(S,\partial S)$ is also clearly non-degenerate and it is pure by the following argument. If, say, $y \in S_J$ then $y = \cphi_J(m) \cap \cphi_L(m)$ for some $m \in M$ and $\bel_J( y ) = \bel_J( \cphi_J(m))$ is a pure cc. Hence if $y$ has minimal rank among elements in $S_J(x)$ then $\cphi_J(m) = x.$ By  condition \ref{cond3col} we can assume that $\rk_M(m) = \rk_J(x)$ therefore
$$ \rk_S(y)  = \rk_M(m) + 1 = \rk_J(x) + 1 = \rk_J^S(x) + 1,$$
and this shows that all elements in $S^-(x)$ have same rank.

Concerning uniformity, for all $\cphi_J(m) \sqcup \cphi_L(m) \in (\cphi_J \sqcup \cphi_L)(M) = S_{\partial S}$ we have that
\begin{align*}
(\cphi_J(m) \sqcup \cphi_L(m))_J &= \{ v \in \cphi_J(m) ~|~ \exists w \in m, ~ \cphi(w) = v \} \\
&= \cphi_J(m) = \left( \cphi_J(m) \sqcup \cphi_L(m)  \right) \cap J^{[0]}
\end{align*}
and similarly for $L.$ Since we also assumed that $J(S) = J \in \mlc$ and $L(S) = L \in \mlc$ it follows that $(S,\partial S)$ is indeed uniform.

Finally, we have that $\cphi_I = \cphi_I^S$ for $I \in \{J,L\}$ by the following argument. It is sufficient to show this for the case $I=J$ in which for any $m \in M$ we have 
$\E_J^x = \{ \cphi_J(v) \sqcup \cphi_L(v) ~|~ v \in m \},$
where $x = \cphi_J(m) \sqcup \cphi_L(m).$ By identifying each vertex $v \in m$ with the edge $\cphi_J(v) \sqcup \cphi_L(v) \in (M^S)^{[0]}$ we obtain $\E_J^x = m$ and therefore
$$ \cphi_J^S(\E_J^x) = x_J = x \cap J^{[0]} =  \cphi_J(m),$$
where in the second equality we used the assumption $J = J(S).$
\end{proof}
\end{thm}

\newpage
\section{Cobordisms} \label{SecCobs}

This final section is based on the results of the previous sections and is organised as follows. We first define cobordisms as a type of relative cc with boundary on which one can generalize the duality map of closed cc, as shown in Theorem \ref{thmdualcob}, an other major result of this work. This is appears to be the most general way to define such a duality that acts as an involution on a sub-class of cobordisms, and in particular it doesn't use the assumption of uniformity.

In part \ref{sseccompcob} we define connecting sequences and show how this notion can be used to define the union of two cobordisms sharing a boundary component. This operation however relies on the assumption of uniformity, hence this composition operation is not compatible with the duality map since the dual of a uniform cobordism is not necessarily uniform. 

The part \ref{sseccatcausalcob} finally introduces a natural class of cobordisms on which both the duality map and the composition operation are defined. These are called causal cobordisms as they are decomposable into a sequence of slices as for CDT. We notice that the resulting category exhibits interesting symmetries.

\subsection{Definition and duality} \label{ssectiondcob}

In order to obtain an involutive duality map on a certain sub-class of cc in $\nsc$, we have to focus on how this duality acts on the cells of the boundary (this property being already satisfied for cells not on the boundary, as a consequence of Lemma \ref{claim2propdual}). The idea here is to look for a canonical way to define a dual set for cells of the boundary, i.e. a notion of dual set which solely depends on the structure of the neighbouring cells. For this purpose, we will use a different type of dual set, noted $\bual{A}$ for $A$ a set of vertices. If $K \in \nsc^R$ and $y \in (\partial K)^{[R-1]} $ is a maximal cell in the boundary, one must have in particular that the dual of $y$ is an edge of the dual of $K,$ so that it becomes again a sub-maximal cell of the double dual of $K.$ Since $\bual{y}$ therefore has to contain at least two elements, it is natural to define
$$ \bual{y} = \{ y , z \},$$
where $z$ is the unique element of $\kR$ containing $y.$ In fact, the "generalized dual" of a cell $x \in \partial K$ will be defined as
$$ \bual{x} := \{ y \in (\partial K)^{[R-1]} ~|~ x \subset y\} \sqcup \{z \in \kR ~|~ x \subset z\} = \cdual{x}{\partial K} \sqcup \cdual{x}{K},$$
where this definition also applies to any set $A \subset \kz,$ with the convention that $\cdual{A}{\partial K} = \emptyset$ whenever $A \not \subset \partial K^{[0]}.$

The first issue one then encounters is that $\bual{K}$ is not a cc, as for example in the case of $y \in (\partial K)^{[R-1]},$ the poset $\bual{K}$ does not contain the vertex $\{y\} = \cdual{y}{\partial K}.$ One can easily fix this by considering the dual of $K$ to be $\bual{K} \sqcup \dual{\partial K}$ but then this notion of dual is not involutive as it "adds" more cells to the primal cc: for example the dual of a graph made of one edge, therefore having two boundary components each made of one vertex, is a graph made of two edges forming a simple path of length 2. The double dual would then be a graph forming a simple path of length 3, etc. The radically opposite alternative would be to "ignore" the cells in $\partial K$ and define the dual of $K$ to be $\dual{K \setminus \partial K},$ with obvious drawbacks when it comes to the involution property, as this definition "removes" cells. It turns out that alternating between these two definitions is essentially the idea behind the construction of the duality for cobordisms: the dual of a path of length 1 is a path of length 2 and the double dual is again a path of length 1.
But then for each boundary component, one has the choice of either applying one or the other definition of dual cell. Relative cc therefore become a handy solution, as one can define the dual of a relative cc $(K,J)$ where $J \leq \partial K$ using the definition that adds cells on the cells in $\partial K \setminus J$ and use the definition that removes cell on the cells in $J.$

This is where the picture with cobordisms arises: an element $(K,J)$ where $K \in \nsc^R$ and $J \in \mlc^{[R-1]}, ~ J \leq \partial K,$ can be interpreted as a cobordism from some "ingoing components" $J$ to some "outgoing components" $L := \partial K \setminus J.$ In this picture, the duality map switches the locations of the ingoing components and the outgoing components.

Let us now make this more precise by introducing our combinatorial notion of cobordism. For this we will use the notation $(K-J)$ to distinguish such an element from a usual relative cc. The sign "$-$" is used to signal that cells in $J$ can be thought of as "removed" in the Definition \ref{defdualcob} of the duality map, since it acts on $(K-J)$ as if the cells in $J$ are not in $K.$

\begin{defi}[Cobordism] \label{deficob}
The set of \textit{cobordisms} is defined as 
$$ \cob := \{ (K-J) ~|~ K \in \nsc, ~ J \leq \partial K, ~ J \in \mlc^{\Rk(K)-1}_\sqcup, ~(K,J) \text{ non-deg. and local}\}.$$
We use $\cob^R$ to denote elements in $\cob$ such that $\Rk(K) = R,$ in which case $R$ is also called the \textit{rank} or \textit{dimension} of the cobordisms in $\cob^R.$ The boundary of a cobordism is simply defined as 
$$\partial (K-J) := \partial K.$$
The boundary components $J$ of a cobordism $(K-J)$ are called \textit{ingoing (boundary) components} or \textit{removed components} of the cobordism and the cells in $J$ are sometimes called the \textit{removed cells}. The boundary components in $\partial K \setminus J$ are called \textit{outgoing (boundary) components} of $(K -J).$ 
We say that $(K-J) \in \cob$ is \textit{exactly collared} if $(K,J)$ is exactly collared.
\end{defi}

\begin{figure}[!h]
\centering
\includegraphics[scale=0.8]{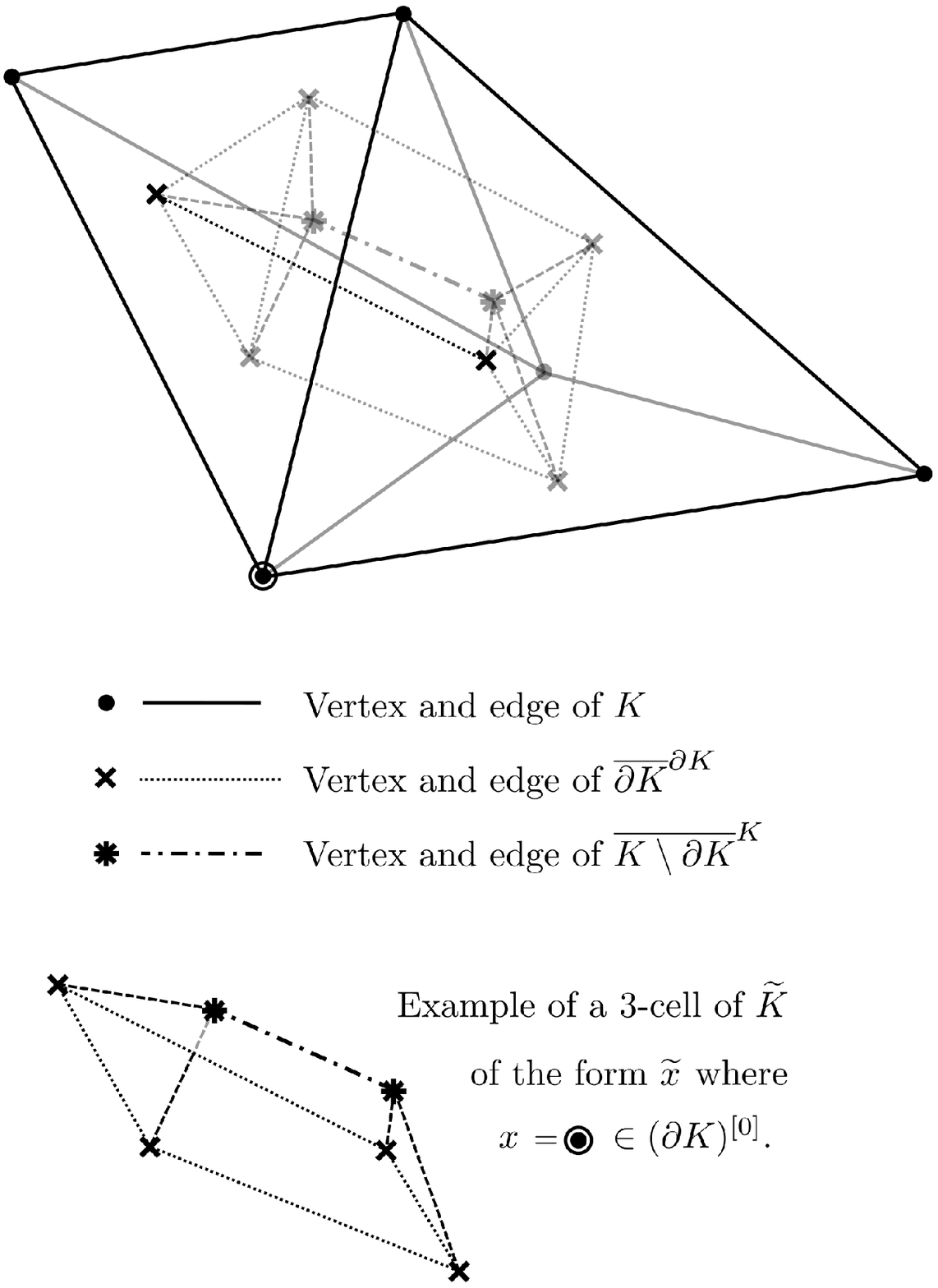}
\caption{\label{figdualbitetrahedra}A cc $K \in \nsc$ composed of two tetrahedra sharing a triangle and seen as a 3-dimensional cobordism $(K - \emptyset),$ i.e. with empty ingoing component. Its dual $\dual{(K - \emptyset)}$ is represented inside $K.$ In this example the entire boundary of the dual is made of removed cells (i.e. cells in the ingoing component). The cobordism $\dual{(K - \emptyset)}$ has only two vertices not lying on the boundary corresponding to the dual of the two tetrahedra in $K.$}
\end{figure}

\begin{figure}[!h]
\centering
\includegraphics[scale=0.8]{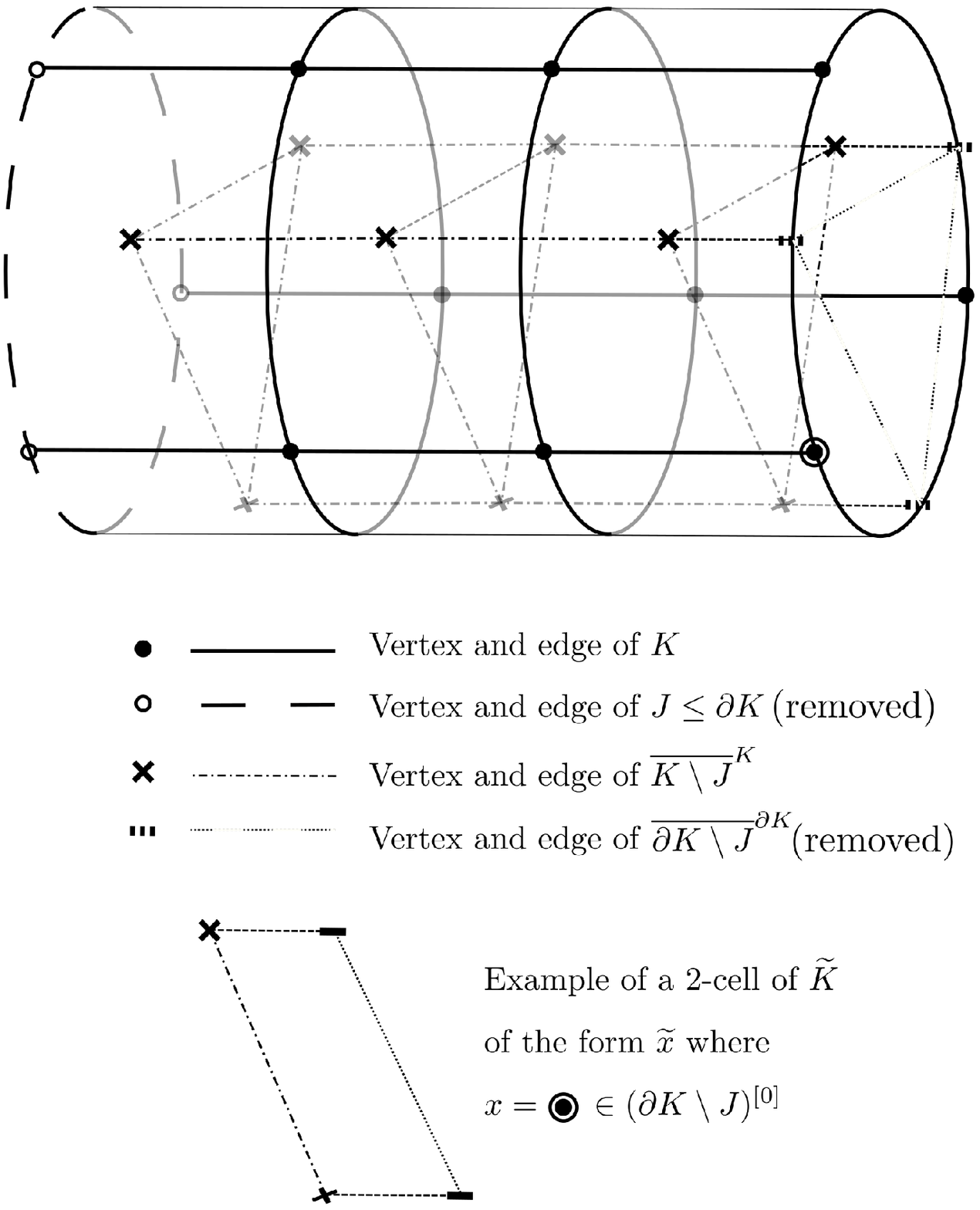}
\caption{\label{figdualcylinder} An example of a 2-dimensional cobordism that can be seen as a discretization of the boundary of a cylinder. In this case the primal and dual cobordisms are isomorphic, with the location of the ingoing and outgoing components exchanged, the ingoing component being represented using cells with white stripes. This is an example of an exactly collared cobordism.}
\end{figure}

We will discuss some examples of cobordisms below, as those shown in Figure \ref{figdualbitetrahedra} and \ref{figdualcylinder}. But let us point out that this is a rather large class of objects including in particular all elements $(K-J)$ where $K$ is a simplicial complex, as the assumption of non-degeneracy and locality are directly satisfied in this cases.

Our next goal is to introduce the duality map on cobordisms. The following definition constitutes a generalization of the Definition \ref{defdualset} of dual cell to the cells on the boundary of a non-singular cc.

\begin{defi}\label{defbdualset}
Let $K$ be a cc and let $A \subset \kz$ be non-empty. We define the \textit{$\sim$-dual of $A$} to be
$$\bdual{A}{K}:= \cdual{A}{K} \sqcup \cdual{A}{\partial K },$$
where by convention $\cdual{A}{\partial K} = \emptyset$ if $A \not\subset (\partial K)^{[0]}.$ We also write $\bual{x}$ for $\bdual{x}{K}$ when there is no ambiguity on $K.$
\end{defi}


For a collection of sets $\A \subset \pws{\kz},$ we use the notation $$ \bdual{\A}{K} := \{ \bdual{A}{K} ~|~ A \in  \A\}.$$
For example if $(K-J) \in \cob,$ we obtain $ \bdual{K \setminus J}{K} = \{ \bdual{x}{K} ~|~ x \in K \setminus J \}.$ We sometimes simply write $\bual{K \setminus J}$ instead of $\bdual{K \setminus J}{K}.$ Note that if $K$ is closed then $\bual{K} = \bual{K \setminus \emptyset} = \dual{K}.$

We would also like to define a rank function for elements in $\bual{K}$ when $K \in \nsc.$ This can be done by mean of the following lemma. 

\begin{lem} \label{leminclbdual}
Let $K$ be a non-singular cc and let $x,y \in K.$ Then 
$$\bdual{x}{K} \subsetneq \bdual{y}{K} \quad \text{ if and only if } \quad y \subsetneq x.$$
\begin{proof}
This is a simple consequence of Lemma \ref{claim2propdual} applied to cells in $K \setminus \partial K$ and $\partial K.$
\end{proof}
\end{lem}

As a consequence of Lemma \ref{leminclbdual}, the map $x \mapsto \bual{x}$ is a bijection from a non-singular cc $K$ to $\bual{K}$ and the rank function $$ \rk_{\bual{K}}( \bual{x} ) := \Rk(K) - \rk_K(x)$$  where $x \in \bual{K}$ is well defined for all $\bual{x} \in \bual{K}.$

Let $K, K'$ be disjoint posets with rank functions $\rk_K$ and $\rk_{K'}.$ We define the disjoint union $K \sqcup K'$ to be a ranked poset with rank function
$$\rk_{K \sqcup K'} := \rk_K + \rk_{K'}.$$
For an element $(K-J) \in \cob,$ this definition allows us to have a rank function on the poset defined as the disjoint union
$$ \bual{K \setminus J} \sqcup \dual{\partial K \setminus J} = \bdual{K \setminus J}{K} \sqcup \cdual{\partial K \setminus J}{\partial K}$$
used below in our definition of dual cobordism.
Note that the latter union is indeed  disjoint by the non-degeneracy of $(K,J):$ every cell $x \in \partial K \setminus J$ satisfies $\cdual{x}{\partial K} \subsetneq \bdual{x}{K},$ which implies in particular that $\cdual{x}{\partial K} \neq \bdual{x}{K}$ for all such $x$ and therefore
$\bdual{K \setminus J}{K} \cap \cdual{\partial K \setminus J}{\partial K} = \emptyset.$

More explicitly, if $(K,J)$ is non-degenerate our notation leads to the following expressions for the rank function on $\bual{K \setminus J} \sqcup \dual{\partial K \setminus J}:$
$$ \rk_{\bual{K \setminus J} \sqcup \dual{\partial K \setminus J}}(x') = \begin{cases}
\rk_{\kb}(x') = R - \rk_K(x) &\textit{if } x' = \cdual{x}{K} \textit{ for } x \in K \setminus \partial K,\\
\rk_{\bual{K}}(x') = R - \rk_K(x) &\textit{if } x' = \bdual{x}{K} \textit{ for } x \in \partial K,\\
\rk_{\dual{\partial K}}(x') = (R - 1) - \rk_{\partial K}(x) &\textit{if } x' = \cdual{x}{\partial K} \textit{ for } x \in \partial K.
\end{cases}$$
where $R = \Rk(K).$ The particular case $J = \emptyset$ yields 
$$ (\bual{K} \sqcup \dual{\partial K})^{[0]} = \cdual{\kR}{K} \sqcup \cdual{(\partial K)^{[R-1]}}{\partial K},$$
for the dual sets of rank $0.$ For dual sets of rank $1 \leq r \leq R-1$ we have
$$(\bual{K} \sqcup \dual{\partial K})^{[r]} = \cdual{(K \setminus \partial K)^{[R-r]}}{K}  \sqcup \cdual{\partial K^{[R-r-1]}}{\partial K} \sqcup \bdual{ (\partial K)^{[R-r]}}{K}.$$
and the dual sets of rank $R$ are
$$(\bual{K} \sqcup \dual{\partial K})^{[R]} = \cdual{(K \setminus \partial K)^{[0]}}{K}  \sqcup \bdual{ (\partial K)^{[0]}}{K}.$$
For example, if $e' \in (\bual{K} \sqcup \dual{\partial K})^{[1]}$ then $e'$ is a dual set having one of the three following forms:
\begin{itemize}[topsep=2pt, parsep=2pt, itemsep=1pt]
\item the "usual" dual of a sub-maximal cell of $K$ i.e. $e' = \cdual{y}{K}$ for $y \in K^{[R-1]},$
\item the "usual" dual of a sub-maximal cell of $\partial K,$ i.e. $e' = \cdual{y}{\partial K}$ for $y \in (\partial K)^{[R-2]},$
\item the $\sim$-dual of a maximal cell in $\partial K$, i.e. $e' = \bdual{y}{K} = \{ y, z\}$ where $y \in (\partial K)^{[R-1]}$ and $z \in \kR$ is the unique maximal cell containing $y.$
\end{itemize}
With this in mind, the next definition of duality map for cobordism is more transparent.

\begin{defi}[Dual cobordism] \label{defdualcob}
Let $(K-J) \in \cob,$ we define the \textit{dual cobordism} to be 
$$ \dual{(K-J)} := \left( ( \bual{K \setminus J} \sqcup \dual{\partial K \setminus J}  ) - \dual{\partial K \setminus J} \right).$$ 
\end{defi}

This definition, above all, raises the following question: is $\dual{(K - J)}$ also a cobordism? This question is answered positively by the  following theorem. 

\begin{thm} \label{thmdualcob}
If $(K- J) \in \cob^R$ then:
\begin{itemize}
\item $\dual{(K - J)} \in \cob^R,$ 
\item $\dual{(K - J)}$ is exactly collared,
\item $\dual{K_J} \in \mlc^{R-1}_\sqcup$ and  
$$\partial \dual{(K -J)} = \dual{\partial K \setminus J} \sqcup \dual{K_J}.$$
\end{itemize}
\begin{proof} \renewcommand{\qedsymbol}{}
In Appendix \ref{appThmcob}.
\end{proof}
\end{thm}

In the proof of Theorem \ref{thmdualcob} it is seen that, as a consequence of Lemma \ref{lemmidsec}, we have $\dual{K_J} \cong \dual{M_J^K}.$ In other words, the dual outgoing component $\dual{K_J}$ is better understood as the dual of the midsection associated to the ingoing components $J.$ We point out that although $\dual{K_J}$ is a closed cc, it is not the case of $K_J$ as cells in the collar are not defined as subsets of edges in the collar (i.e. minimal elements in the collar).

Definition \ref{defdualcob} leads to the interpretation that the action of the duality map on cobordisms in effect exchanges the role of the ingoing and outgoing boundary components. This can be seen particularly well in the case of the cylinder-like cobordism depicted in Figure \ref{figdualcylinder}.

As an example, we can also look at what Definition \ref{defdualcob} implies for the special cases where $J = \emptyset$ and $J = \partial K.$ In the first case we get:
$$ \dual{(K - \emptyset)} =  \left( ( \bual{K} \sqcup \dual{\partial K}  ) - \dual{\partial K} \right).$$
We can think of this case as duality acting on a non-singular cc $K,$ as no other data is specified. Such an example is depicted in Figure \ref{figdualbitetrahedra}. As a consequence of Theorem \ref{thmdualcob}, we have that $\bual{K} \sqcup \dual{\partial K} \in \nsc$ and 
$$ \partial ( \bual{K} \sqcup \dual{\partial K}  )  = \dual{\partial K}.$$
The dual of $(K - \emptyset)$ is a cobordism with ingoing components equal to the entire boundary. For such cobordisms, corresponding to our second case $J = \partial K,$ the duality gives
$$  \dual{(K - \partial K)} =  \left(  \bual{K \setminus \partial K}  - \emptyset \right).$$
Another consequence of Theorem \ref{thmdualcob} is therefore that $\bual{K \setminus \partial K} \in \nsc$ for all $(K - \partial K) \in \cob.$

Finally, the following corollary concerns the special case of exactly collared cobordisms, for which the involution property holds. To state this result, we defined the isomorphism $(K - J) \cong (K' - J')$ between cobordisms as $K \cong K'$ with an isomorphism mapping $J$ to $J'.$ 

\begin{cor} \label{corbidualcob}
If $(K-J) \in \cob$ is exactly collared \footnote{A consequence of this corollary is that $(K-J)$ is exactly collared if and only if there exists $(K'-J') \in \cob$ such that $\dual{(K'-J')} = (K - J),$ a property that motivated us to include the notion of "exactness" in the term used to denote such cobordisms.} then we have
$$\partial \dual{(K -J)} \cong \dual{J} \sqcup \dual{\partial K \setminus J} \quad \text{and} \quad \dual{\dual{(K-J)}} \cong (K - J).$$
\begin{proof}
By Theorem \ref{thmdualcob} and using the assumption that $(K-J)$ is exactly collared, i.e. that $ J \cong M_J^K,$ and the identity $ \dual{M_J^K} \cong \dual{K_J},$ we directly obtain that $\partial \dual{(K -J)} \cong \dual{J} \sqcup \dual{\partial K \setminus J}.$

The identity $\dual{\dual{(K-J)}} \cong (K - J)$ is a consequence of the following computation, using the previous point and the assumption that $(K-J)$ is exactly collared:
\begin{align*}
\dual{\dual{(K-J)}} &= \dual{ ( \bual{ K \setminus J} \sqcup \dual{ \partial K \setminus J} - \dual{ \partial K \setminus J} )}\\
&= \left( \bual{ \left( ( \bual{ K \setminus J } \sqcup \dual{ \partial K \setminus J} ) \setminus \dual{ \partial K \setminus J} \right)} \sqcup \dual{\dual{ K_J }}  - \dual{ \dual{ K_J}} \right)\\
&\cong ( \bual{ \bual{ K \setminus J}} \sqcup \dual{ \dual{ J}} - \dual{\dual{J}} ) \cong ( K - J ),
\end{align*}
where we also used Lemma \ref{leminclbdual} and Proposition \ref{propdual} for the last identity.
\end{proof}
\end{cor}

\newpage

\clearpage 

\subsection{Composition} \label{sseccompcob}

Composing cobordisms will require the assumption of uniformity, simply defined as follows for cobordisms.

\begin{defi}[Uniform cobordism]
A cobordism $(K - J) \in \cob,$ will be called \textit{uniform} if $(K, J')$ is uniform for all connected components $J'$ of $\partial K.$ 
\end{defi}

The idea is then relatively simple: two uniform cobordisms $(K - J),$ $(K' - L)$ such that $\partial K = J \sqcup L$ induce a sequence of the type 
\begin{align} \label{equconnseq}
M_L^K \col_{\cphi_L^K} L(K) \der_{\sphi_L^K} L \red_{\sphi_L^{K'}} L(K') \loc_{\cphi_L^{K'}} M_L^{K'},
\end{align}
which is geometric if the reductions $\sphi_L^K$ and $\sphi_L^{K'}$ are reflective (these maps were introduced in Proposition \ref{proptransit}). Such sequence is thus given the following name.

\begin{defi}[Connecting sequence] \label{defconnseq}
A geometric sequence of the form
$$M \col_{\cphi_J} J \der_{\sphi_J} I \red_{\sphi_L} L \loc_{\cphi_L} M'$$
is called a \textit{connecting sequence.}
More explicitly, the maps $\cphi_J, \sphi_J, \cphi_L, \sphi_L$ in such a connecting sequence satisfy
$$ \sphi_J \cpa \cphi_J, \quad \sphi_L \cpa \cphi_L, \quad  \sphi_J \refl  \sphi_L.$$
\end{defi}

One can then consider the dual of the connecting sequence (\ref{equconnseq}). This produces a slice sequence and allows to make use of the Correspondence Theorem to obtain a slice. Using the duality map one the cells in the slice one can then ensure that the cells in $K \cup K'$ satisfy the axioms of a cc.

For the sake of completeness, we can formulate this in a somewhat more general manner, using a freedom one has in choosing the (relative) reductions used in the connecting sequence. This freedom is incorporated in the next definition.

\begin{defi}[Connecting boundary components] \label{defconnbdry}
Let $K, H \in \nsc^R$ and $J,L \in \mlc^{R-1}$ be connected components of $\partial K$ and $ \partial H,$ respectively, such that $(K,J)$ and $(H,L)$ are uniform. Then we say that \textit{$K$ and $H$ connect through} $I \in \mlc^{R-1}$ if there exist cell complexes $K^{\sphi_J},H^{\sphi_L} \in \nsc^R$ and relative reductions $(I,K^{\sphi_J}) \red_{\sphi_J} (J,K)$ and $(I, H^{\sphi_L}) \red_{\sphi_L} (L,H)$ such that 
\begin{equation} \label{connectsequdef}
M_J^K \col_{\cphi^K_J} J(K) \der_{\sphi_J \circ \sphi_J^K} I \red_{\sphi_L \circ \sphi_L^H} L(H) \loc_{\cphi_L} M_L^H
\end{equation}
is a connecting sequence ($\cphi^K_J, \sphi_J^K, \cphi_L^H$ and $\sphi_L^H$ are defined in Proposition \ref{proptransit}). In this case we say that $\sphi_J$ and $\phi_L$ \textit{are connecting $K$ and $H$ through $I.$}
\end{defi}

Connecting boundary components are then those that admit a lower bound with respect to the partial order on relative cc induced by $\red$ with the additional condition that the corresponding reductions are reflective. 

Two cc having a connecting boundary components can be associated a ranked poset defined as their union in the following way, shown to be a cc in the next corollary.

\begin{defi}[Union of cc]\label{defunioncc}
Let $K, H \in \nsc^R$ and $J,L \in \mlc^{R-1}$ be boundary components of respectively $K$ and $H$ such that $(K,J)$ and $(H,L)$ are uniform. Let 
$$(K^{\sphi_J},I) \red_{\sphi_J} (K, J), \quad (H^{\sphi_L},I) \red_{\sphi_L} (H,L)$$
be two relative reductions connecting $K$ and $H$ through $I$ and such that  $K^{\sphi_J} \cap H^{\sphi_L} = I.$ We define the \textit{union of $K$ and $H$ by $\sphi_J$ and $\sphi_L$ } to be the poset
$$K \ccup{\sphi_J ~ \sphi_L} H := K^{\sphi_J} \cup H^{\sphi_L}$$
with rank function 
$$\rk_{\sphi_J, \sphi_L} = \rk_{K \ccup{\sphi_J ~ \sphi_L} H} := \rk_{K^{\sphi_J}} + \rk_{H^{\sphi_L}} - \rk_I.$$
In the particular case were $J = I = L,$ i.e. $\sphi_J = \sphi_L = \id_I,$ we will simply write
$$K \ccup{\sphi_J ~ \sphi_L} H = K \ccup{I} H.$$
\end{defi}

\begin{cor}\label{corunioncc}
$(K \ccup{ \sphi_J ~ \sphi_L} H, \rk_{\sphi_J, \sphi_L} )$ as in Definition \ref{defunioncc} is an element of $\nsc^R$ uniquely determined by $K,H$ and the reductions $\sphi_J : I \dans J,$ $\sphi_L : I \dans L$ up to cc-isomorphism.
\begin{proof}
As pointed out in Definition \ref{defikphi}, $K' := K^{\sphi_J}$ and $H' := H^{\sphi_L}$ are uniquely determined by $K,H, \sphi_J$ and $\sphi_L$, hence it remains to show that $K' \cup H' \in \nsc^R.$ Notice first that $\rk_{\sphi_J, \sphi_L}|_I = \rk_I.$

The central argument of this proof is the following. By our assumptions, and considering that we are using the same notation as in Definition \ref{defconnbdry}, we obtain the connecting sequence (\ref{connectsequdef}). As a consequence, and by Lemma \ref{lemdualcompnorth}, the following sequence
$$\dual{M_J^K} \red_{\sphi'_J} \dual{J(K)} \loc_{ \cphi'_J} \dual{I} \col_{\cphi'_L} \dual{L(H)} \der_{ \sphi'_L} \dual{M_L^H}$$
where 
$$\sphi'_J := \dual{\cphi_J^K}, \quad \cphi'_J := \dual{\sphi_J \circ \sphi_J^K}, \quad \cphi'_L := \dual{\sphi_L \circ \sphi_L^H}, \quad \sphi'_L := \dual{\cphi_L^H},$$
is a slice sequence. Therefore Theorem \ref{thmcorresp} provides us with a slice $S$ of rank $R$ such that 
$$ S = \dual{M_J^K} \sqcup (\cphi_J' \sqcup \cphi'_L)(\dual{I}) \sqcup \dual{M_L^H}, \quad  \partial S = \dual{M_J^K} \sqcup \dual{M_L^H}, \quad M^S = \dual{I}.$$
Lemmas \ref{lemmidsec} and \ref{lemsqcup} imply that there is an isomorphism of posets
$$ S \cong \dual{K_I^{\sphi_J}} \sqcup \dual{I} \sqcup \dual{H_I^{\sphi_L}}.$$
Hence by Lemma \ref{claim2propdual} the duality map $ x \longmapsto \dual{x}$ defines a poset anti-isomorphism between $S$ and $K_I^{\sphi_J} \sqcup I \sqcup H_I^{\sphi_L},$ i.e. a bijection between the two posets which reverses inclusion relations. Moreover, for any element $x \in (\cphi_J' \sqcup \cphi'_L)(\dual{I})$ we have
$$ \rk_S(x) = \rk_{\dual{I}}(x) + 1 = (R - 1) - \rk_I(\dual{x}) = R - \rk_{\sphi_J, \sphi_L}(\dual{x}),$$
therefore $\rk_{\sphi_J, \sphi_L}(\dual{x}) = R - \rk_S(x)$ for all $x \in S.$

By the same arguments as in the proof of Proposition \ref{propdual} this implies that the axioms of cc as well as all properties defining elements in $\nsc$ (i.e. graph-based, pure, non-branching, non-pinching, connected, cell-connected) are satisfied for the cells in $K_I^{\sphi_J} \sqcup I \sqcup H_I^{\sphi_L}$ as the same properties are true for the cells in $S.$
\end{proof}
\end{cor}

In particular, if $(K - J) , (H - L) \in \cob$ are uniform cobordisms such that $\partial K = J \sqcup L$ and $K \cap H = L$ then
$$(K \ccup{L} H - J) \in \cob.$$

Hence the simpler case $J = I = L$ in Corollary \ref{corunioncc} will be most important to us for the remaining of this section, as the emphasis will be put primarily on the structure of the resulting cc. However if we fix two cobordisms $(K - J), (H - L)$ to start with as above then there remains some freedom in how one can define their union: 
if $\alpha : (K,L) \dans (K_\alpha,L)$ and $ \beta : (H,L) \dans (H_\beta,L)$ are relative cc-isomorphisms (i.e. relative cc-automorphisms) then $K_\alpha$ and $H_\beta$ also connect through $L$ but in general $ K \ccup{ L} H  \not \cong K_\alpha \ccup{\alpha ~ \beta } H_\beta.$
If we consider the case where $L$ has a large class of cc-automorphism, for example the case of a closed square lattice discussed in \ref{exampclosedcc} \ref{exampsqlattice}, this freedom appears to be similar to a gauge symmetry, as it amounts to choose the discrete analogue of a rotation of the space.

\subsection{The category of causal cobordisms}  \label{sseccatcausalcob}

Having defined a composition operation for cobordisms via the definition of the union of two cc connecting through one of their boundary components, it would be straightforward to define a category whose morphisms correspond to cobordisms, and objects $\mlc_\sqcup^R$ for some rank $R.$ However the dual $\dual{(K - J)}$ of a uniform cobordism $(K - J)$ is only uniform if the relative cc $( ( \bual{K \setminus J} \sqcup \dual{\partial K \setminus J}  ), \dual{K_J})$ is uniform, but there is no assumption guaranteeing this to be true in general. The idea of this part is to restrict our attention to a class of cobordisms having a causal structure generalizing that of CDT on which both a composition operation and the duality map are defined. As a result we obtain a category endowed with three non-trivial dualities (defined as functors of categories).

In order to so, it is convenient to introduce what can be considered as fundamental building blocks of geometric sequences that we call semi-sequences. A \textit{semi-sequence} corresponds to a geometric sequence 
$$M \col_\cphi J' \der_\sphi J,$$
for which we use the lighter notation $M \col \der J.$  If $M,J,J' \in \mlc^d$ are as above then we say that the corresponding semi-sequence $M \col \der J$ has \textit{dimension $d$} and $J'$ is called the \textit{transition of $M \col \der J$} (although there is usually no need to specify it in what follows).

Unless stated otherwise, we set the convention that if $(K, J)$ is a uniform relative cc then the semi-sequence $M_J^K \col \der J$ stands for
$$ M_J^K \col_{\cphi_J^K} J(K) \der_{\sphi_J^K} J.$$

It is also suited to borrow Dirac's braket notation in order to talk about states and dual states forming object of a category, as we do thereafter.

\begin{defi}[Braket notation, sets of states $\sts,$ sequences of states $\seqs$] \label{defbraket}
We introduce the following braket notation, where each \textit{ket} $\ket{J_1, J_2, J_3}$ and \textit{bra} $\bra{J_1, J_2, J_3}$ is \textit{labelled} by three elements $J_i \in \mlc, ~i= 1,2,3,$ satisfying the following conditions:
\begin{itemize}
\item $\ket{M, L , M'}$ denotes the \textit{connecting state} corresponding to the connecting sequence $$M \col \der L \red \loc M',$$
\item $\bra{J, M, L}$ denotes the \textit{slice state} corresponding to the slice sequence 
$$J \red \loc M \col \der L,$$
\item $ \braket{J', M, J}{ N , L , N'}$ indicates that $M \col \der J$ is equal to $N \col \der L,$
\item $\ktbr{N', J ,N}{L , M , L'}$ indicates that $J \red \loc N$ is equal to $L \red \loc M.$
\end{itemize}
We define the \textit{set of states} $\sts$ to be the union of the \textit{set of connecting states} $\ists$ and the \textit{set of slice states} $\osts.$ 
We define a \textit{sequence of states} to be an expression $s_0 \dots s_n$ in which $s_i \in \sts, ~i = 0, \dots, n,$ and $n \in \N$ is the \textit{length} of the sequence, satisfying that $s_0 \dots s_n$ defines a geometric sequence. We denote by $\seqs$ the \textit{set of sequences of states} where for each state $s$ we denote by $\id_s := s$ the sequence of states of length 1 containing $s$ called the \textit{identity sequence on $s.$} If $\sigma := s_0 \dots s_n \in \seqs$ we define the \textit{image of $\sigma$} by $\im(\sigma) := s_0$  and the \textit{domain of $\sigma$} by $\Dom(\sigma) := s_n.$ 

Two sequences of the form $\sigma = s_0 \dots s_n, \gamma = s_n \dots s_m,$ where $n,m \in \N, ~n \leq m$ can be \textit{composed} to form the sequence
$$\sigma \circ \gamma := s_0 \dots s_m.$$
By Remark \ref{remrksredncol} all the labels of states belonging to a given sequence of states are elements in $\mlc$ of the same rank.
We define the \textit{set of $d$-dimensional states} $\sts_d$ to be the set of states labelled by elements in $\mlc^d.$ Similarly, the \textit{set of $d$-dimensional sequences} $\seqs_d$ is defined as the set of sequences of states with labels in $\mlc^d.$
\end{defi}

\begin{defi}[Category $\Cob_d$ of causal cobordism of rank $d$]
Let $d \in \N^*.$ The \textit{category of $d$-dimensional causal cobordisms} $\Cob_d$ is defined by having the set of states $\sts_{d-1}$ as its set of objects and set of morphisms from $a$ to $b$ defined by
$$\homc(a,b) = \{ \sigma \in \seqs ~|~ a = \Dom(\sigma), ~ b = \im(\sigma)\}.$$ 
The composition of morphisms in $\Cob_d$ is defined via the composition of sequences and the identity on $s \in \sts_{d-1}$ is given by $\id_s \in \homc(s,s).$
\end{defi}

It is clear that $\Cob_d$ defines a category, hence it remains to explain how it is related to cobordisms and slices. 

Let us first specify that a cobordism $(S - L)$ is called a \textit{slice} if $S$ is a slice and we will denote the set of slices (seen as cobordisms) by $\cob_s.$

By the Correspondence Theorem \ref{thmcorresp}, we can associate each slice state $\bra{ J, M, L}$ with $(S - L ) \in \cob_s$ such that 
$$\partial S = J \sqcup L,~M^S \cong M,$$ 
and where the slice $S$ corresponds, via the theorem, to the slice sequence
$$J \red \loc M \col \der L.$$
Hence in this case we use the notation
$$ \bra{S - L} := \bra{ J, M, L}.$$
If $\bra{S - L}, \bra{S' - L'}$ are slice states such that there exists a connecting states $\ket{ M, L, M'}$ satisfying that
$$ \braket{S - L}{ M, L, M'} \bra{S' - L'} \in \seqs$$
then this is equivalent to stating that $\partial S' = L \sqcup L'$ and  $M \col \der L \red \loc M'$ is the connecting sequence
$$M^S \col_{\cphi_L^S} L(S) \der_{\sphi_L^S} L \red_{\sphi_L^{S'}} L(S') \loc_{\cphi_L^{S'}} M^{S'}.$$ 
Therefore Corollary \ref{corunioncc} implies that $(S \ccup{L} S' - L')$ is a cobordism so that we define the bra
$$  \bra{ S \ccup{L} S' - L'} := \braket{S - L}{ M, L, M'} \bra{S' - L'},$$
where $M = M^S$ and $M' = M^{S'}$ by the conditions of the braket notation.
More generally, we recursively define bras (and the corresponding cobordisms) of the form $\bra{K - J}$ with $K$ being a union of slices $S_1, \dots, S_n$ using the relation:
$$\bra{ \cup^n_{i = 1} S_i - L_n} := \braket{  \cup_{i = 1}^{n-1} S_i - L_{n-1}}{M_{n-1}, L_{n-1} , M_{n}} \bra{ S_n - L_n },$$
for some connecting states $\ket{M_i, L_i, M_{i+1}}, ~i=1, \dots ,n-1.$

With the latter definitions at hand, we are able to give a precise meaning to the notion of causal cobordisms.

\begin{defi}[Causal cobordisms $\cob_c$] \label{defcauscob}
A cobordism $(K - J) \in \cob$ is said to be \textit{causal} if there exist slices $S_1, \dots, S_n$  such that
$$ (K - J) = ( \cup_{i = 1}^n  S_i - L_n)$$
for some connecting states $\ket{M_i, L_i, M_{i+1}}, ~i = 1, \dots, n-1.$
The set of causal cobordisms is denoted by $\cob_c.$ We will also call a bra of the form $\bra{K - J}$ where $(K - J) \in \cob_c$ a causal cobordism.
\end{defi}

Figure \ref{figdualcylinder} in fact gives a simple example of causal cobordism. 

For each $J \in \mlc$ we conventionally define an element called the \textit{empty causal cobordism at $J$} denoted $\bra{J - J}$ to be a causal cobordism with boundary $J$ and rank $\Rk(J) + 1.$ 
Empty cobordisms are simply introduced to make some expressions of sequences of states more general. We will for example use the bra $\bra{J - J}$ in an expression of the form:
$$ \ket{M, J, M'} \braket{J - J}{N, L, N'}  \quad \text{to imply that} \quad \ket{M, J, M'} = \ket{N, L, N'}.$$

In order to more explicitly connect causal cobordisms with the related category it is also convenient to introduce a notation indicating that a slice is either the first or the last slice of a causal cobordism as follows. Let us consider a causal cobordism $(K - J) \in \cob_c$ given by
$$(K - J) = ( \cup_{i = 1}^n S_i - L_n),$$
for some slices $S_1, \dots, S_n$ and connecting states $\ket{M_i, L_i, M_{i+1}}, ~i = 1, \dots, n-1$ such that $J = L_{n-1}.$ We will name $L$ the other boundary component of $K$ so that $\partial K = J \sqcup L.$
We say that a slice $\bra{S - J} = \bra{J',  M^S , J}$  is the \textit{ ingoing slice of $\bra{K - J}$} and use the notation 
$$\bra{J', M^S, J } \ins \bra{K - J},$$ if $(S - J) = (S_n - L_n).$
Similarly, we say that a slice $\bra{S - L'}= \bra{L , M^S , L'}$ is the \textit{outgoing slice of $\bra{K - J}$} and use the notation
$$\bra{L , M^S, L'} \outs \bra{K - J},$$
if $(S - L') = (S_1 - L_1).$

Using this notation, we can give expressions associated to morphisms in $\Cob_d.$ These are of four different kinds depending on whether their image and domain are connecting or slice states:
\begin{align*}
&\homc( \ket{M, J, M'} , \ket{N', L, N}) := \left\lbrace ~ \ket{N', L, N} \braket{K - J}{M, J, M'} ~:~ \bra{K - J} \in \cob_c  \right\rbrace,\\[1em]
&\homc( \bra{J', M, J} , \ket{N', L, N}) := \left\lbrace ~ \ktbr{N', L, N}{K - J} ~:  \begin{array}{c} \bra{K - J} \in \cob_c, \\ \bra{J',M,J} \ins \bra{K - J} \end{array} \right\rbrace,\\[1em]
&\homc( \ket{M, J, M'} , \bra{ L', N, L}) := \left\lbrace ~ \braket{K - J}{M, J, M'} ~:  \begin{array}{c} \bra{K - J} \in \cob_c, \\ \bra{L', N, L} \outs \bra{K - J} \end{array} \right\rbrace, \\[1em]
&\homc( \bra{J', M, J} , \bra{ L', N, L}) := \left\lbrace ~ \bra{K - J} \in \cob_c ~:  \begin{array}{c} \bra{J',M,J} \ins \bra{K - J},\\ 
\bra{L', N, L} \outs \bra{K - J} \end{array} \right\rbrace.
\end{align*}

It is seen that only the morphisms in the last set $\homc( \bra{J', M, J} , \bra{ L', N, L})$ correspond exactly to the notion of causal cobordisms  from Definition \ref{defcauscob}. The other sets of morphisms include additional connecting states corresponding to connecting sequences associated with one or both of their boundary components. The morphisms of the first set $\homc( \ket{M, J, M'} , \ket{N', L, N})$ actually define the \textit{ sub-category of $d$-dimensional connecting states} $\Cob_d^{\top}$ with $\ists_d$ as its set of objects. Similarly, one can single out the \textit{sub-category of $d$-dimensional slice states} $\Cob_d^{\perp}$ having $\osts_d$ as objects and morphisms defined by the last set $\homc( \bra{J', M, J} , \bra{ L', N, L}).$

\begin{defi}[Duality map on $\sts$ and $\seqs$]
The dual of a sequence of states is defined via the following rules:
\begin{itemize}
\item if $s = \ket{M, L, M'}$ then $\dual{s} = \bra{\dual{M'}, \dual{L}, \dual{M}}$ is the slice state characterized by the slice sequence
$$\dual{M'} \red \loc \dual{L} \col \der \dual{M},$$
\item if $s = \bra{J, M , L}$ then $\dual{s} = \ket{\dual{L}, \dual{M}, \dual{J}}$ is the connecting state characterized by the connecting sequence
$$ \dual{L} \col \der \dual{M} \red \loc \dual{J},$$
\item if $ \sigma = \sigma_1 \sigma_2$ is a sequence of states then
$$ \dual{\sigma} = \dual{\sigma_2} ~~ \dual{\sigma_1}.$$
\end{itemize}
\end{defi}

The definition of the duality map on $\sts$ and $\seqs$ for example implies:
\begin{align}\label{exdualmapbrkt}
\dual{ \braket{ J, M , L}{M , L , M'} } &= \dual{ \ket{M, L, M'}} ~~ \dual{\bra{J, M ,L}} = \braket{\dual{M'}, \dual{L}, \dual{M}}{\dual{L}, \dual{M}, \dual{J}}.
\end{align}

\begin{cor}[Duality map on $\Cob^d$]
Let $a,b \in \sts^d$ and let $\sigma$ be a morphism in $\Cob^d.$ We have the following equivalence:
$$\sigma \in \homc(a,b)  \quad \text{if and only if} \quad \dual{\sigma} \in \homc(\dual{b}, \dual{a}).$$
\end{cor}

In other words, $\Cob_d$ is self-dual as a category under the contravariant functor $\charge = \dual{( \cdot )}$ induced by the duality map.  It is also clear that $\charge$ can be restricted to a functor between the sub-categories $\Cob_d^{\top}$ and $\Cob_d^{\perp}.$

Let us furthermore note the following contravariant functor $\tim = \rev{(\cdot)}$ in $\Cob_d.$ For any state $s \in \sts,$ let the \textit{reverse state} $\rev{s}$  be the state defined by the same labels in reversed order, for example
$$ \rev{\ket{M',L,M}} = \ket{M,L,M'}.$$
Then $s \in \ists$ if and only if $\rev{s} \in \ists$ and similarly for $\osts.$ We also define the \textit{reverse sequence of states} $\rev{\sigma}$ of a sequence $\sigma = \sigma_1 ~ \sigma_2$ by 
$$\rev{\sigma} = \rev{\sigma_2}~ \rev{\sigma_1},$$
which evidently implies that
$$ \sigma \in \homc(a,b) \quad \text{if and only if} \quad \rev{\sigma} \in \homc(b,a).$$

Finally, it is worth mentioning a third (covariant) functor $\parity = (\cdot)^*$ defined as follows. The functor $\parity$ acts on states as a map $s \longmapsto s^*,$ where $s^*$ is the state corresponding to the sequence dual to the sequence characterizing $s$ labelled by the dual of the labels of $s$ in the same order. For example,
$$ \ket{M', L, M}^* = \bra{ \dual{M'}, \dual{L}, \dual{M}}.$$
The functor $\parity$ acts on morphisms in $\Cob_d$ simply by acting on each states of the corresponding sequence of states, i.e. if $\sigma = \sigma_1 ~ \sigma_2$ then 
$$\sigma^* = \sigma_1^* ~ \sigma_2^*.$$
This in turn implies the equivalence:
$$ \sigma \in \homc(a,b) \quad \text{if and only if} \quad \sigma^* \in \homc(a^*, b^*).$$

These functors satisfy $\tim^2 = \charge^2 = \parity^2 = \id_{\Cob_d}$ and are in particular related by the relation $\charge =  \parity ~ \tim ,$ as can be seen for the specific example used in (\ref{exdualmapbrkt}):
\begin{align*}
\left( \rev{\left( \braket{ J, M, L}{M, L, M'} \right)} \right)^* &= \left( \ktbr{M',L,M}{M,L,J} \right)^*\\
&= \braket{ \dual{M'}, \dual{L}, \dual{M}}{\dual{L}, \dual{M}, \dual{J}} = \dual{ \braket{ J, M, L}{M, L, M'} }.
\end{align*}


\newpage
\begin{appendices}

\section{Lemmas \ref{lemcface} to \ref{lemredbdry}} \label{apppflemconcomp}


\begin{lem}\label{lemcface}
Let $x$ be a cell of a cc $K$ such that $\abs{\cface{x}} \geq 2$ then 
\begin{equation*}
x = \bigwedge \cface{x} = \bigcap_{y \in \cface{x}} y.
\end{equation*}
\begin{proof}
Since $\abs{\cface{x}} \geq 2$ Axiom \ref{cccinter} implies $x' :=  \bigcap_{y \in \cface{x}} y \in K.$ If $y'$ is a lower bound of $\cface{x}$ then $y' \subset y$ for all $y \in \cface{x}$ and therefore $y' \subset x'.$ The latter implies $$x' = \bigcap_{y \in \cface{x}} y = \bigwedge \cface{x}.$$ We also have that $x \subset x'$ since $x$ is a lower bound of $\cface{x}.$ If by contradiction $x' \neq x$ then by Axiom \ref{cccrank} $\rk(x') > \rk(x),$ hence $\rk(\bigwedge \cface{x}) = \rk(x) + 1.$ But if $y \in \cface{x}$ then $ \bigwedge \cface{x} \subset y$ and $\rk(y) = \rk( \bigwedge \cface{x})$ so that $y = \bigwedge \cface{x}$ by Axiom \ref{cccrank}, which implies $\abs{\cface{x}} = 1$ and contradicts $\abs{\cface{x}} \geq 2.$
\end{proof}
\end{lem}

\begin{lem} \label{rembdrycc}
If $K$ is a non-singular cc such that there is no $K$-pinch in $\partial K$ then $\partial K$ is closed.
\begin{proof}
Let $R = \Rk(K).$ It is sufficient to show that if $\partial K$ contains no $K$-pinch then 
$$\abs{\cface{x} \cap \partial K} = 2, \quad \forall x \in (\partial K)^{[R-2]}.$$ 
We start by showing that $\abs{ \cface{x} \cap \partial K} \geq 2,$ which in particular implies that $\partial ( \partial K) = \emptyset.$ Since $x \in \partial K$ there is $y_0 \in (\partial K)^{[R-1]}$ and $z_0 \in \kR$ such that $x \subset y_0 \subset z_0$ and by the diamond property \ref{cccdiamond} there exists $y_1 \in K^{[R-1]}$ such that $ \face{z_0} \cap \cface{x} = \{y_0, y_1\}.$ If $y_1 \not\in \partial K$ then there exist $z_1 \in \kR, ~y_2 \in K^{[R-1]}$ such that $\cface{y_1} =\{z_0, z_1\}$ and $\face{z_1} \cap \cface{x} =\{y_1,y_2\}.$ By repeating this argument we can construct a sequence of elements in $K^{[R-1]},$ distinct by non-singularity of $K,$ until at some step $k$ we get $y_k \in \partial K$ and this implies $\{y_0, y_k\} \subset \cface{x} \cap \partial K.$ If by contradiction we assume that there is more that two elements in $\cface{x} \cap \partial K$ then the previous construction would show that it contains at least four elements and that the corresponding two sequences of elements in $K^{[R]}$ sharing an element of $K^{[R-1]}$ are not connected, implying that the local figure of $x$ is not strongly connected. This would imply that $x \in \partial K$ is a $K$-pinch, a contradiction with our hypothesis.
\end{proof}
\end{lem}

\begin{lem} \label{lemconcomp}
Let $K \in \nsc^R.$ Then $J \leq \partial K$ and $J \in \mlc^{R-1}_\sqcup$ if and only if $J$ is a (possibly empty) union of connected components of $\partial K.$
\begin{proof}
Let $R = \Rk(K).$ A union of boundary components of $K$ is clearly an element in $\mlc^{R-1}_\sqcup,$ by Lemma \ref{rembdrycc}. In order to prove the reverse implication it is sufficient to prove the case where $\partial K$ is connected and to show that $J \neq \emptyset$ implies $J = \partial K.$ If $J \in \mlc^{R-1}$ is non-empty then it is a closed $(R-1)$-cc. If by contradiction there exists a vertex in $\partial K \setminus J,$ the connectedness of $\partial K$ implies that this vertex is linked to a vertex in $J$ via a path in $(\partial K)^{(1)}.$ Hence there is an edge $e = \{v,w\}$ in $\partial K$ such that $v \in J$ and $w \in \partial K \setminus J.$ Now we have that $\dual{J} \leq \dual{\partial K}$ and $\cdual{e}{\partial K}$ (the dual of $e$ in $\partial K$) is a $(R-2)$-cell of $\dual{\partial K}$ contained in $\cdual{v}{\partial K},$ an $(R-1)$-cell of $\dual{J}.$ Therefore it implies that $\cdual{e}{\partial K}=\cdual{e}{J} \in J^{[R-2]}$ but since $\dual{J}$ is closed we have that $\dual{e}$ is contained in two maximal cells of $J.$ Since $\cdual{w}{\partial K}$ is a maximal cell of $\dual{\partial K}\setminus \dual{J}$ containing $\cdual{e}{J}$ this contradicts the non-singularity of $\dual{\partial K}.$
\end{proof}
\end{lem}

\begin{lem} \label{lemmidsec}
Let $K$ be a local cell complex and $J \leq K.$ Then the map
\begin{equation}\label{mapmidsec}
x \longmapsto \E_J^x, \quad x \in K_J
\end{equation} 
is an injective poset homomorphism such that the inverse is a poset homomorphism.
\begin{proof}
We first notice that since $K$ is cell-connected we have that $\E^x_J \neq \emptyset$ for all $x \in K_J.$ 
 In order to prove that (\ref{mapmidsec}) is injective, one considers $x,y \in K_J$ such that $\E_J^x = \E_J^y$ and show that $x=y.$ As consequence of the definition of collar we have $\E_J^x = \E_J^y = \E_J^{x \cap y}$ hence we can without loss of generality assume that $x \subset y.$

We then proceed by induction on $r = \rk(x) \geq 1.$
If $\rk(x)=1$ and $x \subsetneq y$ then by Axiom \ref{cccenough} we can assume that $\rk(y) = \rk(x) + 1.$ This means that $y$ is a 2-cell which, as discussed before, corresponds to a cycle of edges. One can then easily show that there must be a edge in $\E_J^y$ other than $x$ and since $\E_J^x ={x}$ this contradict $\E_J^x = \E_J^y.$

Assume by induction that the claim is true for all $x, y$ such that $x \subset y$ and $\rk(x) \leq r-1.$
Take $x \in K_J^{[r]}$ and $y \in K_J$ such that $\E_J^x = \E_J^y$ and $x \subsetneq y.$ Like before we can without loss of generality assume that $\rk(y) = r + 1.$ By induction we have that $\E_J^w \subsetneq \E_J^x$ for all $w \subsetneq x,$ and, using this inclusion with $w = x \cap x',$ this implies in particular that $\E_J^{x'} \not \subset \E_J^x$ for all $x' \in K_J \cap K_x$ (since $\E_J^{x'} \neq \emptyset$ by locality of $K$). Since $\E_J^x \neq \emptyset$ there exists $w \in \face{x} \cap K_J.$ By the diamond property \ref{cccdiamond}, there exists $x' \in K^{[r]}$ such that $ \cface{w} \cap \face{y} = \{x,x'\}.$ But $x' \in K_J$ since so does $w$ and $w \subset x',$ therefore $\emptyset \neq \E_J^{x'} \subset \E_J^y = \E_J^x.$ This is a contradiction with our induction hypothesis and therefore $x = y.$

The definition of $\E_J^x$ and the injectivity proven above imply that if $\E_J^x \subset \E_J^y$ then 
$$\E_J^{(x \cap y)} = \E_J^x \cap \E_J^y = \E_J^x$$
and therefore $x = x \cap y$ or equivalently $x \subset y$ and this concludes the proof.
\end{proof}
\end{lem}

\begin{lem} \label{lembdiv}
The barycentric subdivision of a cc $K$ is a subdivision of $K,$ i.e. $\bdiv{K} \red K.$
\begin{proof}
Define $\sphi : \bdiv{K} \dans K$ by $\sphi( \{x_0 \subsetneq \dots \subsetneq x_r\} ) := x_r,$ clearly a surjective poset homomorphism and injective on $\sphi^{-1}(\kz) = \{ \{v\} ~|~ v \in \kz\}.$

Let $\{x_0 \subsetneq \dots \subsetneq x_r\} \in \bdiv{K}$ and $y \in K$ such that $\sphi(\{x_0 \subsetneq \dots \subsetneq x_r\}) \subset y$ then by \ref{cccenough} there exists a cell of $\bdiv{K}$ of the form $ \sigma = \{x_0 \subsetneq \dots \subsetneq x_r \subset \dots \subset y \}$ such that $\rk_{\bdiv{K}}(\sigma) = \rk_K(y)$  and $\sphi(\sigma) = y$ and therefore $\sphi$ also satisfies condition \ref{cond3red} of a reduction.
For condition \ref{cond4red}, if $r := \rk_{\bdiv{K}}(\{x_0 \subsetneq \dots \subsetneq x_r\}) = \rk_K(x_r) + 1$ then there exists a unique $0 \leq i \leq r$ such that $\rk_K(x_i) = \rk_K(x_{i+1})-2$ and therefore
\begin{align*}
& \abs{\cface{\{x_0 \subsetneq \dots \subsetneq x_r\}} \cap \sphi^{-1}(x_r) } \\
&= \abs{\{ \{x_0 \subsetneq \dots \subsetneq x_i \subsetneq y \subsetneq x_{i+1} \subsetneq \dots \subsetneq x_r \} ~|~ y \in \cface{x_i} \cap \face{x_{i+1}} \}}\\
&= \abs{\cface{x_i} \cap \face{x_{i+1}}} = 2.
\end{align*}

Finally, for condition \ref{cond5red}, if $ r:=  \rk_{\bdiv{K}}( \{x_0 \subsetneq \dots \subsetneq x_r \}) = \rk_K(x_r)$ then $\rk(x_i) = i $ for all $0 \leq i \leq r.$ Hence for all $x_{r+1} \in \cface{x_r},$ we have
\begin{align*}
\abs{\cface{\{x_0 \subsetneq \dots \subsetneq x_r\}} \cap \sphi^{-1}(x_{r+1}) }
= \abs{\{ \{x_0 \subsetneq \dots \subsetneq x_{r+1} \} \} } = 1.
\end{align*}
\end{proof}
\end{lem}

\begin{lem}\label{lemunifrelcc}
Let $(K,J)$ be a pure relative cc satisfying (\ref{conduniform}) and let $x \in J(K).$ Then for all $x' \in K_J^-[x]$ and all maximal cells $w \in \bel_J(x')$ we have
$$  \rk_J^K(x) := \rk_K(x') - 1 = \rk_K(w) .$$ 
Moreover,  for all $x' \in K_J^-[x]$ we have that $K_J(x) = K_J(x' \cap J^{[0]})$ and
$$K_J^-(x) = K_J^-[x] = K_J^-(x' \cap J^{[0]}),$$
which implies that for any given $y' \in K_J(x)$ there exists $x' \in K_J^-[x]$ such that $x' \subset y'.$
\begin{proof}
First, it is clear that $K_J[x] \subset K_J(x).$ An element $x' \in K_J^-[x]$ satisfies that $x \subset x',$ which implies the inclusion 
$$K_J(x) \supset K_J(x' \cap J^{[0]}).$$
The reverse inclusion is also valid by the following argument. Note that an element $y' \in K_J(x)$ satisfies
$$ x'_J = x \subset y'_J$$ 
and (\ref{conduniform}) applied to $(K,J)$ gives
$$ x' \cap J^{[0]} \subset y' \cap J^{[0]},$$
which in particular implies that $y \in K_J(x' \cap J^{[0]}).$ Therefore we indeed have $K_J(x) = K_J(x' \cap J^{[0]}).$

Moreover, a minimal element in $K_J[x]$ is also a minimal element in $K_J(x).$ This is a consequence of the fact that if $x' \in K_J[x]$ and  $y' \in K_J(x)$ satisfy $y' \subset x'$ then $$x \subset y'_J \subset x'_J = x$$
and therefore $y'_J = x$ i.e. $y \in K_J[x].$
As a consequence of the results above, the purity of $(K,J)$ implies that for all $x' \in K_J^-[x]$ we have
$$K_J^-[x] \subset K_J^-(x) = K_J^-(x' \cap J^{[0]}).$$

Therefore if $x' \in K_J^-[x]$ then $x' \in K_J^-(x' \cap J^{[0]})$ and since $J \cap x'$ is pure, Axiom \ref{cccenough} implies that a maximal cell $w$ of $J \cap x'$ satisfies
$$\rk_K(w) = \rk_K(x') - 1.$$ 

Suppose by contradiction that there exists $w' \in K_J^-(x) \setminus K_J^-[x]$ and let $x' \in K_J^-[x].$ Then $w' \cap x'$ cannot be an element of $K_J$ as it would in particular contradict the minimality of $x',$ hence $w' \cap x' \in J.$ But this implies that $J^{[0]} \cap w' = J^{[0]} \cap x'$ which by condition (\ref{conduniform}) implies $x = x'_J = w'_J.$ This contradicts our assumptions and concludes the proof.
\end{proof}
\end{lem}

\begin{lem} \label{lemredcell}
If $(K,J)$ is uniform and $x \in J$ then there exists a unique cell $x_J \in J(K)$ such that
$$K_J^-[x_J] = K_J^-(x) \quad \text{and} \quad \rk_{J(K)}(x_J) = \rk_J^K(x).$$
\begin{proof}
In order to prove the first statement, it is sufficient to show that if $y,y'$ are distinct elements in $K_J^-(x)$ then $y_J = y'_J.$ 
Since $y$ and $y'$ are minimal elements in $K_J(x)$ then $y \cap y'$ cannot be contained in $K_J,$ so $y \cap y' \in J.$ Hence $J^{[0]} \cap y = J^{[0]} \cap y'$ and condition (\ref{conduniform}) implies that $y_J = y'_J.$

Also, if $y \in K_J^-(x)$ then by definition $\rk_J^K(x) = \rk_K(y) - 1$ and by Lemma \ref{lemunifrelcc} we have $\rk_K(y) = \rk_K(w) + 1$ for all maximal cells $w$ in $\bel_J(y).$ The second statement then follows since $y \in K_J^-[x_J]$ by the first statement and therefore
$$\rk_{J(K)}(x_J) = \rk_{J(K)}(y_J) = \rk_K(w)  = \rk_J^K(x).$$
\end{proof} 
\end{lem}

\begin{lem}\label{lemredbdry}
Let $K \in \nsc$ and let $J'$ be a connected component of $\partial K.$ If $\sphi: J \dans J'$ is a reduction (where $J$ is disjoint from $K$) then the map from $K_{J'}$ to $\pws{\left(\kz \setminus(J')^{[0]}\right) \sqcup J^{[0]}}$ defined by
\begin{equation} \label{maplemredbdry}
 y' \longmapsto \left( y' \setminus (J')^{[0]} \right) \sqcup \left( \bigcup_{x \in J, ~ \sphi(x) \subset y'} x \right)
\end{equation}
is an injective poset homomorphism such that the inverse is also a poset homomorphism.
\begin{proof}
Let us denote by $\phi$ the map (\ref{maplemredbdry}), which clearly is a poset homomorphism since so is $\sphi.$

The injectivity of $\phi$ is a direct consequence of condition \ref{cond1red} and the observation that any vertex $v \in \sphi^{-1}( (J')^{[0]} \cap y')$ satisfies that $\phi(v) \subset y'$ hence $v$ is included in $\left( \bigcup_{x \in J, ~ \sphi(x) \subset y'} x \right).$

From the last observation one also get that the inverse of $\phi$ is a poset homomorphism since if $x',y' \in K_{J'}$ satisfy that $\phi(x') \subset \phi(y')$ then it implies that $x' \setminus (J')^{[0]} \subset y' \setminus (J')^{[0]}$ and
$$ \sphi^{-1}( (J')^{[0]} \cap x') \subset \sphi^{-1}( (J')^{[0]} \cap y').$$
Therefore \ref{cond1red} implies that $ (J')^{[0]} \cap x' \subset (J')^{[0]} \cap y'$ and we can indeed conclude that $x' \subset y'.$
\end{proof}
\end{lem}

\section{Proof of Proposition \ref{propprecpartialorder}} \label{appPropPrec}

We will show that $\red$ is a partial order on the set of closed cc. The proof applies also to $\col$ using the same arguments while reversing inclusion relations between cells and replacing vertices with maximal cells. 

\begin{proof}
The relation $\red$ is clearly reflexive. The main part of this proof focuses on showing that $\red$ is transitive. For this, we assume that $J \red_{\sphi_1} K,$ $K \red_{\sphi_2} L$ and check the five conditions for $\sphi_2 \circ \sphi_1.$

Condition \ref{cond1red} for $\sphi_2 \circ \sphi_1$ is checked easily: if $y \in L^{[0]}$ we have $\abs{\sphi_2^{-1}(y)} = 1$ since $\sphi_2$ is a reduction, i.e. $ v = \sphi_2^{-1}(y) \in \kz$ and we have $\abs{\sphi_1^{-1}(v)} = 1$ since $\sphi_1$ is a reduction, hence $\abs{(\sphi_2 \circ \sphi_1)^{-1}(y)} = 1.$

Proving condition \ref{cond3red} for $\sphi_2 \circ \sphi_1$ is relatively direct as well. If $y \in L^{[r]}, x \in J$ are such that $\sphi_2 \circ \sphi_1 (x) \subset y$ then since $\sphi_2$ is a reduction there exists $w' \in \sphi_2^{-1}(y)^{[r]}$ such that $\sphi_1(x) \subset w'.$ Also, since $\sphi_1$ is a reduction there exists $w \in \sphi^{-1}(w')^{[r]} \subset (\sphi_2 \circ \sphi_1)^{-1}(y)^{[r]}$ such that $x \subset w$ and this proves that $$A( \sphi_2(\sphi_1(x)))^{[r]} \subset \sphi_2(\sphi_1(A(x)^{[r]})).$$

We will show condition \ref{cond5red} for $\sphi_2 \circ \sphi_1$ and finish the proof of transitivity with condition \ref{cond4red} below. 
If we assume that $\rk_J(x) = \rk_L(\sphi_2(\sphi_1(x))),$ it implies that $\rk_K(\sphi_1(x)) = \rk_J(x)$ since the rank is increasing under subdivision as observed in Remark \ref{remrksredncol}. Hence conditions \ref{cond5red} for $\sphi_1$ and $\sphi_2$ imply respectively the two following conditions:

\begin{equation}\label{enumcond5_1}
\abs{ \cface{x} \cap \sphi_1^{-1}(y)} = 1 \quad \text{for all } y \in \cface{ \sphi_1(x)};
\end{equation}

\begin{equation}\label{enumcond5_2}
\abs{ \cface{\sphi_1(x)} \cap \sphi_2^{-1}(y)} = 1 \quad \text{for all } y \in \cface{ \sphi_2(\sphi_1(x))}.
\end{equation}

Using that 
\begin{equation}
f( A \cap f^{-1}(B)) = f(A) \cap B
\end{equation}
for a function $f : S \dans T$ between two sets $S,T$ and $A \subset S,~ B \subset T$ and the fact that condition \ref{cond3red} implies $\cface{\sphi_1(x)} \subset \sphi_1(\cface{x}),$ we get the following inequality for all $y \in \cface{ \sphi_2( \sphi_1(x))}:$
\begin{align*}
\abs{ \cface{x} \cap (\sphi_2 \circ \sphi_1)^{-1}(y)} &\geq \abs{ \sphi_1( \cface{x} \cap (\sphi_2 \circ \sphi_1)^{-1}(y))} \\
&= \abs{ \sphi_1(\cface{x}) \cap \sphi_2^{-1}(y)}\\
&\geq \abs{ \cface{x} \cap \sphi_2^{-1}(y)} = 1,
\end{align*}
where the last equality is due to condition (\ref{enumcond5_1}).

Moreover if we suppose that for some $y \in \cface{\sphi_2(\sphi_1(x))}$ there are two distinct elements 
$$w,w' \in \cface{x} \cap (\sphi_2 \circ \sphi_1)^{-1}(y)$$ 
then since $\cface{ \sphi_2(\sphi_1(x)} \subset \sphi_2(\cface{\sphi_1(x)})$ by condition \ref{cond3red}, there exists $y' \in \cface{\sphi_1(x)} \cap \sphi_2^{-1}(y)$ and this element is unique by condition (\ref{enumcond5_2}). This implies that
$$ \{w,w'\} \subset \abs{ \cface{x} \cap \sphi_1^{-1}(y')}$$ which contradicts condition (\ref{enumcond5_1}) and therefore we have proven \ref{cond5red} for $\sphi_2 \circ \sphi_1.$ 

Finally, we show condition \ref{cond4red} for $\sphi_2 \circ \sphi_1$ as follows. If $x \in J$ satisfies 
$$\rk_J(x) = \rk_L(\sphi_2(\sphi_1(x))) - 1$$
then we can have the two following possible cases.

The first possible case is
$$rk_J(x) = \rk_K(\sphi_1(x)), \quad \rk_K(\sphi_1(x)) = \rk_L(\sphi_2(\sphi_1(x))) - 1,$$
which respectively implies condition (\ref{enumcond5_1}) as well as condition \ref{cond4red} for $\sphi_2,$ i.e.
\begin{equation}\label{equsphi2proppreord}
\abs{ \cface{\sphi_1(x)} \cap \sphi_2^{-1}(\sphi_2(\sphi_1(x)))} = 2.
\end{equation}
Every element $w \in \cface{x} \cap (\sphi_2 \circ \sphi_1)^{-1}(\sphi_2(\sphi_1(x)))$ satisfies $\sphi(x) \subset \sphi(w)$ and $\sphi_2(\sphi_1(w)) = \sphi_2(\sphi_1(x)),$ hence
$$r + 1 = \rk_J(w) \leq \rk_K(\sphi_1(w)) \leq \rk_L(\sphi_2(\sphi_1(w))) = \rk_L(\sphi_2(\sphi_1(x))) = r + 1.$$
Therefore $\rk_K(\sphi_1(w)) = r + 1$ and $\sphi_1(w) \in \cface{\sphi_1(x)}.$  Condition (\ref{enumcond5_1}) implies that $\sphi_1^{-1}(\sphi_1(w)) = \{w\}$ for every such $w$ and using (\ref{equsphi2proppreord}) we get
$$ \abs{ \cface{x} \cap (\sphi_2 \circ \sphi_1)^{-1}(\sphi_2(\sphi_1(x)))} = \abs{ \cface{\sphi_1(x)} \cap \sphi_2^{-1}( \sphi_2(\sphi_1(x)))} = 2.$$

The second possible case is
$$rk_J(x) = \rk_K(\sphi_1(x)) - 1, \quad \rk_K(\sphi_1(x)) = \rk_L(\sphi_2(\sphi_1(x))).$$
This case implies condition \ref{cond4red} for $\sphi_1:$
\begin{equation}\label{equsphi1proppreord}
\abs{\cface{x} \cap \sphi_1^{-1}(\sphi_1(x))} = 2.
\end{equation}
If $w \in \cface{x} \cap (\sphi_2 \circ \sphi_1)^{-1}(\sphi_2(\sphi_1(x)))$ then $\sphi(x) \subset \sphi(w)$ and $\rk_K(\sphi(w)) = r + 1 = \rk_K(\sphi_1(x))$ by the same argument as above, therefore $\sphi_1(w) = \sphi_1(x).$
This implies that
$$\cface{x} \cap (\sphi_2 \circ \sphi_1)^{-1}(\sphi_2(\sphi_1(x))) = \cface{x} \cap \sphi_1^{-1}(\sphi_1(x))$$
and we get the desired result by (\ref{equsphi1proppreord}). This concludes the proof that $\sphi_2 \circ \sphi_1$ is a reduction and therefore $\red$ is transitive.

At last, the relation $\red$ is antisymmetric by the following argument. Suppose $ J \red_\phi K$ and $K \red_\gamma J$  (which by Remark \ref{remrksredncol} implies $\Rk(J) = \Rk(K)$) and define $\phi_r := \phi|^{K^{[r]}}$ and $\gamma_r := \gamma|^{J^{[r]}}.$ We will prove by induction on $r$ that $\phi_r : J^{[r]} \dans \kr$ is a bijection for all $0 \leq r \leq R$ implying that $\phi$ is an isomorphism by Lemma \ref{lemisom}. For the case $r=0$ we know that $\phi_0$ and $\gamma_0$ are surjective and this implies that $\abs{J^{[0]}} = \abs{\kz}.$ Hence $\phi_0$ and $\gamma_0$ are also injective and therefore bijective. Suppose that $\phi_s : J^{[s]} \dans K^{[s]}$ and $\gamma_s : K^{[s]} \dans J^{[s]}$ are bijective for all $s = 0,1 \dots r-1.$ As noted in Remark \ref{remcond3redncol} since $\phi$ is a reduction we have $\phi^{-1}( \kr ) \subset J^{(r)}$ and $J^{(r-1)} = \phi^{-1}(K^{(r-1)}).$ Hence $\phi^{-1}(\kr) \subset J^{[r]}$ and $\abs{\kr} \leq \abs{J^{[r]}}.$ By the same argument for $\gamma$ we also get the reverse inequality and therefore $\phi_r : J^{[r]} \dans \kr$ and $\gamma_r : \kr \dans J^{[r]}$ are indeed bijections. 
\end{proof}

\section{Proof of Proposition \ref{proptransit}} \label{appproptransit}

In this appendix we show the following result: let $K \in \nsc^{R+1}$ and $J \leq \partial K,$ $J \in \mlc^R$ such that $(K,J)$ is uniform and local.
The two following statements hold.
\begin{enumerate}[label=\arabic*) , topsep=2pt, parsep=2pt, itemsep=1pt]
\item  We have $J \red J(K)$ as the poset homomorphism $\sphi_J^K: J \dans J(K)$ defined by
$$\sphi_J^K(x) = x_J$$
is a reduction.
\item We have $M_J^K \col J(K)$ as the poset homomorphism $ \cphi_J^K: M_J^K \dans J(K)$ defined by $$\cphi_J^K(\E^x_J) = x_J$$ is a collapse.
\end{enumerate}

\begin{proof}
In order to prove the point \ref{proptransitred} we first prove that $\sphi_J^K: J \dans J(K)$ is a reduction. The map $\sphi_J^K$ is clearly a surjective poset homomorphism. Condition \ref{cond1red} follows from the fact that if a vertex $v \in J^{[0]}$ satisfies $v_J \in J(K)^{[0]}$ then we have $v_J = \{v\}.$

Condition \ref{cond3red} is obtained in the following way. Suppose $x \in J$ satisfies
$$ x_J \subset y \in J(K)^{[r]}.$$
For $y' \in K_J^-[y]$ we then have that $\Rk(\bel_J(y')) = r$ and $x \in \bel_J(y').$ By purity of $(K,J)$ there exists  an $r$-cell $x'$ in $\bel_J(y')$ such that $x \subset x'.$ If we name $w = x_J'$ then Lemma \ref{lemredcell} implies that
$$K_J^-[w] = K_J^-(x').$$
Since $x'$ is a maximal cell of $\bel_J(y')$ we have that $y' \in K_J^-(x')$ and by the previous relation this implies
$$ y = y'_J = w = x'_J,$$
which proves condition \ref{cond3red} for $\sphi_J^K.$

We then turn to condition \ref{cond4red}. Let $x \in J$ such that $\rk_J(x) = \rk_{J(K)}(x_J) - 1$ and let $x' \in K_J^-[x_J].$ Then $\rk_K(x') = \rk_{J(K)}(x_J) + 1,$ hence $\rk_K(x') = \rk_K(x) - 2.$ Since $x \subset x',$ Axiom \ref{cccdiamond} for $K$ provides us with two cells $y,y'$ such that 
$$ \cface{x} \cap \face{x'} = \{y,y'\}.$$
By minimality of $x',$ we have that $y$ and $y'$ are in $J$ and 
$$x' \in K_J^-(y) \cap K_J^-(y') = K_J^-[y_J] \cap K_J^-[y'_J],$$
where the equality is due to Lemma \ref{lemredcell}. Also, if $w \in \cface{x}$ satisfies $w_J = x_J$ then $w \subset x'$ because $x' \in K_J^-[x_J].$ Since $\rk_J(w) = \rk_J(x) + 1 = \rk_K(x') - 1,$ we have $w \in \face{x'}$ and therefore $w \in \{y,y'\}.$ This completes the proof that
$$ \abs{ \cface{x} \cap (\sphi_J^K)^{-1}(x_J)} = 2.$$

Then condition \ref{cond5red} can be proven by taking $x \in J$ such that $\rk_J(x) = \rk_{J(K)}(x_J).$ This implies that $x$ is a maximal cell of $\bel_J(x')$ for all $x' \in K_J^-[x_J].$ If we take $y \in \cface{x_J},$  $y' \in K_J^-[y]$ and $x' \in K_J^-[x_J]$ such that $x' \subset y'$ then we obtain $$\rk_K(y') = \rk_{J(K)}(y) + 1 = \rk_K(x) + 2.$$ 
Therefore there exists $x'' \in \cface{x}$ such that
$$ \cface{x} \cap \face{y'} = \{x',x''\}.$$
If $x'' \in K_J$ then $x' \cap J^{[0]} = x'' \cap J^{[0]} = x$ which implies $y' \cap J^{[0]} = x$ but this contradicts the fact that $\Rk(\bel_J(y')) = \rk_J(x) + 1.$
Therefore $x'' \in J,$ hence  $x''_J = y'_J = y$ and we have $$\cface{x} \cap (\sphi_J^K)^{-1}(y) = \{x''\}.$$ 

Next we prove that $\cphi_J^K : M_J^K \dans J(K)$ is a collapse. Condition \ref{cond1col} is clear since $\abs{\cface{z}} = 1$ for all $z \in J^{[R-1]}$ and therefore $K_J^-[z] = K_J^-(z)$ contains a unique cell $z'$ which is a maximal cell of $K,$ in other words $$(\cphi_J^K)^{-1}(z) = \{ \E_J^{z'}\}.$$

Condition \ref{cond3col} is a consequence of Lemma \ref{lemunifrelcc}, since if $x \in J(K)$ and $\E_J^{y'} \in M_J^K$ are such that $x \subset y_J'$ then there exists $x' \in K_J^-[x]$ such that $x' \subset y'.$ Hence we have  $\rk_{M_J^K}(\E_J^{x'}) = \rk_{J(K)}(x)$ and Lemma \ref{lemmidsec} implies that $\E_J^{x'} \subset \E_J^{y'}.$

In order to prove condition \ref{cond4col}, we take $\E_J^{x'} \in M_J^K$ such that
$$\rk_{M_J^K}(\E_J^{x'}) = \rk_{J(K)}(x'_J) + 1 = \rk_K(x') - 1.$$
If $w \in \bel_J(x')$ is maximal cell then Axiom \ref{cccdiamond} for $K$ gives
$$ \cface{w} \cap \face{x'} = \{y, y'\},$$
for $y,y' \in K.$ The maximality of $w$ then implies that $\{y,y'\} \subset K_J.$ 
We therefore have that 
$$w \subset y \cap y' \in J,$$ 
which by Axiom \ref{cccrank} implies that $w = y \cap y'.$  Therefore $y_J = y'_J = x_J$ and we indeed have
$$\abs{ \face{\E_J^{x'}} \cap (\cphi^{-1})(x)} = 2.$$

Finally, we can prove condition \ref{cond5col} in the following way. Suppose that $\E_J^{x'} \in M_J^K$ satisfies $ \rk_{M_J^K}(\E_J^{x'}) = \rk_{J(K)}(x'_J).$ If $w$ is a maximal cell in $\bel_J(x')$ we have that $\rk_K(w) = \rk_K(x') - 1,$ i.e. $w \in \face{x'}.$ By Lemma \ref{lemredcell} we also have that
$$x' \in K_J^-(w) = K_J^-[w_J].$$
If we fix an element $y \in \face{x}$ where $x := x'_J = w_J$ then by Lemma \ref{lemunifrelcc} there exists $y' \in K_J^-[y]$ such that $y' \subset x'$ and
$$\rk_K(y') = \rk_{J(K)}(y) + 1 = \rk_{J(K)}(x)  = \rk_K(x') - 1.$$
This implies that $y' \in \face{x'},$ i.e. $\E_J^{y'} \in \face{\E_J^{x'}}$  by Lemma \ref{lemmidsec}. Moreover if $\E_J^{w'} \in \face{\E_J^{x'}} \setminus \{\E_J^{y'}\}$ satisfies $w'_J = y$ then by minimality of $y'$ we have $y' \cap w' \in J.$ Hence it must be that $\bel_J((y'\cap w'))$ has a unique maximal cell $u$ and
$$  \{y' , w'\} \subset \cface{u} \cap \face{x'} .$$
This is a contradiction with Axiom \ref{cccdiamond} for $K$ since $\bel_J(x')$ is a pure cc such that
$$ \Rk(\bel_J(x')) = \rk_K(x') - 1 = \rk_K(y') = \rk_K(u) + 1$$ and therefore there exists a maximal cell $w$ of $\bel_J(y')$ such that $w \in \cface{u} \cap \cface{x'}\setminus \{y',w'\}.$
Thus we can indeed conclude that
$$ \face{\E_J^{x'}} \cap (\cphi_J^K)^{-1}(y) = \{\E_J^{y'}\}.$$
\end{proof}

\section{Proof of Proposition \ref{propredbdry}} \label{apppropredbdry}

In this appendix we show the following result: let $K \in \nsc$ and let $J'$ be a connected component of $\partial K$ such that $(K , J')$ is uniform. If $\sphi : J \dans J'$ is a reduction such that $\sphi \circ \sphi_{J'}^K : J \dans J'(K)$ is compatible with $\cphi_{J'}^K : M_{J'}^K \dans J'(K)$ and $J$ is disjoint from $K$ then the cc $K^\sphi$ (introduced in Definition \ref{defikphi}) is an element in $\nsc$ and $J$ is a connected component of $\partial K^\sphi,$ in other words 
$$( K^\sphi, J) \red_\sphi (K, J')$$
defines a relative reduction. Moreover, we have that $\sphi_J^{K^\sphi} = \sphi \circ \sphi_{J'}^K$ and $(K^\sphi, J)$ is uniform.

\begin{proof}
The transitivity of $\red$ (Proposition \ref{propprecpartialorder}) implies that $\sphi \circ \sphi_{J'}^K$ is also a reduction hence we can without loss of generality assume that $\sphi_{J'}^K = \id_{J'(K)},$ i.e. $J' = J'(K)$ and $\sphi = \sphi_{J'}^K \circ \sphi.$

Define $(J')^c := \bel_K ( \kz \setminus (J')^{[0]}),$ i.e. the sub-cc of $K$ containing the cells in $K$ not intersecting $J'.$ Let $\phi$ be the map defined by (\ref{maplemredbdry}) which by Lemma \ref{lemredbdry} defines a poset isomorphism 
$$ \phi : K_{J'} \dans \phi(K_{J'}).$$
We define $(K^\sphi, \rk_{K^\sphi})$ by
\begin{align*}
K^\sphi &:= (J')^c \sqcup J \sqcup \phi(K_{J'}) \\
\rk_{K^\sphi} &:= \rk_K|_{(J')^c} + \rk_J + \rk_K|_{K_{J'}} \circ \phi^{-1}.
\end{align*}
It then follows that $K_J^\sphi = \phi(K_{J'}).$ Also, since $J' = \cphi_{J'}^K(M_{J'}^K)$ as $\sphi_{J'}^K = \id_{J'}$ we have that
\begin{equation}\label{condapropredbdry}
\text{for all } y' \in K_{J'}, (y')_{J'} = y' \cap ({J'})^{[0]}.
\end{equation}

Let's now check the axioms of cc for $K^\sphi.$ To make our notations more explicit, for a cell $x$ in a cc $L$ we will denote by $\kcface{x}{L}$ the set of co-faces of $x$ in $L$ and similarly for faces. Since these axioms are satisfied for $K$ they are automatically satisfied when all cells involved are in $J$ or $(J')^c \sqcup \phi(K_J)$ by the poset isomorphism property of $\phi.$ This allows us to only check the axioms in the case of inclusions of cells between the sets $J$ and $(J')^c \sqcup \phi(K_J).$

To check Axiom \ref{cccrank} we only need to consider the case where $x \in J,$ $y \in K^\sphi_{J}$ and $x \subsetneq y.$ In this case we have that $\sphi(x) \subsetneq \phi^{-1}(y)$ and therefore
$$ \rk_{K^\sphi}(x) \leq \rk_{K}(\sphi(x)) < \rk_K( \phi^{-1}(y) ) = \rk_{K^\sphi}(y).$$

To check Axiom \ref{cccinter} for the non-direct cases amounts to show that $x \cap y \in K^\sphi$ for $x \in K_J^\sphi \sqcup J$ and $y \in K_J^\sphi.$ If $x \in K_J^\sphi$ and $x',y' \in K_{J'}$ are such that $\phi(x') = x $ and $\phi(y') = y$ then we have two possible cases: either $\E_{J'}^{x'} \cap \E_{J'}^{y'} \neq \emptyset$ or $\E_{J'}^{x'} \cap \E_{J'}^{y'} = \emptyset.$ In the first case one has that $x' \cap y' \in K_{J'}$ and $$\phi(x' \cap y') = \phi(x') \cap \phi(y') = x \cap y \in K^\sphi_J.$$
For the case $\E_{J'}^{x'} \cap \E_{J'}^{y'} = \emptyset,$ then by property (\ref{condapropredbdry}), we have that $$x' \cap y' = (x')_{J'} \cap (y')_{J'} \in J'(K) = J'.$$
By Proposition \ref{proptransit} we know that $\cphi_{J'}^K$ is a collapse, hence by \ref{cond3col} we can pick $x'_0 \in \bel(x')^{[r+1]}$ and $y'_0 \in \bel(y')^{[r+1]}$ where $r = rk(x' \cap y') = \rk_{M_{J'}^K}(\E_{J'}^{x'_0}) = \rk_{M_{J'}^K}(\E_{J'}^{y'_0}).$ By Lemma \ref{lemunifrelcc} this implies that $x'_0 ,y'_0 \in K^-_{J'}[x' \cap y'],$ and since $\E_{J'}^{x'} \cap \E_{J'}^{y'} = \emptyset,$ we also have that $\E_{J'}^{x_0'} \cap \E_{J'}^{y_0'} = \emptyset.$ The latter implies that $\abs{(\cphi_{J'}^K)^{-1}(x'\cap y')}>1$ which by condition \ref{cpa} of compatibility implies that there exists $x \in J$ such that $\sphi^{-1}(x' \cap y')=\{w_0\}.$ Therefore we obtain $$\phi(x') \cap \phi(y') = \bigcup_{w\in \sphi^{-1}(B(x' \cap y'))} w = w_0 \in J.$$
For the case where $x \in J$ and if $y' = \phi^{-1}(y) \in K_{J'}$ then one has to show that $$x \cap \left(  \bigcup_{w \in J ~:~ \sphi(w) \subset y'} w \right) \in J.$$ Let $w_0 := (y')_{J'},$ then $\{w \in J ~|~ \sphi(w) \subset y'\} = \sphi^{-1}(w_0),$ hence we need to show that there exists a unique maximal cell in $\bel(x) \cap \sphi^{-1}(w_0).$ Suppose by contradiction that $w_1,w_2$ are two distinct such maximal cells. Since we have $w_i \subset x$ for $i=1,2,$ we have that $w_1 \vee w_2 \neq \all.$ Condition \ref{cpau} of compatibility then implies that $$\sphi_{J'}^K(w_1 \vee w_2) = \sphi(w_1) \vee \sphi(w_2) \subset w_0,$$ where the inclusion is due to the fact that $\sphi(w_i) \subset w_0$ for $i=1,2.$ But this implies that $w_1 \vee w_2 \in \sphi^{-1}(\bel(w_0))$ and contradicts the maximality of $w_1$ and $w_2.$ This finishes the proof of Axiom \ref{cccinter} for $K^\sphi.$

Axiom \ref{cccenough} can be inferred from the case $J \ni x \subsetneq y \in K^\sphi_J$ which is proven as follows. By the poset isomorphism property of $\phi$ we have that $\sphi(x) \subset \phi^{-1}(y) \in K_{J'}.$ Therefore by Axiom \ref{cccenough} for $K$ we can suppose without loss of generality that $\rk_K(y) = \rk_K(\sphi(x)) +1.$ If $\rk_J(x) = \rk_K(\sphi(x))$ then we have that $y \in \kcface{x}{K^\sphi}$ as desired. If $\rk_J(x) < \rk_K(\sphi(x))$ then, as noted in Remark \ref{remcond3redncol}, there exists $w \in \sphi^{-1}(\sphi(x))$ such that $\rk_J(w) = \rk_K(\sphi(x)).$ Therefore using \ref{cccenough} for $J$ we can find a co-face of $x$ contained in $w$ which is then also contained in $y.$

For Axiom \ref{cccdiamond} it is also sufficient to focus on the case where $x \in J, y \in K_J^\sphi,$ $x \subsetneq y$ and 
$$\rk_{K^\sphi}(x) = \rk_{K^\sphi}(y) - 2.$$
Since $\rk_{K^\sphi}(y) = \rk_K(\phi^{-1}(y)),$ this leaves two possible cases: either we have $\rk_K(\sphi(x))= \rk_{K^\sphi}(x)$ or $\rk_K(\sphi(x)) = \rk_{K^\sphi}(x) +1.$ In the first case where $\rk_K(\sphi(x)) = \rk_{K^\sphi}(x),$ Axiom \ref{cccdiamond} for $K$ implies that 
$$\abs{ \kcface{\sphi(x)}{K} \cap \kface{\phi^{-1}(y)}{K}} = 2.$$
To conclude this first case, it remains to show that there is a bijection between $\kcface{\sphi(x)}{K} \cap \kface{\phi^{-1}(y)}{K}$ and $\kcface{x}{K^\sphi} \cap \kface{y}{K^\sphi}.$
If $w' \in \kcface{\sphi(x)}{K} \cap \kface{\phi^{-1}(y)}{K}$ then either $w' \in K_{J'}$ or $w' \in J'.$ Since $w' \in K_{J'}$ if and only if
$ \phi(w') \in \kcface{x}{K^\sphi} \cap \kface{y}{K^\sphi}$ and $\phi$ is bijection, we are good for this case.
If $w '\in J'$ then $w' \in \kcface{\sphi(x)}{J'}$ and \ref{cond5red} implies that the map 
$$ \kcface{x}{J} \cap \sphi^{-1}(\kcface{\sphi(x)}{K} ) \ni w \longmapsto \sphi(w) \in \kcface{\sphi(x)}{J'} $$
is a bijection, and this concludes the case  $\rk_K(\sphi(x)) = \rk_{K^\sphi}(x).$ 

In the second case where $\rk_K(\sphi(x)) = \rk_{K^\sphi}(x) + 1$ condition \ref{cond4red} implies that
$$\abs{ \kcface{x}{J} \cap \sphi^{-1}(\sphi(x))} = 2.$$
Moreover, if $w \in \kcface{x}{K^\sphi} \cap \kface{y}{K^\sphi}$ then $\sphi(w) \subset \phi^{-1}(y)$ and 
$$\rk_J(x) + 1 = \rk_J(w) \leq \rk_K(\sphi(w)) < \rk_K(y) = \rk_J(x) + 2.$$
Therefore $\rk_K(\sphi(w)) = \rk_K(\sphi(x))$ and $\sphi(w) = \sphi(x).$ Thus $w \in \kcface{x}{J} \cap \sphi^{-1}(\sphi(x))$ and this also proves that $$\abs{\kcface{x}{K^\sphi} \cap \kface{y}{K^\sphi}}=2,$$
which concludes the proof of \ref{cccdiamond} and shows that $K^\sphi$ is a cc

The cc $K^\sphi$ is graph-based if $K^\sphi_J$ is graph-based, which is the case since $e \in (K^\sphi_J)^{[1]}$ if and only if $e' := \phi^{-1}(e) \in \ko_{J'}$ and therefore 
$$\abs{e} = \abs{e' \setminus (J')^{[0]} \sqcup \sphi^{-1}(J^{[0]} \cap e)} = 2$$
by condition \ref{cond1red}. 

The previous argument also implies that  $K^\sphi_J$ is cell-connected. Since $K$ is connected and cell-connected, this shows that $K^\sphi$ is connected and cell-connected.

One has that $K^\sphi$ is clearly pure and $R := \Rk(K^\sphi) = \Rk(K).$ And $K^\sphi$ is non-branching if $\abs{\cface{y'}} = 2$ for all $y' \in (K^\sphi_J)^{[R-1]},$ which is directly verified as a consequence of the poset isomorphism property of $\phi.$

A $K^\sphi$-pinch cannot occur in $(J')^c$ nor $K^\sphi_J$ since the poset isomorphism property of $\phi$ would produce a $K$-pinch. If $x \in J$ then $x \subsetneq y \in K^\sphi$ if and only if $\sphi(x) \subsetneq \phi^{-1}(y) \in K$ hence we have
$$ M_{\bel(x)}^{K^\sphi} \cong K_x^\sphi \cong K_{\sphi(x)} \cong M^K_{\bel(\sphi(x))},$$
where $\cong$ means isomorphic as ranked posets, the second isomorphism being $\phi^{-1}.$
Therefore there also cannot be a $K^\sphi$-pinch in $J$ since this would contradict that $K$ has no $K$-pinch on $J'.$
This in addition shows that $K^\sphi \in \nsc.$

By Remark \ref{remrksredncol} we have $\Rk(J) = \Rk(J') = R - 1$ and every cell in $J$ is included in a cell $y \in J^{[R-1]}.$ By Remark \ref{remcond3redncol} such a $y \in J^{[R-1]}$ satisfies that $\sphi(y) \in (J')^{[R-1]}.$ Therefore we have
$$\abs{\kcface{y}{K^\sphi}} = \abs{\kcface{\sphi(y)}{K}} = 1,$$
hence $J$ is indeed a connected component of $\partial K^\sphi.$

We have that $\sphi_J^{K^\sphi} = \sphi$ by the following arguments. 
If $v'$ is the vertex in $e \cap (J')^{[0]}$ for $e \in  \E'_{J'}$ then by \ref{cond1red} we have that $\sphi^{-1}(\{v'\})$ contains exactly one vertex $v$ from $J.$ We will therefore consider the vertices $v'$ and $v$ as identical to simplify the discussion. With this convention, a cell $y \in J'(K)$ can be seen as having its vertices in $J.$ Moreover, if $y' \in K_{J'}^-[y]$ then $\phi(y') \in (K^\sphi)_J^-(x)$ for all $x \in J$ such that $\sphi(x) \subset y.$ Since we also assume that $J' = J'(K)$ and $\sphi_{J'}^K = \id_{J'}$ we have that $y = y_{J'}$ for all $y \in J'.$ Therefore, if we take $x \in J$ and $y' \in K_{J'}^-[\sphi(x)_{J'}]$ then
$$ x_J = \phi(y')_J = y'_{J'} = \sphi(x)_{J'} = \sphi(x)$$
and this proves that $\sphi = \sphi_J^{K^\sphi}.$
This also directly implies that $J(K^\sphi) = J'(K)$ and that $(K^\sphi, J)$ is uniform.
\end{proof}

\section{Proof of Theorem \ref{thmdualcob}} \label{appThmcob}

This section simply contains the proof of Theorem \ref{thmdualcob} stating that if $(K- J) \in \cob^R$ then:
\begin{itemize}
\item $\dual{(K - J)} \in \cob^R,$ 
\item $\dual{(K - J)}$ is exactly collared,
\item $\dual{K_J} \in \mlc^{R-1}_\sqcup$ and  $\partial \dual{(K -J)} = \dual{\partial K \setminus J} \sqcup \dual{K_J}.$
\end{itemize}

\begin{proof}
Let $R = \Rk(K).$ We split the proof into three statements \ref{thmdualcobspta}, \ref{thmdualcobsptb} and \ref{thmdualcobpt2} as follows.

Suppose that $(K - J) \in \cob.$
\begin{enumerate}
\item \label{thmdualcobpt1} Then we have $K' := \bual{K \setminus J} \sqcup \dual{\partial K \setminus J} \in \nsc,$ 
\begin{enumerate}
\item \label{thmdualcobspta} for the special case $J = \emptyset,$
\item \label{thmdualcobsptb} and for $J$ arbitrary as a consequence of the following statement: if $K' \in \nsc$ and $J_0$ is a connected component of $\partial K \setminus J$ such that $(K, J \sqcup J_0)$ is local then
$$K'_0 := \bual{K \setminus (J \sqcup J_0)} \sqcup \dual{\partial K \setminus (J \sqcup J_0)} \in \nsc;$$
\end{enumerate}
\item \label{thmdualcobpt2} If we define $ J' := \dual{\partial K \setminus J},$ then $ J' \in \mlc^{R-1}_\sqcup,$ $(K',J')$ is exactly collared and $$\partial K' = J' \sqcup \dual{K_J}.$$
\end{enumerate}
To prove the point \ref{thmdualcobspta}, we show that $K' := \bual{K} \sqcup \dual{\partial K}$ satisfies Axioms \ref{cccrank}, \ref{cccinter}, \ref{cccenough} and \ref{cccdiamond} and is a local non-singular non-pinching cell complex of rank $R.$

Axiom \ref{cccrank} is a consequence of Lemma \ref{leminclbdual} for the cells of $K'$ included in $\bual{K}$ and is a consequence of Lemma \ref{claim2propdual} for the cells included in $\dual{\partial K}.$ To prove \ref{cccinter}, we take $x', y' \in K'$ such that $x' \cap y' \neq \emptyset.$ If $x'= \bual{x}$ and $y' = \bual{y}$ are included in $\bual{K}$ then there exists a cell $w \supset x \cup y $ of minimal rank, and by Lemma \ref{leminclbdual} we get $\bual{w} = \bual{x} \cap \bual{y}$ (using the same arguments as in the proof of Proposition \ref{propdual}). If $x' = \dual{x}$ and $y' = \dual{y}$ are cells in $\dual{\partial K}$ then \ref{cccinter} is a consequence of Proposition \ref{propdual}. If say $x' = \bual{x} \in \bual{K}$ and $y' = \dual{y} \in \dual{\partial K}$ then
$$ \bual{x} \cap \dual{y} = (\cdual{x}{K} \cup \cdual{x}{ \partial K}) \cap \cdual{y}{ \partial K} = \cdual{x}{\partial K} \cap \cdual{y}{\partial K},$$
so $\cdual{x}{\partial K} \cap \cdual{y}{\partial K} \neq \emptyset.$ Therefore $x,y \in \partial K$ and
$$\cdual{x}{\partial K} \cap \cdual{y}{\partial K} = \cdual{w}{\partial K},$$ where $w = \dual{ \dual{x} \cap \dual{y}} \in \partial K.$ Axiom \ref{cccenough} is satisfied for elements in $\bual{K \setminus \partial K}$ or in $\dual{\partial K}$ as a consequence of Axiom \ref{cccenough} for $K$ and $\partial K$ and by Lemma \ref{leminclbdual} and Proposition \ref{propdual} respectively. The case $\dual{x} \subsetneq \bual{y}$ is shown by finding a cell $x' \in K'$ included in $\bual{y}$ and having $\dual{x}$ as a face by the following arguments. We have $$ \cdual{x}{\partial K} \subsetneq ( \cdual{y}{K} \cup \cdual{y}{\partial K})$$ which in particular implies $\cdual{x}{\partial K} \subset \cdual{y}{\partial K}.$ This in turn implies that $\cdual{y}{\partial K} \neq \emptyset$ and thus $y \in \partial K.$ If $\dual{x} \subsetneq \dual{y}$ then the existence of $x'$ is given by Axiom \ref{cccenough} for $\dual{\partial K}.$ If $\dual{x} = \dual{y}$ then $x = y$ and since $\rk_{K'}(\bual{y}) = \rk_{\dual{\partial K}}(\dual{y}) + 1$ then $\dual{x}$ is a face of $\bual{y}$ and $x'= \bual{y}$ satisfies the desired conditions. The diamond property \ref{cccdiamond} is proven in a similar way. The only non-direct case to check is when $\dual{x} \in \dual{\partial K},$ $\bual{y} \in \bual{K},$ $\dual{x} \subset \bual{y}$ and 
\begin{equation} \label{diamondrkprfpropbdual}
\rk_{K'}(\dual{x}) = \rk_{K'}(\bual{y}) - 2.
\end{equation}
By the same argument as before, $y \in \partial K$ and (\ref{diamondrkprfpropbdual}) implies that $y$ is a face of $x.$ We also have that $\bual{x}$ is a coface of $\dual{x}$ and a face of $\bual{y}$ by (\ref{diamondrkprfpropbdual}). Since $y \subset x$ this implies that 
\begin{equation} \label{diamondfacepropbdual}
\{\dual{y}, \bual{x}\} \subset \cface{\dual{x}} \cap \face{\bual{y}}.
\end{equation}
If $z \in \cface{\dual{x}} \cap \face{\bual{y}}$ then either $z \in \dual{\partial K}$ and $$\dual{x} \subsetneq z \subset \dual{y}$$ which implies $z= \dual{y}$ or $z \in \bual{K}$ and $$\bual{x} \subset z \subsetneq \bual{y}$$ which implies $z = \bual{x}.$ Therefore the inclusion in (\ref{diamondfacepropbdual}) is in fact an equality and this shows Axiom \ref{cccdiamond} as well as that $K'$ is a cc. 

We then prove that $K'$ is cell-connected.  $K'$ is graph-based since the non-singularity of $K$ implies that the edges of $K'$ are either composed of two maximal cells of $K$ or of one maximal cell $y$ of $\partial K$ and the unique maximal cell of $K$ containing $y.$ The cc $K'$ is cell-connected by the following argument. If $x' \in \bual{K}$ is a cell of the form $x' = \cdual{x}{K}$ where $x \in K \setminus \partial K$ then $x'$ is connected since $K$ is non-pinching. If $x' = \bual{x}$ where $x \in \partial K$ then $x'$ is connected if and only if there is and edge $e'$ of $K'$ with a vertex in $\cdual{x}{K}$ and in $\cdual{x}{\partial K},$ as both of these sets are connected since $K$ is non-pinching. Let $y$ be a maximal cell of $\partial K$ containing $x.$ The cell $y$ is then  included in a maximal cell $z$ of $K,$ we can then take $e' = \{y,z\}.$ Cells in $\dual{\partial K}$ are connected since $\partial K$ is non-pinching and this shows that $K'$ is cell-connected. 

The cc $K'$ is connected by the following argument. If $v_1',v_2'$ are vertices of $K'$ then by purity of $K$ these vertices are included in maximal cells $z_1'$ and $ z_2'$ of $\bual{K}.$ These maximal cells are duals to vertices $v_1,v_2$ of $K$ which are connected by a path in $\kg.$ This path provides a finite sequence of maximal cells of $\bual{K}$ starting with $z_1'$ and ending with $z_2'$ such that each element in the sequence shares a sub-maximal cell with the next element in the sequence and since the cells in $\bual{K}$ are connected this ensures that there is a path linking $v_1'$ to $v_2'$ in $\bual{K}.$ 

The following arguments show that $K'$ is a non-pinching non-singular cell complex of rank $R.$ In order to show that $K'$ is non-pinching, we first note that $\partial K' = \dual{\partial K}$ therefore there is no $\partial K'$-pinch in $K'$ since $\partial K$ is cell-connected. Let us assume by contradiction that there is a $K'$-pinch in $K',$ i.e. a cell $x' \in K'$ such that $\dual{K'}^{(1)} \cap \dual{x'}$ is not connected. If we take $x \in K $ such that $ \bual{x} = x',$ we can see that a path between two vertices $v,w$ in $x$ corresponds to a path between two vertices $\dual{\bual{v}}$ and $\dual{\bual{w}}$ in $\dual{K'}^{(1)} \cap \dual{x'}.$ Therefore if $\dual{K'}^{(1)} \cap \dual{x'}$ is not connected then so is $\kg \cap x,$ which contradicts the assumption that $K$ is cell-connected.  Every cell in $K'$ is included in a $R$-cell since this is clear for the cells in $\bual{K}$ and every $(R-1)$-cell $\dual{v} \in \dual{\partial K}$ is contained in $\bual{v} \in (K')^{[R]},$ which is then pure of dimension $R.$ Finally $K'$ is also non-singular since $K$ is graph-based.

We then turn to the point \ref{thmdualcobsptb}. Proving this statement amounts to verify that all the properties shown for the point \ref{thmdualcobspta} are still valid when restricted to cells not included in the sets $\bual{J_0} \sqcup \dual{J_0}.$ Axioms \ref{cccrank}, \ref{cccenough} and \ref{cccdiamond} are simply a consequence of the cc structure of $K'$ and the fact that whenever $x',y' \in K'_0$ are two cells such that $x' \subset y'$ and $z' \in K'$ satisfies $x' \subset z' \subset y'$ then $ z' \in K'_0.$ Axiom \ref{cccinter} is also verified for $K'_0$ since if $x',y' \in K'_0$ then $x' \cap y'$ contains no maximal cell of $J_0$ and therefore cannot be included in $\bual{J_0} \sqcup \dual{J_0}.$ We have that $K'_0$ is clearly graph-based since so is $K'$ and the set of edges of $K'_0$ is a subset of the set of edges of $K'.$ 
Also $K_0'$ is cell-connected since so is $K'.$ The purity of $K'_0$ is verified if we can show that for every cell $x'$ of $K'_0$ contained in a maximal cell of $K'$ of the form $\bual{v}$ where $v \in J_0^{[0]},$ we have that $x'$ is also contained in a maximal cell of the form $\bual{w},$ where $w \in \kz \setminus (J_0^{[0]} \sqcup J^{[0]}).$ This is ensured by the non-degeneracy of $(K,J_0 \sqcup J)$ together with the connectedness of the cells in $K_0'.$
The cc $K'_0$ is also non-singular since the set of $(R-1)$-cells of $K_0'$ is a subset of the set of $(R-1)$-cells of $K'$ and their set of cofaces in $K_0'$ is identical to their set of cofaces in $K'.$

Now comes a key point of this proof where the locality of $(K,J)$ enters crucially: showing that $K'_0$ is non-pinching. First we can see that there is no $K'_0$-pinch in $K_0'.$ Indeed $K'$ has no $K'$-pinch and $\dg{K'_0} \cap \bual{x'} = \dg{K'} \cap \bual{x'}$ for all $x' \in K'_0$ since if two maximal cells of $K'_0$ contain $x'$ and share a sub-maximal cell $y' \in K'$ then $y'$ is also in $K'_0.$ It is therefore sufficient to show that $\partial K'_0$ contains no $\partial K'_0$-pinch. For this we first prove the following relation:

\begin{equation} \label{claim6thmdual}
\partial K_0' = \dual{\partial K \setminus (J \sqcup J_0)} \sqcup \dual{K_J} \sqcup \dual{K_{J_0}}.
\end{equation}

In order to prove (\ref{claim6thmdual}) we first observe that $\dual{\partial K \setminus (J \sqcup J_0)} \leq \partial K_0'$ as a consequence of the following: every vertex $v \in (\partial K \setminus (J \sqcup J_0)^{[0]}$ satisfies that $y' := \cdual{v}{\partial K} \in (\partial K \setminus (J \sqcup J_0))^{[R-1]}$ and $\bual{v}$ is the only $R$-cell of $K'$ containing $y'.$ Since $(K, J \sqcup J_0)$ is non-singular, every $(R-1)$-cell of $K_0'$ not in $\dual{ \partial K \setminus (J \sqcup J_0)}$ is of the form $y' = \cdual{e}{K}$ for some $e \in \ko$ such that $\abs{e \cap (J \sqcup J_0)^{[0]}} \leq 1.$ Hence $\abs{\cface{y'}} = 1$ if and only if $e \in K_{(J \sqcup J_0)}$ and this proves (\ref{claim6thmdual}).

We can now turn back to showing that there is no $\partial K'_0$-pinch on $\partial K'_0.$ Since we assume that $K' \in \nsc,$ the relation (\ref{claim6thmdual}) implies that it is sufficient to prove that $\dual{K_{J_0}}$ is non-pinching. By Lemma \ref{lemmidsec}, we have that the map $ (M_{J_0}^K) \ni \E_{J_0}^x\mapsto x \in K_{J_0}$ is an isomorphism of posets. By Proposition \ref{propdual}, $(\dual{M_{J_0}^K}, \rk_{\dual{M_{J_0}^K}})$ is a closed cc and for all $x \in K_{J_0}$ we have
$$\rk_{\dual{M_{J_0}^K}}( \dual{\E_{J_0}^x} ) = R - 1 - \rk_{M_{J_0}^K}(\E_{J_0}^x) = R - \rk_K(x) = \rk_{\dual{K}}(\dual{x}) =: \rk_{\dual{K_{J_0}}}(\dual{x}).$$ 
Hence we have that $(\dual{K_{J_0}}, \rk_{\dual{K_{J_0}}})$ is a cc and $\dual{K_{J_0}} \cong \dual{M_{J_0}^K}.$ Therefore $\dual{K_{J_0}}$ is non-pinching since $M_{J_0}^K$ is cell-connected by locality of $(K - J).$
We can then conclude that the point \ref{thmdualcobpt1} is verified for all possible $J.$

The point \ref{thmdualcobpt2} is shown using (\ref{claim6thmdual}) (with $J_0 = \emptyset$) and observing that $J':= \dual{ \partial K \setminus J}$ is an element of $\mlc^{R-1}_\sqcup$ since it is the dual of a union of local non-pinching closed cell complexes. Finally $\dual{(K -J)}$ is exactly collared since, using the notation $L = \partial K \setminus J,$ the map 
$$ \{ \{y,z\} ~|~ L^{[R-1]} \ni y \subset z \in \kR\} \ni \{y,z\} \longmapsto y \in L^{[R-1]}$$
is a bijection and we have the isomorphism $\cong$ in the following derivation: 
\begin{align*}
M_{\dual{L}}^{\bual{K \setminus J} \sqcup \dual{L}} &= \{ \E_{\dual{L}}^{\bual{x}} ~|~ x \in L \}\\
 &= \{ \{ \dual{e} \in \bual{K}^{[1]}_{\dual{L}} ~|~ \dual{e} \subset \bual{x} \} ~|~ x \in L \} \\
 &= \{ \{ \{y,z\} ~|~ y \in L^{[R-1]}, ~ z \in \kR_L ~:~ x \subset y \subset z \} ~|~ x \in L \} \\
 &\cong \{ \cdual{x}{L} ~|~ x \in L \} = \dual{L}.
\end{align*}
\end{proof}

\end{appendices}

\newpage
\paragraph*{Acknowledgements} The author was supported by the ERC SG CONSTAMIS.

\addcontentsline{toc}{section}{References}

\bibliographystyle{alpha}
\bibliography{biblio}

\end{document}